\documentclass[11pt]{article}
\pdfoutput = 1

\usepackage[utf8]{inputenc}
\usepackage{color,graphicx}
\usepackage{epsfig}
\usepackage{amsmath}
\usepackage{verbatim}
\usepackage{amssymb}
\usepackage{mathabx}
\usepackage[mathcal]{eucal}
\usepackage{stmaryrd}
\usepackage{esint}
\usepackage{physics}
\usepackage{amsfonts}
\usepackage{cite}
\usepackage{array}
\usepackage{setspace}
\usepackage{braket}
\usepackage{float}
\usepackage{url}
\usepackage{mathtools}
\usepackage{tensor}
\usepackage{bbold}
\usepackage{slashed}
\usepackage{tikz}
\usepackage{graphicx}
\usepackage{epstopdf}
\usepackage{subfigure}
\usepackage[labelfont=bf,font={sf}]{caption}
\usepackage{pgfplots}
\usepackage{mathrsfs}
\usepackage{fancybox}
\usepackage{eurosym}
\usepackage{tcolorbox}
\usepackage{tensor}
\allowdisplaybreaks

\usepackage[margin = 2.2cm]{geometry}
\usepackage[ragged]{footmisc}
\usepackage[bookmarks=true,bookmarksnumbered=false,
hyperindex=true,bookmarksopen=true,hyperfigures=true,
colorlinks=true,linkcolor=webblue,citecolor=webgreen,urlcolor=webblue,breaklinks]{hyperref}
\definecolor{webgreen}{rgb}{0, 0.5, 0}
\definecolor{webblue}{rgb}{0, 0, 0.5}
\definecolor{webred}{rgb}{0.5, 0, 0}
\definecolor{darkgreen}{rgb}{0,0.5,0}

\newcommand{\+}{\scalebox{.4}{$+$}} 
\newcommand{\mm}{\scalebox{.4}{$-$}} 
\newcommand{\pmt}{\scalebox{.4}{$\pm$}}

\newcommand{\Uqsu}{\text{SU}_q(1,1)}
\newcommand{\Uqsual}{U_q(\mathfrak{su}(1,1))}
\newcommand{\Uqsl}{U_q(\mathfrak{sl}(2,\mathbb{R}))}

\newcommand{\dpi}{\mathcal{D}}

\def\ben{\begin{equation}}
\def\een{\end{equation}}

 \let\b=\beta \let\g=\gamma \let\d=\delta 
  \let\q=\theta 
    \let\p=\phi \let\r=v

\def\be{\begin{equation}}
\def\ee{\end{equation}}
\def\ba{\begin{array}}
\def\ea{\end{array}}

\def\dalemb#1#2{{\vbox{\hrule height .#2pt
       \hbox{\vrule width.#2pt height#1pt \kern#1pt
               \vrule width.#2pt}
       \hrule height.#2pt}}}

\newcommand{\bea}{\begin{eqnarray}}
\newcommand{\eea}{\end{eqnarray}}

\let\tilde=\widetilde

\renewcommand{\d}{\mathrm{d}}
\renewcommand{\i}{\mathrm{i}}

\numberwithin{equation}{section}

\begin{document}

\thispagestyle{empty}
\begin{center}
    ~\vspace{5mm}

     {\LARGE \bf 
    
   Dynamical actions and q-representation theory}
   
   \vspace{0.2cm}
   
   {\LARGE \bf for double-scaled SYK}
    
   \vspace{0.4in}
    
    {\bf Andreas Blommaert$^1$, Thomas G. Mertens$^2$, Shunyu Yao$^3$}

    \vspace{0.4in}
    {$^1$SISSA and INFN, Via Bonomea 265, 34127 Trieste, Italy\\
    $^2$Department of Physics and Astronomy\\
Ghent University, Krijgslaan, 281-S9, 9000 Gent, Belgium\\
    $^3$Department of Physics, Stanford University, Stanford, CA 94305, USA}
    \vspace{0.1in}
    
    {\tt ablommae@sissa.it, thomas.mertens@ugent.be, shunyu.yao.physics@gmail.com}
\end{center}

\vspace{0.4in}

\begin{abstract}
\noindent We show that DSSYK amplitudes are reproduced by considering the quantum mechanics of a constrained particle on the quantum group SU$_q(1,1)$. We construct its left-and right-regular representations, and show that the representation matrices reproduce two-sided wavefunctions and correlation functions of DSSYK. We then construct a dynamical action and path integral for a particle on SU$_q(1,1)$, whose quantization reproduces the aforementioned representation theory. By imposing boundary conditions or constraining the system we find the $q$-analog of the Schwarzian and Liouville boundary path integral descriptions. This lays the technical groundwork for identifying the gravitational bulk description of DSSYK. We find evidence the theory in question is a sine dilaton gravity, which interestingly is capable of describing both AdS and dS quantum gravity.

\end{abstract}

\pagebreak
\setcounter{page}{1}
\tableofcontents

\newpage
\section{Introduction}\label{sect:intro}
The SYK model \cite{Sachdev:1992fk,kitaev2015simple} has attracted enormous attention in the field ever since the realization \cite{kitaev2015simple,Maldacena:2016hyu} that it is arguably the simplest quantum mechanical system with a gravitational dual. In particular, the SYK model is a quantum mechanical system of $N$ fermions with a $p$-local Hamiltonian with (Gaussian) random interactions $J$\footnote{Our convention for $q$ differs from the one used for instance in \cite{Lin:2022rbf} by $q^2=q_\text{there}$, we follow the group theory literature \cite{Klimyk:1997eb}.}
\begin{equation}
    H=\i^{p/2}\sum_{i_1\dots i_p}J_{i_1\dots i_p}\psi_{i_1}\dots \psi_{i_p}\,,\quad -\log q=\frac{p^2}{N}\,,\quad q<1\,.\label{1.1}
\end{equation}
For fixed finite $p$, the low-energy physics is governed by Schwarzian quantum mechanics \cite{kitaev2015simple,Maldacena:2016hyu}, which is in turn gravitationally dual to JT gravity \cite{Jackiw:1984je,Teitelboim:1983ux,Maldacena:2016upp,Engelsoy:2016xyb,Jensen:2016pah} upon introducing near AdS$_2$ boundary conditions.\footnote{For a recent review on JT gravity, see e.g.\cite{Mertens:2022irh}.} This duality has taught us a lot about quantum black holes and quantum gravity in AdS$_2$. However, it also has its shortcomings, one of the most notable ones being that JT gravity is not UV complete (since it describes only the IR limit of SYK).
The holographic bulk dual of the full SYK model is unknown at this time, and remains one of the big unsolved problems in this field. Some proposals have been made, see for instance \cite{Gross:2017hcz,Das:2017pif,Das:2017hrt,Goel:2021wim}, but no unified and accepted conclusion has been reached.

Fortunately, besides the low energy limit of SYK, there are other interesting limits. In particular one interesting regime called double-scaled SYK \cite{Cotler:2016fpe,Berkooz:2018jqr,Berkooz:2018qkz} (or DSSYK for short) is obtained by sending $p\to\infty$ whilst simultaneously sending $N\to\infty$ with $q$ ($0<q<1$) defined in \eqref{1.1} held fixed. The JT or Schwarzian regime is recovered by afterwards sending $q\to 1^-$ whilst zooming in on low energies. Using Hamiltonian methods all the amplitudes of DSSYK have been calculated \cite{Berkooz:2018jqr,Berkooz:2018qkz}. These share a remarkable degree of structural similarity with the amplitudes computed for JT gravity \cite{Mertens:2017mtv,Yang:2018gdb,Mertens:2018fds,Kitaev:2018wpr,Blommaert:2018oro,Iliesiu:2019xuh,Saad:2019pqd}. We review both results in \textbf{section \ref{sect:review}}.

This begs the question whether perhaps DSSYK also has an understandable bulk description, some type of minimal UV completion of JT gravity which captures some of the more stringy features of the full SYK model \cite{Maldacena:2016hyu}. It has been pointed out in recent work \cite{Lin:2022rbf} that the answer - at least to some degree - is yes.\footnote{For other interesting recent work involving DSSYK and chords see for instance \cite{Jafferis:2022wez,Susskind:2022bia,Bhattacharjee:2022ave,Okuyama:2022szh,Susskind:2023hnj,Mukhametzhanov:2023tcg,Berkooz:2023cqc,Okuyama:2023bch,Lin:2022nss,Susskind:2021esx,Berkooz:2020xne,Berkooz:2022mfk,Goel:2023svz}.}

The quantum amplitudes calculated in \cite{Berkooz:2018jqr,Berkooz:2018qkz} use so-called ``chord diagrams'',\footnote{We will not review chord diagrams here, some excellent reviews exist already (see for instance \cite{Lin:2022rbf}), moreover chords will not play a major role in our story.} where the emerging bulk picture that appears is that quantum mechanically one can think of the states of the bulk Hilbert space (on any slice between two asymptotic boundaries) as states with a fixed number of chords $n$. As such, quantum mechanically one can roughly think of spacetime as being ``discretized'' at a fundamental level.

Although this chord picture is attractive, it still leaves us slightly unsatisfied. Namely, chords are a highly quantum mechanical picture, and much of our insights and intuition in physics arises from quantizing some classical dynamical system. The goal of our work is to build towards a bulk description of this kind.

Classically JT gravity has a first-order formulation as an SL$(2,\mathbb{R})$ BF theory, derived in the early literature in \cite{Fukuyama:1985gg, Isler:1989hq, Chamseddine:1989yz, Jackiw:1992bw}, and more recently in e.g. \cite{Grumiller:2017qao,Gonzalez:2018enk}. At the quantum level, the relation holds up to subtleties related to the precise choice of algebraic structure \cite{Blommaert:2018iqz,Blommaert:2018oro,Mertens:2018fds,Iliesiu:2019xuh,Saad:2019lba}.\footnote{These subtleties will not be important for our main story as they are mainly there for JT and Liouville gravity, whereas DSSYK seems to avoid them automatically.} The SL$(2,\mathbb{R})$ symmetry allows one to derive WdW wavefunctions and various correlation functions, via representation theory techniques \cite{Blommaert:2018oro}.  BF theory is topological - it can be reduced entirely to boundary dynamics. As we quickly review in \textbf{section \ref{sect:review}}, for JT gravity this boundary dynamics is a constrained particle on an SL$(2,\mathbb{R})$ group manifold, where the constraints arise by imposing asymptotic (nearly) AdS$_2$ boundary conditions \cite{Maldacena:2016upp,Engelsoy:2016xyb,Jensen:2016pah}. These effectively reduce the dynamics to Schwarzian quantum mechanics.

In this work, we study some aspects of the representation theory of the quantum group SU$_q(1,1)$. We reproduce the two-sided wavefunction and correlation functions of DSSYK by implementing the correct boundary conditions on the constructed representation matrices. We clarify the relevant techniques in \textbf{section \ref{Sec:rep}}. 

We then derive a path integral description (involving in particular a classical action) for a particle traveling on the SU$_q(1,1)$ (quantum) group manifold, which we supplement with constraints generalizing the asymptotic AdS$_2$ boundary conditions. The resulting restricted boundary action is dubbed the \textbf{q-Schwarzian} and derived in \textbf{section \ref{sect:pathintegrals}}. By construction, quantization of this system reproduces DSSYK amplitudes (representation theory of SU$_q(1,1)$).

This description should enable one to reverse engineer a topological bulk theory akin to BF gauge theory, whose boundary dynamics is precisely the $q$-Schwarzian. Such a model would then be the first-order formulation of the bulk dual to DSSYK. Building in part on \cite{Goel:2022pcu,HVerlindetalk}, we propose in \textbf{section \ref{sect:discbulkdual}} that the bulk model in question is a particular ``Poisson sigma model'' \cite{Cattaneo:2001bp}, which can be rewritten as a 2d \textbf{dilaton gravity} \cite{Ikeda:1993aj,Ikeda:1993fh} with potential
\begin{equation}
    V(\Phi)=\frac{\sin(2\log q \Phi)}{\log q}\,,\label{1.2}
\end{equation}
which results in classical solutions with dS$_2$ and AdS$_2$ regions. It would be interesting to proceed along this lines to check the relation between dS$_2$ quantum gravity and DSSYK \cite{Susskind:2021esx,Susskind:2022bia,Lin:2022nss,Rahman:2022jsf,Susskind:2022dfz}.\footnote{Or perhaps we should call it dSSYK?} We will present more details on this bulk description elsewhere \cite{wip}, see also \cite{Blommaert:2023wad} for the related Liouville gravity picture.

This \textbf{web of dualities} is summarized below diagrammatically in \eqref{fig:JTduality}, \eqref{fig:DSSYKduality}, \eqref{fig:JTduality2} and \eqref{fig:DSSYKduality2}.

\subsection*{Summary and structure}

The remainder of this work is structured as follows.

In \textbf{section \ref{sect:review}} we review and rephrase the structure of amplitudes in JT gravity that follow from the SL$(2,\mathbb{R})$ BF description, and review how DSSYK amplitudes have a similar structure \cite{Berkooz:2018jqr}.

In \textbf{section \ref{Sec:rep}} - the technical heart of this work - we construct the right- and left-regular realization of the quantum group SU$_q(1,1)$ and use this to compute observables in the Hamiltonian description of a quantum particle on SU$_q(1,1)$. We compare our construction to the one that appeared in \cite{Berkooz:2022mfk}. We then constrain this system, and show that the amplitudes of the resulting constrained quantum mechanics system are equivalent to those of DSSYK.

In \textbf{section \ref{sect:pathintegrals}} we construct a continuum dynamical system describing a particle on SU$_q(1,1)$, which gives rise to the aforementioned Hamiltonian system upon using canonical quantization, and impose the constraints \eqref{constraints} which lead to a $q$-Schwarzian phase space path integral \eqref{4.30} for the case with one asymptotic boundary, and to a $q$-Liouville phase space path integral 
for the case of a Cauchy slice with two asymptotic boundaries:
\begin{equation}
    \int\dpi \phi \dpi p_\phi\exp\bigg(\i\int\d t\bigg(p_\phi \,\phi' +(1- e^{-2\phi})\,\frac{e^{-\i\log q\,p_\phi}}{(\log q)^2}+\frac{e^{\i\log q\,p_\phi}}{(\log q)^2}\bigg)\bigg)\,.  \label{1.3}
\end{equation}

In the concluding \textbf{section \ref{s:conclu}} we propose a (topological) gravitational bulk dual that is equivalent to these boundary theories, and discuss potential generalizations.

We provide more mathematical details on the quantum group $\Uqsu$ and its relation to $\Uqsual$ in \textbf{Appendix \ref{app:gauss}}.

Our main technical results are the following:
\begin{itemize}
    \item The (quantum) group manifold SU$_q(1,1)$ is a non-commutative three-dimensional space with Gauss coordinates $(\g,\phi,\b)$. The non-commutativity arises from quantization; the fields $(\g,\phi,\b)$ in the path integral are ordinary classical fields. In fact, via a simple coordinate transformation in phase space, one can obtain an equivalent quantum mechanical description with commutative coordinates.
   Upon constraining the system one finds that $\g$ and $\b$ are redundant, hence one can gauge-fix $\g=\beta=0$, resulting in the two-dimensional phase space path integral \eqref{1.3}. 
   \item The Laplacian on this quantum group manifold can be written down explicitly as the Casimir in the regular representation of the quantum algebra. It is physically interpreted as the Hamiltonian (time-independent) Schr\"odinger operator whose corresponding classical action describes a particle moving on the quantum group. Importantly, this Hamiltonian is non-Hermitian. Upon implementing the boundary conditions, the Hamiltonian becomes the (non-Hermitian) Wheeler-de Witt Hamiltonian describing evolution of two-boundary (or wormhole) gravitational wavefunctions.
    \item The chord wavefunctions \cite{Lin:2022rbf, Berkooz:2018jqr,Berkooz:2018qkz} are representation matrix elements of SU$_q(1,1)$
    \begin{equation}
        \braket{n\rvert \theta}=R_{\theta\, \i\i}(n)=\frac{q^n}{(q^2;q^2)_n}H_n(\cos(\theta)\rvert q^2)\,,
    \end{equation}
    determined as eigenfunctions of the Casimir operator $\mathbf{H}$, 
    where the gravitational boundary conditions \eqref{constraints} determine the ``row-and-column'' labels $\i$ of the matrix elements \eqref{gravmatintinter}.\footnote{We follow the notation of \cite{Blommaert:2018iqz} for gravitational states.} Also, the natural bilocal operator insertions in DSSYK correspond with the following representation matrix elements of SU$_q(1,1)$ \eqref{3.76}
    \begin{equation}
        \mathcal{O}_\Delta=R_{\Delta\, 00}(n)=q^{2 n \Delta}\,,
    \end{equation}
    in analogy with the JT gravity \cite{Blommaert:2018oro} and the Liouville gravity story \cite{Fan:2021bwt} (equation (2.48)).
    \item Our choice of Haar measure on SU$_q(1,1)$, upon gauge-fixing to $\beta=\gamma=0$, fixes the inner product on wavefunctions 
     \begin{equation}
        \braket{\theta_1\rvert \theta_2}=\sum_{n=0}^{\infty}q^{-2 n}\braket{\theta_1\rvert n }\braket{n\rvert \theta_2},
    \end{equation}
    which reproduces the correct DSSYK amplitudes \eqref{ipfinal}. Here the left eigenfunctions \cite{yao2018edge} of the Hamiltonian $\braket{\theta_1\rvert n }$ differ from the right eigenfunctions by en important factor $(q^2;q^2)_n$. This inner product has the important feature that it samples the wavefunction at discretely separated points, which avoids infinite degeneracies in the physical Hilbert space and effectively discretizes the chord number $n$. This discretization is hence a quantum effect.
\end{itemize}

Our current understanding of the dual descriptions of DSSYK in this language can be summarized by the following picture for the two-sided model:
\begin{equation}
    \begin{tikzpicture}[baseline={([yshift=-.5ex]current bounding box.center)}, scale=0.7]
 \pgftext{\includegraphics[scale=1]{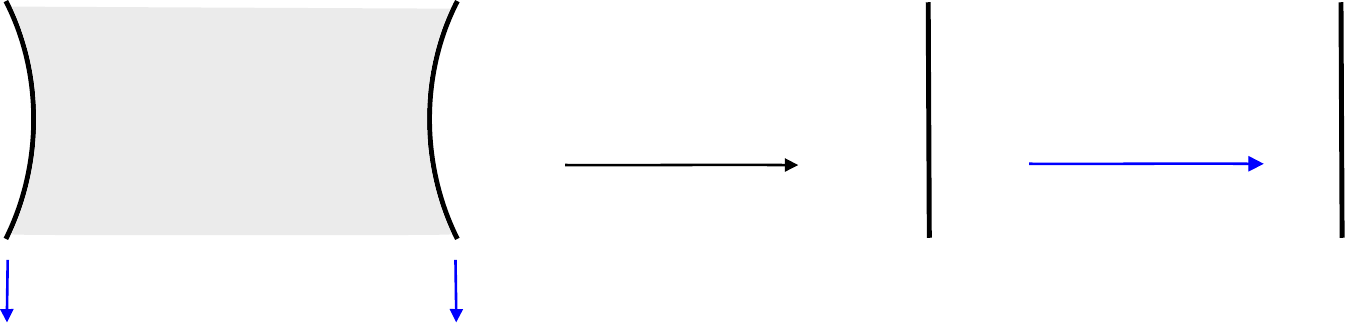}} at (0,0);
    \draw (8.4,2.2) node {$q$-Liouville \eqref{q-liouville}};
    \draw (2.25,2.2) node {particle on SU$_q(1,1)$ \eqref{pathintegralbeforeconstraints}};
    \draw (0.35,0.8) node {$g=$ ?};
    \draw (0,-0.8) node {holography};
    \draw (4.6,-0.8) node {\color{blue}$q^{-L_H} L_F$ fixed};
        \draw (4.7,-1.5) node {\color{blue}$q^{R_H} R_E$ fixed};
    \draw (-4.4,-2.2) node{\color{blue}bdy conditions ?};
    \draw (-4.5,1) node {PS model \eqref{PSMbos}};
    \draw (-4.3,-0.3) node {$V(\Phi)$ \eqref{1.2}};  \end{tikzpicture}\label{fig:DSSYKduality2}
\end{equation}
and the one-sided model:
\begin{equation}
    \begin{tikzpicture}[baseline={([yshift=-.5ex]current bounding box.center)}, scale=0.7]
 \pgftext{\includegraphics[scale=1]{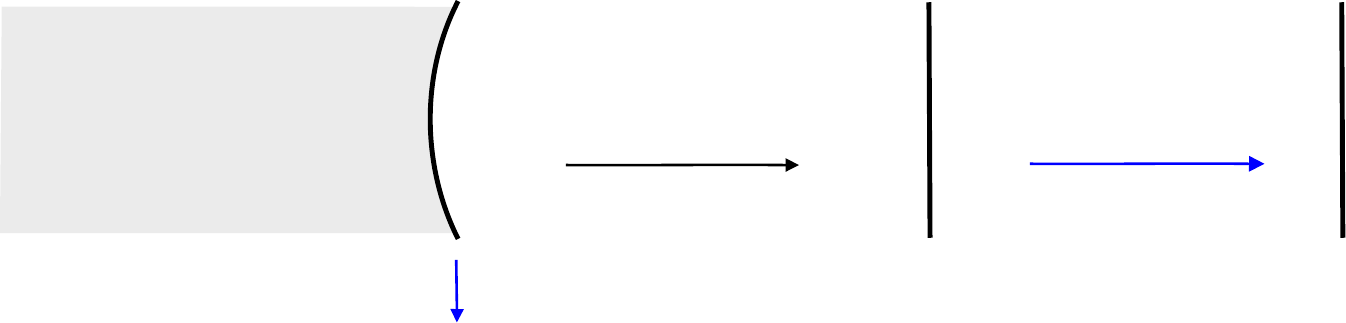}} at (0,0);
     \draw (8.4,2.2) node {$q$-Schwarzian \eqref{4.30}};
     \draw (2.25,2.2) node {particle on SU$_q(1,1)$ \eqref{pathintegralbeforeconstraints}};
   \draw (0.35,0.8) node {$g=$ ?};
    \draw (0,-0.8) node {holography};
    \draw (4.7,-0.8) node {\color{blue}$q^{-L_H} L_F$ fixed};
    \draw (-3,-2.2) node{\color{blue}bdy condition ?};
    \draw (-4.5,1) node {PS model \eqref{PSMbos}};
    \draw (-4.3,-0.3) node {$V(\Phi)$ \eqref{1.2}};  
  \end{tikzpicture}\label{fig:DSSYKduality}
\end{equation}
The leftmost part of this diagram (the holographic bulk) and how to transfer in detail to the boundary description (the middle of the diagram) will not be addressed here, and is not fully developed yet. We make some comments in section \ref{sect:discbulkdual}, but postpone a deeper study to future work \cite{wip}.

\section{Background material}\label{sect:review}
We start by reviewing and rephrasing some material necessary to follow our logic in the main sections \ref{Sec:rep} and \ref{sect:pathintegrals}. We first discuss aspects of the SL$(2,\mathbb{R})$ BF description of JT gravity \cite{Blommaert:2018iqz,Blommaert:2018oro, Mertens:2018fds,Iliesiu:2019xuh,Saad:2019lba,Jackiw:1992bw,Grumiller:2017qao,Gonzalez:2018enk}.

\subsection{Representation theory for JT gravity}\label{sect:JTreview}
One can rewrite the Euclidean JT action (we'll set $8\pi G_N = 1$) in first-order variables \cite{Fukuyama:1985gg, Isler:1989hq, Chamseddine:1989yz} including the boundary term \cite{Mertens:2018fds} as
\begin{equation}
    I = -\frac{1}{2}\int_{ \mathcal{M}} \d^2 x\sqrt{g}\, \Phi(R+2)-\int_{\partial \mathcal{M}}\d u \sqrt{h}\, \Phi K \quad \sim \quad -\i\int_{ \mathcal{M}} \Tr(\chi F)+\frac{\i}{2}\int_{\partial \mathcal{M}} \Tr(\chi A)\,,\label{2.1}
\end{equation}
with $A$ and $\chi$ valued in the $\mathfrak{sl}(2,\mathbb{R})$ algebra:
\begin{equation}
    E=\begin{pmatrix}
        0&1\\0&0
    \end{pmatrix}\,,\quad F=\begin{pmatrix}
        0&0\\1&0
    \end{pmatrix}\,,\quad H=\frac{1}{2}\begin{pmatrix}
        1&0\\0&-1
    \end{pmatrix}\,,\label{eq:slal}
\end{equation}
as $A=A_E F+A_F E+A_H 2 H$ and $\chi=\chi_E F+\chi_F E+\chi_H 2 H$ and with the trace in the 2d representation. The boundary conditions equate $\chi$ and $A$, and additionally constrain one component of both fields:
\begin{align}
    \chi+\i A_u \vert_{\partial \mathcal{M}}=0\,, \label{bc}\\
    \i \chi_F \vert_{\partial \mathcal{M}}=\frac{1}{\varepsilon}\,,\label{bc2}
\end{align}
gravitationally corresponding to fixing the total boundary length $\ell=\beta/\varepsilon$ in terms of the renormalized length $\beta$, and the dilaton asymptotics $\Phi \vert_{\partial \mathcal{M}}=1/2\varepsilon$. 
For the second boundary condition \eqref{bc2}, the precise value we choose for $\chi_F$ at the boundary is somewhat arbitrary, as long as it is non-zero. This can be appreciated as follows. As in the earlier literature \cite{Coussaert:1995zp}, one can perform a gauge transformation in the bulk BF model which has support on the boundary, changing $\chi \to b^{-1}\chi b$ and $A \to b^{-1} A b$ with $b = e^{\log a H}$. This causes $\chi_F$'s value at the boundary to multiply with $a$. This is usually done in the aAdS context to rescale the boundary condition to a natural value, e.g. $\i \chi_F \vert_{\partial \mathcal{M}}=1$. Here though, for comparison with DSSYK later on, it is more insightful to not do this and leave it as \eqref{bc2}.

One proceeds by path integrating out $\chi$ in the bulk BF action of \eqref{2.1}. This localizes the model to flat SL$(2,\mathbb{R})$ connections $A=\d g g^{-1}$. Plugging this back into \eqref{2.1}, and using only the first boundary condition in \eqref{bc}, the model reduces to the dynamics of a non-relativistic particle moving on the group manifold:
\begin{equation}
I = \frac{1}{2} \int \d u\, \text{Tr}(g^{-1}\partial_u g)^2,
\end{equation}

Now consider the specific case of a manifold $\mathcal{M}$ that is (locally) a strip with two timelike boundaries. The boundary action contains naively two pieces, on each part of the boundary:
\begin{equation}
I = \frac{1}{2} \int \d u \text{Tr}(g_1^{-1}\partial_u g_1)^2 + \frac{1}{2} \int \d u \text{Tr}(g_2^{-1}\partial_u g_2)^2,
\end{equation}
These are however not totally independent due to a common zero-mode. The more precise derivation leads to a path integral
\begin{equation}
\int \d\lambda\int_{\text{mod }G} \dpi g_1 \dpi g_2 \, e^{-I_\lambda[g_1] - I_\lambda[g_2]},
\end{equation}
where one mods out constant common transformations $g_{1,2} \sim g_{1,2}h, \, h \in G$. This represents two twisted versions of $I$, with an integral over the twist/defect parameter $\lambda$, which we will not specify further (for details we refer e.g. to appendix C of \cite{Blommaert:2018oro}).
Using path integral manipulations, one can rewrite this as a single particle on a group model as\footnote{To see this \cite{Coussaert:1995zp}, one uses the field redefinition $(g_1,g_2) \to (g,\pi_g)$ with $g = g_2g_1^{-1}$ and $\pi_g = g_1g_2^{-1}\partial_u g_2 g_1^{-1} + \partial_u g_1 g_1^{-1}$, after which the $\pi_g$ path integral decouples and is Gaussian. The remaining action is again the particle on a group action for $g$. One way to quickly appreciate this procedure is to realize that this is the dimensional reduction of the well-known argument on how to transfer from two chiral WZW models into a single non-chiral WZW model. The dimensional reduction of chiral and non-chiral WZW model both yield the particle on a group system.} 
\begin{equation}
\int \d\lambda\int_{\text{mod }G} \dpi g_1 \dpi g_2 \, e^{-I_\lambda[g_1] - I_\lambda[g_2]}\,\sim\, \int \dpi g \, e^{-I[g]}\,,\quad g=g_2\, g_1^{-1}\,.
\end{equation}

As a phase space integral, and using the explicit coordinatization \eqref{gausssclassical}, we can write
\begin{align}
    &\int \dpi g\exp\bigg(-\frac{1}{2}\int\d u\,\Tr( g^{-1}\partial_t g)^2\,\bigg)\nonumber\\&\qquad =
    \int \dpi \g \dpi \phi \dpi \b \dpi p_\g \dpi p_\phi \dpi p_\b\exp\bigg(\i\int \d t\,\bigg(p_\phi\,\phi'+p_\b\, \b'+p_\g\, \g'-\frac{p_\phi^2}{4}-p_\g p_\b e^{-2\phi}\bigg)\bigg)\,.\label{2.4}
\end{align}
This system has a Hamiltonian that is equal to the Laplacian on the group manifold SL$(2,\mathbb{R})$, which is mathematically given by the quadratic Casimir $L_C$ in the regular representation. The latter is diagonalized by the irrep matrix elements, which hence form the physical wavefunctions of the model. Let us work this out more explicitly. A SL$(2,\mathbb{R})$ group element can be parameterized by the Gauss decomposition with coordinates ($\g,\phi,\beta$):
\begin{equation}
    g=e^{\g F}e^{2\phi H}e^{\beta E}\,,\label{gausssclassical}
\end{equation}
In these coordinates, one can write down the left-regular representation of first-order differential operators:
\begin{align}
    L_F&=-\partial_\g\,,\quad L_H=-\frac{1}{2}\partial_\phi+\g\partial_\g\,,\quad L_E=\g^2\partial_\g-\g\partial_\phi-e^{-2\phi}\partial_\b\,.\label{2.3}
\end{align}
and analogously the right-regular realization which we will not write down explicitly. The quadratic Casimir in this representation is a quadratic differential operator:
\begin{equation}
L_C = -\frac{1}{4} \partial_\phi^2 - e^{-2\phi}\partial_\beta\partial_\gamma.
\end{equation}

Next we implement the asymptotic boundary condition \eqref{bc2}. Since both $g_2$ and $g_1$ are subject to this constraint, one finds that $g$ is subject to two constraints (left transformations on $g_1$ become right transformations on $g$):
\begin{equation}
    L_F=-\frac{\i}{\varepsilon}=-ip_\gamma,\quad R_E=\frac{\i}{\varepsilon}=\i\, p_\b\,.\label{2.10}
\end{equation}
Let us first circle back to the argument that one can rescale the precise value of the eigenvalue. From the Gauss decomposition \eqref{gausssclassical}, we can derive the mathematical identity:
\begin{equation}
\label{eq:shift}
g(\g,\phi + \log a , \beta) \equiv  e^{\g F}e^{2(\phi +\log a)H}e^{\beta E} = b \,\, e^{\g F a }e^{2\phi H}e^{\beta E a} \,\, b,
\end{equation}
where $b= e^{\log a H }$ as before. Now, changing $g \to b^{-1} g b^{-1}$ induces a gauge transformation on $A$ because $A= dg g^{-1} \to b^{-1} A  b$. This means that rescaling the eigenvalues of $E$ and $F$ by $a$ is gauge-equivalent to changing the origin of the $\phi$-coordinate on the group manifold. This will be explicit in the equations that follow.

We can write down a constrained version of the action \eqref{2.4} by plugging the asymptotic constraints \eqref{2.10} into \eqref{2.4}. From a quantum mechanics perspective, these are first-class constraints (their Poisson bracket vanishes) \cite{dirac2001lectures}. The path integral description \eqref{2.4} then reduces to
\begin{equation}
\int \dpi \phi \dpi p_\phi \exp\bigg(\i\int \d t\,\bigg(p_\phi\,\phi'-\frac{p_\phi^2}{4}-\frac{1}{\varepsilon^2}e^{-2\phi}\bigg)\bigg)=\int \dpi \phi \exp\bigg(\i\int \d t\,\bigg(\phi'^2-\,\frac{1}{\varepsilon^2}e^{-2\phi}\bigg)\bigg)\,,\label{liouvilleaction}
\end{equation}
where both $\beta$ and $\gamma$ have become redundant fields, and can be gauge-fixed to $\beta=\gamma=0$. This last equation is the Liouville quantum mechanics description of the Schwarzian model (see for instance \cite{Bagrets:2016cdf,Harlow:2018tqv}). Liouville QM has the famous KPZ shift property where we can define $\phi_{\text{r}} \equiv \phi + \log \varepsilon$ to absorb the $1/\varepsilon^2$ prefactor from the potential term, confirming the argument above in eq. \eqref{eq:shift}.
The physical wavefunctions of this constrained quantum mechanical system are thus the 
Hamiltonian eigenfunctions:
\begin{equation}
\left(-\frac{1}{4}\partial_\phi^2 + \frac{1}{\varepsilon^2}e^{-2\phi}\right) \braket{\phi\rvert E} = E \braket{\phi\rvert E},
\end{equation}
which are constrained representation matrices of SL$(2,\mathbb{R})$, computed explicitly in \cite{Blommaert:2018oro}:\footnote{In representation theory one usually computes inner products with the (reduced) Haar measure $e^{2\phi}\d\phi$, whereas in the current description the measure is naturally flat $\d\phi$. This causes an additional prefactor $e^{-\phi}$ in the representation matrix elements compared to \eqref{2.11}.}
\begin{equation}
    \braket{\phi\rvert E}=R_{E\,\i\i}(\phi) = K_{2 \i E^{1/2}}(2e^{-\phi}/\varepsilon^2) = K_{2 \i E^{1/2}}(2e^{-\phi_{\text{r}}})
    \,.\label{2.11}
\end{equation}
The spectral density then follows from the orthogonality relation of the wavefunctions (up to constant prefactors)
\begin{equation}
    \int_{-\infty}^\infty \d\phi \braket{E_1\rvert \phi}\braket{\phi\rvert E_2}=\frac{\delta(E_1-E_2)}{\rho(E_1)}\,,\quad \rho(E_1)=\frac{1}{\Gamma(\pm 2\i E^{1/2})}\,,\label{dosJT}
\end{equation}
These wavefunctions of the ``boundary theory'' in fact have a direct bulk interpretation as wavefunctions of the gauge-invariant data: the line integral of $A$ along a spatial slice (evaluated on the flat connection $A=\d g g^{-1}$):
\begin{equation}
    g= g_2g_1^{-1} = \mathcal{P}\exp \int_{x_1}^{x_2} A\,.\label{map}
\end{equation}
Both labels of the representation matrix are constrained to $\i$ simply because we impose the boundary conditions \eqref{bc2} at both boundaries $x_1$ and $x_2$.

Natural operator insertions in this model \eqref{liouvilleaction} have zero charges (so as not to disrupt the boundary conditions \cite{Blommaert:2018oro}):
\begin{equation}
    \label{eq:opgr}
    \mathcal{O}_\Delta=R_{\Delta\,00}(\phi)=e^{-2\Delta\phi}\,.
\end{equation}
In the BF language, these correspond to a boundary-anchored Wilson line. Operator matrix elements are readily explicitly computed as \cite{Blommaert:2018oro}
\begin{equation}
    \bra{E_1}\mathcal{O}_\Delta \ket{E_1}=\int_{-\infty}^\infty\d\phi \braket{E_1\rvert \phi}e^{-2\Delta\phi}\braket{\phi\rvert E_2}=\varepsilon^{2\Delta}\,\frac{\Gamma(\Delta\pm\i E_1^{1/2}\pm \i E_2^{1/2})}{8\Gamma(2\Delta)}\,. \label{matintJT}
\end{equation}
The above relation between JT on an interval (times time) and Liouville quantum mechanics \eqref{liouvilleaction} can be summarized pictorially as
\begin{equation}
    \begin{tikzpicture}[baseline={([yshift=-.5ex]current bounding box.center)}, scale=0.7]
 \pgftext{\includegraphics[scale=1]{dssyk2v2.pdf}} at (0,0);
    \draw (8.4,2.2) node {Liouville QM \eqref{liouvilleaction}};
    \draw (2.25,2.2) node {particle on SL$(2,\mathbb{R})$ \eqref{2.4}};
    \draw (0.35,0.8) node {$g=\mathcal{P}\exp\int_{x_1}^{x_2} A$};
    \draw (0,-0.8) node {holography};
    \draw (4.7,-0.8) node {\color{blue}$L_F,R_E$ fixed};
    \draw (-4.4,-2.2) node{\color{blue}asymptotically AdS bdy};
    \draw (-4.3,1) node {SL$(2,\mathbb{R})$ BF};
    \draw (-4.3,-0.3) node {JT};
  \end{tikzpicture}\label{fig:JTduality2}
\end{equation}

One can also write down a single-sided path integral. The boundary condition \eqref{bc2} then imposes only $L_F=-\frac{\i}{\varepsilon}=-ip_\gamma$. The phase space path integral \eqref{2.4} becomes
\begin{equation}\label{sch4d}
     \int \dpi \b \dpi \phi \dpi p_\b \dpi p_\phi \exp\bigg(\i\int \d t\,\bigg(p_\phi\,\phi'+p_\b\, \b'-\frac{p_\phi^2}{4}-\frac{1}{\varepsilon}\,  p_\b e^{-2\phi}\bigg)\bigg)\,.
\end{equation}
This is the Schwarzian phase space path integral \cite{Engelsoy:2016xyb}. Indeed, integrating out first $p_\phi$, and then $p_\b$, one recovers the more familiar Schwarzian path integral
\begin{equation}
    \int\dpi\b\dpi\phi\,\delta\bigg(\b'-\frac{1}{\varepsilon}e^{-2\phi}\bigg)\exp\bigg(\i\int\d t\,\phi'^2\bigg)=\int \frac{\dpi \b}{\b'}\exp\bigg(\i\int\d t\,\frac{1}{4}\bigg(\frac{\b''}{\b'}\bigg)^2\,\bigg)\,,
\end{equation}
in terms of the dynamical time reparametrization $\beta(t)$ of the wiggly boundary curve. This relation between JT and Schwarzian quantum mechanics can be summarized as follows:
\begin{equation}
    \begin{tikzpicture}[baseline={([yshift=-.5ex]current bounding box.center)}, scale=0.7]
 \pgftext{\includegraphics[scale=1]{dssyk1v2.pdf}} at (0,0);

    \draw (8.4,2.2) node {Schwarzian QM \eqref{sch4d}};
    \draw (2.25,2.2) node {particle on SL$(2,\mathbb{R})$ \eqref{2.4}};
    \draw (0.35,0.8) node {$g=\mathcal{P}\exp\int^{x} A$};
    \draw (0,-0.8) node {holography};
    \draw (4.7,-0.8) node {\color{blue}$L_F$ fixed};
    \draw (-4,-2.2) node{\color{blue}asymptotically AdS bdy};
    \draw (-4.3,1) node {SL$(2,\mathbb{R})$ BF};
    \draw (-4.3,-0.3) node {JT};
  \end{tikzpicture}\label{fig:JTduality}
\end{equation}
In the Schwarzian language with reparametrization $\beta(t)$, the operator \eqref{eq:opgr} corresponds to a bilocal operator:
\begin{equation}
\label{eq:schw}
\mathcal{O}_\Delta = \left(\frac{\varepsilon^2\dot{\beta}_1\dot{\beta}_2}{(\beta_1-\beta_2)^2}\right)^\Delta,
\end{equation}
which is directly related to the bare (unrenormalized) geodesic length $\ell$ computed in an AdS$_2$ geometry between two boundary endpoints on the wiggly boundary curve:
\begin{equation}
\label{eq:geod}
\ell = \log \frac{(\beta_1-\beta_2)^2}{\varepsilon^2\dot{\beta}_1\dot{\beta}_2}.
\end{equation}
Comparing \eqref{eq:opgr} with \eqref{eq:schw} and \eqref{eq:geod} directly shows that we identify $\phi = \ell/2$. The quantity $2\phi_{\text{r}} = 2\phi + 2\log 
 \varepsilon$ introduced earlier is then precisely the renormalized geodesic distance, as the subscript $_{\text{r}}$ already suggested. Working only with renormalized coordinates and renormalized operator insertions, we would strip off the $\varepsilon^{2\Delta}$ prefactor in \eqref{matintJT}, which is the usual choice one makes when discussing the correlation functions in JT gravity \cite{Mertens:2022irh}.

\subsection{Amplitudes of DSSYK}
The amplitudes of the double-scaled regime of the SYK model (DSSYK) were computed in \cite{Berkooz:2018jqr,Berkooz:2018qkz} using a technique based on so-called chord diagrams; these calculations are nicely summarized and reviewed in several places \cite{Lin:2022rbf,Jafferis:2022wez} and we will merely state the results.

The disk partition function of DSSYK is:\footnote{We've rescaled length parameters (like $\beta$) to have our energies measured naturally as the Casimir also in this quantum group case; it is easy to scale them back to match with the usual SYK parameters.\label{fn:9}}
\begin{equation}
    \begin{tikzpicture}[baseline={([yshift=-.5ex]current bounding box.center)}, scale=0.5]
 \pgftext{\includegraphics[scale=1]{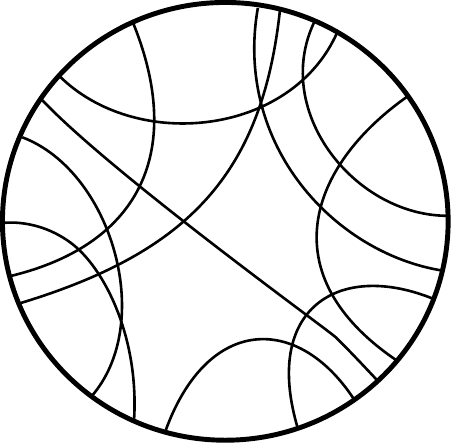}} at (0,0);
    \draw (2.6,-1) node {$\beta$};
  \end{tikzpicture}= \int_0^\pi \d\theta\,\rho(\theta)\,e^{-\beta\,\frac{1}{(q-1/q)^2}2 \cos(\theta)}\,,\quad \rho(\theta)=(e^{\pm 2\i\theta};q^2)_\infty\,.\label{dos}
\end{equation}
Here the $q$-Pochhammer symbol is
\begin{equation}
    (a;q)_n = \prod_{k=0}^{n-1} (1-a q^{k}).
\end{equation}
We have drawn an example of several chords within the bulk of the diagram.
One can also derive the two-sided wavefunction analogous to \eqref{2.11} as
\begin{equation}
    \braket{n\rvert \theta}=R_{\theta\, \i\i}(n)=\frac{q^n}{(q^2;q^2)_n}H_n(\cos(\theta)\rvert q^2)\,,\quad E(\theta)=\frac{1}{(q-1/q)^2}2\cos(\theta)\,,
\end{equation}
where $H_n(a \vert q)$ is the continuous $q$-Hermite polynomial. 

In DSSYK, physicists have computed correlators of specific operators involving the product of $s$ fermions where $s \sim \sqrt{N}$ for $N\to\infty$ \cite{Berkooz:2018qkz,Berkooz:2018jqr,Lin:2022rbf}. The answer for a single such operator pair is
\begin{equation}
\label{eq:2pt}
    \begin{tikzpicture}[baseline={([yshift=-.5ex]current bounding box.center)}, scale=0.5]
 \pgftext{\includegraphics[scale=1]{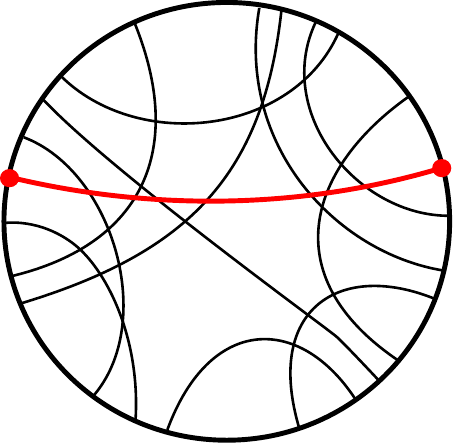}} at (0,0);
    \draw (2.6,-1) node {$\beta_1$};
    \draw (-2.3,1.6) node {$\beta_2$};
    \draw (-2.9,0.5) node {\color{red}$\mathcal{O}_\Delta$};
    \draw (2.9,0.5) node {\color{red}$\mathcal{O}_\Delta$};
  \end{tikzpicture}=\int_0^\pi \d\theta_1\,\rho(\theta_1)\,e^{-\beta_1 \,\frac{1}{(q-1/q)^2}2\cos(\theta_1)}\int_0^\pi \d\theta_2\,\rho(\theta_2)\,e^{-\beta_2 \,\frac{1}{(q-1/q)^2}2\cos(\theta_2)}\bra{\theta_1}\mathcal{O}_\Delta\ket{\theta_2}\, ,
\end{equation}
with the explicit expression for the operator ``matrix element'':
\begin{align}
\bra{\theta_1}\mathcal{O}_\Delta\ket{\theta_2}&=\frac{(q^{4\Delta};q^2)_\infty}{(q^{2\Delta}e^{\pm \i\theta_1\pm \i \theta_2};q^2)_\infty}\,.\label{matint}
\end{align}
For the purposes of this work we want to point out \cite{Berkooz:2018jqr} that these amplitudes have the same structure as the JT amplitudes. In particular, upon taking the scaling limit:
\begin{equation}
    \theta=\pi - 2\log q\, E^{1/2}\,,\quad q\to 1^-\,,
\end{equation}
the spectral density \eqref{dos} of DSSYK reduces to the JT density \eqref{dosJT}.\footnote{The infinite product of the $q$-Pochhammer symbols in $\rho(\theta)$ limits (up to prefactors) to the Euler product formula of the sinh function, which is the JT spectral density.} It was moreover shown in \cite{Berkooz:2018jqr} in eq. (5.13) that the JT scaling limit, upon including some spurious prefactors, of \eqref{matint} becomes
\begin{equation}
(q^2;q^2)_{\infty} (1-q^2)^3 \, \, \frac{(q^{4\Delta};q^2)_\infty}{(q^{2\Delta\pm 2\i \theta_1\pm 2\i \theta_2};q^2)_\infty} \,\, \to \,\, \varepsilon^{2\Delta}\frac{\Gamma(\Delta\pm\i E_1^{1/2}\pm \i E_2^{1/2})}{\Gamma(2\Delta)}.
\end{equation}
This matches the JT matrix element \eqref{matintJT} with the identification
\begin{equation}
    \varepsilon=-2\log q\to 0\,.
\end{equation}
Notice the appearance of $\varepsilon^{2\Delta}$, illustrating that the JT limit of DSSYK produces the bare (unrenormalized) quantities.

This motivates (as alluded to in section \ref{sect:intro}) to search for a representation theoretic interpretation of DSSYK amplitudes, and the associated boundary path integral formulations which generalize the Schwarzian \eqref{sch4d} and Liouville models \eqref{liouvilleaction}.

\section{Representation theory for DSSYK }\label{Sec:rep}
In this section we demonstrate how the amplitudes of double-scaled SYK derived in \cite{Berkooz:2018qkz,Berkooz:2018jqr}, are composed from the representation theory of the quantum group SU$_q(1,1)$. This decomposition of the amplitudes, in parallel to those of JT gravity, was first suggested in \cite{Berkooz:2018jqr}. Aspects of the associated algebra $\Uqsual$ were discussed in \cite{Berkooz:2018qkz,Berkooz:2018jqr,Lin:2022rbf,Lin:2023trc}, and especially recently by \cite{Berkooz:2022mfk}, where they find a realization of the algebra that reproduces the transfer matrix of DSSYK.

We will derive the relevant representation theory of SU$_q(1,1)$ in a systematic manner by first defining and explicitly constructing the right-and left-regular realizations of the quantum group of interest. The outcome of this calculation was reported also in \cite{Berkooz:2022mfk} (sections 5 and 6, and in particular equation (6.25)) while this work was in progress. Here we derive these results within a different calculational scheme.
Our strategy contains two key points:
\begin{enumerate}
    \item 
    We parametrize our quantum group element using the (not widely appreciated) Gauss decomposition \cite{Fronsdal:1991gf,jaganathan2000introduction,Mertens:2022aou}. This formula is the bridge between the infinitesimal level (algebra) and the global level (group), which is key in deriving (and solving) the quantum mechanics of a particle on the manifold SU$_q(1,1)$. This coordinatization allows us to give a very explicit and down-to-earth derivation of the relevant representation theory, and simultaneously making more explicit the parallels with the story in JT gravity as reviewed above. This point is physically crucial. The Gauss decomposition allows one to easily implement Brown-Henneaux \cite{brown1986central} asymptotically AdS$_2$ boundary conditions \cite{Saad:2019lba,Mertens:2017mtv,Grumiller:2017qao,Gonzalez:2018enk,Coussaert:1995zp,Mertens:2018fds,Blommaert:2018oro,Blommaert:2018iqz} in the JT model. We will impose similar boundary conditions that will allow us to make contact with the DSSYK model.
    \item From the regular representation, the Casimir eigenvalue problem is solved by the representation matrices. These physically represent the wavefunctions of this quantum mechanics. Given this interpretation, we will have the full machinery of usual group theory at our disposal. In particular, we can find a natural inner product (specified by the Haar measure), and we know that operator insertions in this theory are representation matrices. This enables us to pinpoint the embedding in group theory of the bilocal operators in DSSYK \cite{Berkooz:2018jqr}, and the inner product allows us to carefully compute correlators, which we match with the known DSSYK answers in section \ref{sect:correlators}. On a technical level, one important step in the calculation is to properly deal with the fact that the Casimir (or Hamiltonian) is a non-Hermitian operator, therefore the left-and right eigenstates (wavefunctions) are different \cite{mostafazadeh2002pseudo,yao2018edge}.\footnote{However there are other interpretations, for instance \cite{Lin:2022rbf} attempts to give a different interpretation for the factor $(q^2;q^2)_n$ properly explained by this distinction between left-and right eigenstates (which is an algebraic fact about non-Hermitian matrices).}
\end{enumerate}

We now define the quantum group SU$_q(1,1)$, and derive the regular realizations of the associated algebra $\Uqsual$.

\subsection{Left-and right-regular realizations}
The quantum group SU$_q(1,1)$ is a deformation of the Lie group SU$(1,1)$. Similarly, one can also perform a deformation at the level of the (universal enveloping) algebra $\Uqsual$. The generators $E,F$ and $H$ satisfy the $\Uqsual$ algebra
\begin{equation}
\label{eq:qalg}
[H,E]=E\,,\quad [H,F]=-F\,,\quad [E,F] = \frac{q^{2H}-q^{-2H}}{q-q^{-1}}\,.
\end{equation}
The Fronsdal-Galindo result for the Gauss-decomposition of SU$_q(1,1)$ \cite{Fronsdal:1991gf,jaganathan2000introduction} relates these two concepts together:
\begin{equation}
\label{UqGE}
g = e_{q^{-2}}^{\gamma F} \, e^{2\phi H} \, e_{q^2}^{\beta E}\,,\quad 0<q<1\,,
\end{equation}
providing a $q$-deformation of the usual exponentiation that transfers between Lie algebra and Lie group. The $q$-exponential function is defined as
\begin{equation}
e_q^x =\sum_{n=0}^{\infty} \frac{x^n}{[n]_q!}\,, \quad [n]_q = \frac{1-q^{n}}{1-q}\,.
\end{equation}
We provide more details and background on \eqref{UqGE} (which is key for what follows) in appendix \ref{app:gauss}.

The coordinates on the (quantum) group manifold SU$_q(1,1)$ form a non-commutative algebra
\begin{equation}
    \comm{\g}{\phi}=-\log q\,\g\,,\quad  \comm{\b}{\phi}=-\log q\,\b\,,\quad\comm{\b}{\g}=0\,.\label{cooalg}
\end{equation}
It is convenient to also introduce the exponentiated operator $K=q^H$. 
The ``quadratic'' Casimir of the algebra $\Uqsual$ is (up to an overall constant)
\begin{equation}
    C=\frac{q^{2H+1}+q^{-2H-1}}{(q-q^{-1})^2}+F E\,,\label{cas}
\end{equation}
which one checks explicitly by calculating $\comm{C}{X_A}=0$ for $X_A=E,F$ or $H$.

We want to construct the regular realizations of the algebra $\Uqsual$ that follow from the Gauss decomposition \eqref{UqGE}. The idea is that for any one of the generators $X_A=E,F,H$ one can write $X_A\, g$ as a differential operator in the coordinates $\gamma, \phi, \b$ acting on $g$. We exemplified this already for SL$(2,\mathbb{R})$ around \eqref{2.3}. By construction, these differential operators (which we call $L_A$) have the same algebra \eqref{eq:qalg} as $X_A$. We should think of these $L_A$ as naturally acting on functions on the (quantum) group manifold $f(g)$, where they correspond with an infinitesimal action:\footnote{This sign is necessary for $L_A$ to have the same algebra as $X_A$. 
}
\begin{equation}
    L_A\cdot f(g)=f(-X_A\, g).\label{defleft}
\end{equation}
In the non-commutative setting, one should think about the functions $f(g)$ as defined by their Taylor coefficients $f_{nmp}$ in an ordered expansion in $\g, \phi$ and $\b$, eliminating by hand ordering ambiguities\footnote{We should note that the non-commutativity of the coordinates also implies a non-trivial algebra of the derivatives with respect to the coordinates. Namely, besides the obvious
\begin{equation}\label{dercomm1}
    \comm{\g}{\frac{d}{d\g}}=-1\,,\quad \comm{\phi}{\frac{d}{d\phi}}=-1\,,\quad \comm{\b}{\frac{d}{d\b}}=-1\,,
\end{equation}
one immediately also derives \begin{equation}\label{dercomm2}
    \comm{\frac{d}{d\g}}{\phi}=\log q \frac{d}{d\g}\,, \quad \comm{\frac{d}{d\b}}{\phi}=\log q \frac{d}{d\b}\,.
\end{equation}
Another useful fact is $\comm{\g\frac{d}{d\g}}{\phi}=0$. Another similar identity for $\b$ also holds.
}
\begin{equation}
    f(\gamma,\phi,\beta)=\sum_{n,m,p}f_{nmp}\, \gamma^n \phi^m \beta^ p\,.\label{taylor}
\end{equation}

Now we obtain explicit expressions for these operators. By definition we search a differential operator $L_F$ so that
\begin{equation}
L_F\cdot e_{q^{-2}}^{\gamma F} \, e^{2\phi H} \, e_{q^2}^{\beta E}= -F\,e_{q^{-2}}^{\gamma F} \, e^{2\phi H} \, e_{q^2}^{\beta E}\,.
\end{equation}
Introducing the standard notion of a $q$-derivative and rescaling operator:
\begin{equation}
    \left(\frac{d}{dx}\right)_{\hspace{-0.1cm}q} f(x) =\frac{1}{x}\,\frac{f(qx)-f(x)}{q-1}=\frac{1}{x}\frac{R_q^x-1}{q-1}\,f(x)\,,\quad  R_q^x f(x)= f(q x)=\exp\bigg(\log q\, x \frac{d}{d x}\bigg) f(x)\,,\label{qder}
\end{equation}
the $q$-exponential diagonalizes the $q$-derivative
\begin{equation}
    \left(\frac{d}{d\g}\right)_{\hspace{-0.1cm}q^-2} e_{q^-2}^{\g F} = F\,e_{q^2}^{\g F}\,,
\end{equation}
such that
\begin{equation}
        \boxed{L_F=- \left(\frac{d}{d\g}\right)_{\hspace{-0.1cm}q^{-2}}}\,.\label{LF}
\end{equation}

Using the algebra of coordinates \eqref{cooalg} and derivatives \eqref{dercomm2}, one finds that the rescaling operators $R^\beta_a$ and $R^\g_a$ commute with everything (except $\b$ respectively $\g$). This makes them convenient operators for what follows. Next, we calculate $L_H$. Using $HF=F(H-1)$ one obtains $H \g^n F^n=-n \g^n F^n +\g^n F^n H$, such that
\begin{equation}
    L_H\cdot e_{q^{-2}}^{\gamma F} \, e^{2\phi H} \, e_{q^2}^{\beta E}= -H\,e_{q^{-2}}^{\gamma F} \, e^{2\phi H} \, e_{q^2}^{\beta E}=\g \frac{d}{d \g}\, e_{q^{-2}}^{\gamma F} \, e^{2\phi H} \, e_{q^2}^{\beta E}-e_{q^{-2}}^{\gamma F} \, H\,e^{2\phi H} \, e_{q^2}^{\beta E}\,.
\end{equation}
This results in
\begin{equation}
    \boxed{L_H=-\frac{1}{2}\frac{d}{d\phi}+\g \frac{d}{d\g}}\,,\label{LH}
\end{equation}
which happens to be identical to the classical SL$(2,\mathbb{R})$ case \eqref{2.3}. In terms of the exponentiated operator $K=q^H$, one finds
\begin{equation}
        L_K= T^\phi_{\log q /2} R^\gamma_{q}=q^{L_H}\,,
\end{equation}
where we defined the shift operator:
\begin{equation}
    T^\phi_a \equiv \exp\bigg( a\, \frac{d}{d \phi}\bigg)\,.\label{shiftop}
\end{equation}
The final generator takes some more work. Repeatedly using the commutator of $E$ and $F$ one eventually finds
\begin{align}
    -L_E\cdot e_{q^{-2}}^{\gamma F} \, e^{2\phi H} \, e_{q^2}^{\beta E}&= E\, e_{q^{-2}}^{\gamma F} \, e^{2\phi H} \, e_{q^2}^{\beta E}\nonumber\\&= e_{q^{-2}}^{\gamma F} \, E\,e^{2\phi H} \, e_{q^2}^{\beta E}+\left( \gamma \frac{q^{2H}}{q-q^{-1}}e_{q^{-2}}^{q^{2}\gamma F} - \gamma \frac{q^{-2H}}{q-q^{-1}} e_{q^{-2}}^{\gamma F}\right) e^{2\phi H} \, e_{q^2}^{\beta E}\,.\label{3.19}
\end{align}
Furthermore by carefully commuting coordinates and generators passed each other at each step we find
\begin{align}
    e_{q^{-2}}^{\gamma F} \, E\,e^{2\phi H} \, e_{q^2}^{\beta E}= e_{q^{-2}}^{\gamma F}\, e^{-2\phi}\, e^{2\phi H}\, E\,e_{q^2}^{\beta E}&=e^{-2\phi} R^\g_{q^2}\, e_{q^{-2}}^{\gamma F}\, e^{2\phi H}\,\left(\frac{d}{d\beta}\right)_{\hspace{-0.1cm}q^{2}} e_{q^2}^{\beta E}\nonumber\\&=e^{-2\phi} R^\gamma_{q^{2}} \left(\frac{d}{d\beta}\right)_{\hspace{-0.1cm}q^{2}} T^\phi_{-\log q}\,e_{q^{-2}}^{\gamma F} \, e^{2\phi H} \, e_{q^2}^{\beta E}\,.
\end{align}
Combining this with the second piece of \eqref{3.19} one arrives at
\begin{equation}
\boxed{L_E = -e^{-2\phi} R^\gamma_{q^{2}} \left(\frac{d}{d\beta}\right)_{\hspace{-0.1cm}q^{2}} T^\phi_{-\log q} - \gamma\, \frac{T^\phi_{\log q}- R^\gamma_{q^2}T^\phi_{-\log q}}{q-q^{-1}}}\,.\label{LE}
\end{equation}
These expressions indeed satisfy the correct algebra \eqref{eq:qalg}:
\begin{equation}
    \comm{L_H}{L_F}=-L_F\,,\quad \comm{L_H}{L_E}=L_E\,,\quad \comm{L_E}{L_F}=\frac{q^{2L_H}-q^{-2L_H}}{q-q^{-1}}\,.
\end{equation}
The Casimir operator can now be explicitly computed by inserting these expressions in \eqref{cas}, leading to:
\begin{equation}
    \boxed{L_C = \frac{q^{2L_H+1}+q^{-2L_H-1}}{(q-q^{-1})^2} + L_F L_E = \frac{q T^\phi_{\log q}+q^{-1}T^\phi_{-\log q}}{(q-q^{-1})^2} + \left(\frac{d}{d\g}\right)_{\hspace{-0.1cm}q^-2}e^{-2\phi}R^\g_{q^2}\left(\frac{d}{d\beta}\right)_{\hspace{-0.1cm}q^{2}} T^\phi_{-\log q}}\,.\label{3.24}
\end{equation}
Mathematically, this difference operator is the Laplacian on the (quantum) group manifold SU$_q(1,1)$. This expression was previously encountered in \cite{Mertens:2022aou} in the context of Liouville gravity.\footnote{It was shown there that the so-called hyperbolic representation matrices (as known from \cite{Ip}) are solutions of this eigenvalue problem of $L_C$.}
Physically, this Casimir $L_C$ is the Hamiltonian $\mathbf{H}=L_C$ of quantum mechanics on SU$_q(1,1)$, and the Casimir eigenvalue problem is the (time-independent) Schr\"odinger equation of this dynamical system, which we will study in more detail in section \ref{sect:pathintegrals}.

One can repeat the logic for the right-regular representation, defined by: 
\begin{equation}
    R_A\cdot f(g)=f(g\, X_A)\,.\label{defright}
\end{equation}
Again, by construction, the differential operators $R_A$ satisfy the same algebra \eqref{eq:qalg} as the generators $X_A$ and generate infinitesimal transformations on functions on SU$_q(1,1)$. 

Explicit expressions are derived in a similar manner, e.g.:
\begin{equation}
    R_F\cdot e_{q^{-2}}^{\gamma F} \, e^{2\phi H} \, e_{q^2}^{\beta E}= e_{q^{-2}}^{\gamma F} \, e^{2\phi H} \, e_{q^2}^{\beta E}\,F\,.
\end{equation}
Because of the non-commutative nature of the coordinates $(\g,\phi,\beta)$, it is natural to define the right-regular realization as differential operators that act \emph{from the right} as\footnote{Such considerations have to made whenever the variables do not commute. A simpler case is that of a supergroup where the fermionic variables are Grassmann numbers, see e.g. \cite{cmp/1103908695}.}
\begin{equation}
    f(g)\cdot \overleftarrow{R_A}=f(g\,X_A)\,.
\end{equation}
The downside is that one has to go through the mental gymnastics of reading from right to left. One eventually finds:
\begin{equation}\label{rightrealization}
\overleftarrow{R_E} = \overleftarrow{\left(\frac{d}{d\beta}\right)_{\hspace{-0.1cm}q^2}}\,,\quad 
\overleftarrow{R_K} = \overleftarrow{T^\phi_{\log q/2}} \overleftarrow{R^\beta_{1/q}}\,,\quad \overleftarrow{R_F}
= \overleftarrow{T^\phi_{\log q}}\overleftarrow{\left(\frac{d}{d\gamma}\right)_{\hspace{-0.1cm}q^{-2}}} \overleftarrow{R^\beta_{q^{-2}}} e^{-2\phi} + \frac{\overleftarrow{R^\beta_{q^{-2}}}\overleftarrow{T^\phi_{\log q}} -  \overleftarrow{T^\phi_{-\log q}}}{q-q^{-1}} \beta\,,
\end{equation}
which mirrors the expressions for $L_A$. 
For ease of calculations, it is sometimes more convenient to have all operators acting from one side. This leads to the equivalent rewriting of \eqref{rightrealization}:
\begin{equation}
    R_E=T^\phi_{-\log q}\left(\frac{d}{d\beta}\right)_{\hspace{-0.1cm}q^{2}}\,,\quad R_H=\frac{1}{2}\frac{d}{d\phi}-\beta \frac{d}{d\b}\, , \quad R_F=\left(\frac{d}{d\g}\right)_{\hspace{-0.1cm}q^-2}e^{-2\phi}R^\g_{q^2}+\beta\,\frac{T^\phi_{2\log q}R^\beta_{q^{-2}}-1}{q-q^{-1}}\,.\label{RA}
\end{equation}
One can check explicitly that these generators $R_A$ satisfy the algebra \eqref{eq:qalg}, the Casimir is identical to \eqref{3.24} $L_C=R_C=\mathbf{H}$, and that they commute with the left-regular realization $L_A$:
\begin{equation}
    \comm{R_H}{R_F}=-R_F\,,\quad \comm{R_H}{R_E}=R_E\,,\quad \comm{R_E}{R_F}=\frac{q^{2R_H}-q^{-2R_H}}{q-q^{-1}}\,,\quad \comm{L_A}{R_B}=0\,.\label{3.27}
\end{equation}
The latter property is an immediate consequence of the definition of these operators since $L_A\cdot R_B\cdot f(g)=f(-X_A\,g\,X_B)=R_B\cdot L_A\cdot f(g)$.

Finally, let us provide an explict comparison to the recent work \cite{Berkooz:2022mfk} (section 6, equation (6.25)), which appeared while this work was in progress. Their realization of the quantum algebra was constructed by Wick rotating from Euclidean signature as the algebra acting on functions defined on the hyperbolic 3-plane $H_3 \simeq \text{SL}(2,\mathbb{C})/\text{SU}(2)$ and its non-commutative counterpart. Instead of using the explicit description \eqref{UqGE}, they use the underlying Hopf duality (or pairing) to deduce the correct form of the generators. They use two languages: that of a discretized lattice, and of non-commutative variables. The specific realization (in their (6.25)) found there is written in terms of commuting variables defined on a lattice, whereas we directly found the non-commutative description. In terms of the latter, our right regular realization \eqref{rightrealization} is essentially the same as e.g. their eq. (7.2) by identifying their generators $A',B',C',D'$ as\footnote{These values of $a$ and $\tilde{a}$ are also the ones that they argue are relevant to match directly with the DSSYK problem. We also need the coordinates $z_{\text{there}} =\beta$ and $H_{\text{there}}=e^{\phi}$, and a slight change of convention for the non-commutative coordinate algebra $(\beta,\phi,\gamma)$, which in their case satisfy the modified relation $e^\phi \beta= q^{2} \beta e^\phi$.}
\bea
A'=q^{\overleftarrow{R_H}}\,, \quad D'=q^{-\overleftarrow{R_H}}\,, \quad B'=\overleftarrow{R_E}\,,\quad C'=\overleftarrow{R_F}\,,\quad a=0\,,\quad \tilde{a}=2\,.
\eea

\subsection{Gravitational matrix elements }\label{sect:matint}
In this section we compute ``gravitationally constrained'' representation matrix elements of SU$_q(1,1)$. Let us first explain what we mean by ``gravitationally constrained'' representation matrix elements, and why they are physically relevant.

We are interested in solving a quantum mechanical system in the coordinates system $(\g, \phi, \beta)$ with Hamiltonian $\mathbf{H}=L_C$ \eqref{3.24}. This system has a left-right $\Uqsu$ symmetry with generators $L_A, R_A$ and identical Casimirs $L_C=R_C=\mathbf{H}$. As usually in quantum mechanics, this means the spectrum is organized in (irreducible) representations of SU$_q(1,1)$. Indeed, one can simultaneously diagonalize $\mathbf{H}$ and one generator from each set $(L_A, R_B)$, thus we have the set of eigenstates $\ket{\theta,\,\mu_1 \mu_2}$ with
\begin{equation}
    \mathbf{H}\ket{\theta,\,\mu_1 \mu_2}=E(\theta)\ket{\theta,\,\mu_1 \mu_2}\,,\quad L_A \ket{\theta,\,\mu_1 \mu_2}=-\mu_1 \ket{\theta,\,\mu_1 \mu_2}\,,\quad R_B \ket{\theta,\,\mu_1 \mu_2}=\mu_2 \ket{\theta,\,\mu_1 \mu_2}\,.
\end{equation}
We can translate this into difference equations on $L^2(G)$ (as in  \eqref{taylor}) by defining $
    f^\theta_{\mu_1\mu_2}(g)=\braket{g\rvert\theta\,\mu_1 \mu_2}$,
leading to the eigenvalue equations
\begin{equation}
    \mathbf{H}\cdot f^\theta_{\mu_1\mu_2}(g)=E(\theta) f^\theta_{\mu_1\mu_2}(g)\,,\quad L_A\cdot f^\theta_{\mu_1\mu_2}(g)=-\mu_1 f^\theta_{\mu_1\mu_2}(g)\,,\quad R_B\cdot f^\theta_{\mu_1\mu_2}(g)=\mu_2 f^\theta_{\mu_1\mu_2}(g)\,.
\end{equation}
The point is that the solutions to these equations are representation matrix elements of $\Uqsu$:
\begin{equation}
    R_{\theta\,\mu_1\mu_2}(g)=\bra{\theta\,\mu_1}g\ket{\theta\,\mu_2}\,,\label{repmat}
\end{equation}
Indeed, using the definitions of the left-and right-regular realizations \eqref{defleft} and \eqref{defright}, the eigenvalue equations become
\begin{equation}
    f^\theta_{\mu_1\mu_2}(-X_A\,g)=-\mu_1f^\theta_{\mu_1\mu_2}(g)\,,\quad f^\theta_{\mu_1\mu_2}(g\,X_B)=\mu_2 f^\theta_{\mu_1\mu_2}(g)\,,
\end{equation}
which we see are indeed solved by the representation matrix elements \eqref{repmat}: $    f^\theta_{\mu_1\mu_2}(g)=R_{\theta,\,\mu_1\mu_2}(g)$.

As we quickly reviewed in section \ref{sect:JTreview}, in the SL$(2,\mathbb{R})$ BF formulation of JT gravity, special cases of these wavefunctions have a bulk interpretation as the gravitational WdW wavefunction on an interval stretching between two boundaries with asymptotically AdS$_2$ boundary conditions \cite{Blommaert:2018iqz,Blommaert:2018oro,Harlow:2018tqv,Iliesiu:2019xuh,Saad:2019pqd}: imposing the gravitational constraints \eqref{2.10} leads to the constrained wavefunctions $R_{E,\,\i\i}(\phi)$ \eqref{2.11}:

\begin{equation}
    \begin{tikzpicture}[baseline={([yshift=-.5ex]current bounding box.center)}, scale=0.7]
 \pgftext{\includegraphics[scale=1]{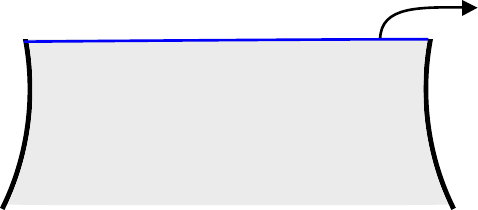}} at (0,0);
    \draw (0,1) node{$\phi$};
    \draw (7.3,1) node{wavefunctions $\braket{\phi\rvert E}=R_{E,\,\i\i}(\phi)$ \eqref{2.11}};
  \end{tikzpicture}\label{fig:338}
\end{equation}

Here we seek the generalization of those asymptotically AdS$_2$ boundary conditions to the DSSYK context. Contrary to the JT case, we do not have a first principles derivation in gravity, since we do not yet know the gravitational dual of DSSYK. The purpose (in part) of this section is to claim the generalization of the group theoretic formulation \eqref{2.10} of the gravitational boundary conditions to the DSSYK context and check that with those constraints, quantum mechanics on $\Uqsu$ reproduces (the amplitudes of) DSSYK. Once a first-order bulk theory is identified that reduces to quantum mechanics on $\Uqsu$, one could take these group theoretic constraints and translate them back to gravitational boundary conditions in the bulk gravitational dual of DSSYK.

We claim that the correct generalization of \eqref{2.10} is\footnote{These constraints were inspired by a similar discussion in \cite{Fan:2021bwt}, where it was observed that the amplitudes of 2d Liouville gravity have an embedding in the representation theory of the so-called modular double $\Uqsl \otimes U_{\tilde{q}}(\mathfrak{sl}(2,\mathbb{R}))$. In that context it would also be interesting to understand how the constraints translate back to gravitational boundary conditions. Since in that case both sides of the duality are better understood it would be a good place to gain momentum for tackling the current DSSYK problem.} 
\begin{equation}
    \boxed{L_F=\i \frac{q^{1/2}}{q-q^{-1}}q^{L_H}\,,\quad R_E= -\i \frac{q^{1/2}}{q-q^{-1}}q^{-R_H}}\,.\label{constraints}
\end{equation}
It may seem strange to have the Cartan elements $q^{L_H}$ respectively $q^{-R_H}$ appear on the RHS of \eqref{constraints}, but in fact this is consistent and necessary. Physically, in quantum mechanics one can constrain any combination of operators as long as the constraint commutes with the Hamiltonian $\mathbf{H}$.\footnote{One checks explicitly using \eqref{LF}, \eqref{LH} and \eqref{RA} that the two constraints in \eqref{constraints} commute with one another and with the Hamiltonian \eqref{3.24}, though this is automatic from the definition.} 

Mathematically, such types of constraints were shown to be required to define constrained representation matrices for higher rank quantum groups \cite{sevostyanov1999quantum}.\footnote{This requires in fact a further slight generalization where one uses a 2-parameter generalization as $q^{\alpha_1 L_H}$ and $q^{\alpha_2 R_H}$. The resulting irrep matrix elements only depend on the difference $\alpha_1-\alpha_2$. We chose simply the most symmetric option.\label{fn:19}} For rank one (our case), it is still a possibility one can introduce. For instance, in a quantum integrability context, this freedom can result in different quantizations of the same underlying classical system \cite{Kharchev:2001rs}. 

Notice that in the $q\to 1^-$ classical limit, the constraints \eqref{constraints} reduce to \eqref{2.10}, with the identification
\begin{equation}
    \varepsilon=-2\log q\to 0\,.
\end{equation}

Analogous to the discussion around \eqref{repmat}, one can solve these constraints by proposing that the wavefunctions are particular representation matrix elements
\begin{equation}
    f^\theta_{\i\i}(g)= R_{\theta,\,\mu_1\mu_2}(g) = \bra{\theta\,\i} g \ket{\theta\,\i}\,,\label{gravmatintinter}
\end{equation}
where the left- and right eigenstates of the respective constraints should now satisfy
\begin{equation}
    E\ket{\i}=-\i \frac{q^{1/2}}{q-q^{-1}}\, q^{-H} \ket{\i}\,,\quad \bra{\i} F=-\i \frac{q^{1/2}}{q-q^{-1}} \bra{\i} q^{-H} \,.
\end{equation}

To determine the gravitational matrix elements, we then solve the Casimir equation \eqref{3.24}, which for our purposes can be conveniently rewritten as
\begin{align}
    \mathbf{H} =-e^{-2\phi} R^\gamma_{q^2} L_F\,R_E++\frac{q T^\phi_{\log q}+q^{-1}T^\phi_{-\log q}}{(q-q^{-1})^2}=\frac{-q e^{-2\phi} T^\phi_{-\log q} R^\gamma_{q^3}R^\beta_q+q T^\phi_{\log q}+q^{-1}T^\phi_{-\log q}}{(q-q^{-1})^2}\,.\label{cascas}
\end{align}
The resulting representation matrices $R_{\theta\,\i \i}(g)$ are still a function of all coordinates $(\g,\phi,\beta)$. However, the dependence on both $\g$ and $\beta$ is relatively simple since it factors out. The ``reduced'' matrix element which you obtain by setting $\g=\beta=0$ are called Whittaker functions \cite{backgammon1967functions,hashizume1979whittaker,hashizume1982whittaker} in the mathematical literature. We claim that those Whittaker functions are actually the physical wavefunctions of our constrained system and study them from hereon.

Physically, the simplest way to prove this step is to use the ``constrain first'' approach to quantization of constrained quantum mechanical systems with some first class (gauge) constraints \cite{dirac2001lectures}.\footnote{Representation theory corresponds with a ``quantize first'' approach. Unfortunately this is often the more complicated trajectory, numerous examples in gauge theory illustrate the efficiency of ``constrain first'' quantization. The most known example is the solution of Chern-Simons theory \cite{witten1989quantum,elitzur1989remarks}, but also 2d Yang-Mills \cite{witten1991quantum,Witten:1992xu,Cordes:1994fc,Blommaert:2018oue} and 3d gravity \cite{Cotler:2020ugk,Eberhardt:2022wlc} are noteworthy. The point is that ``quantize first'' can quickly becomes tedious, one needs a nice inner product and compute the overlaps of all states, identify all null states and mod them out. These final two steps sound much simpler than they are in practice, for instance even in 2d Maxwell this takes some work \cite{Blommaert:2018rsf}. In our set-up, we have an inner product (see section \ref{sect:haar}) so in principle one could go through this. However, it is inefficient therefore we will not do so.}
With the purpose of identifying a dynamical action and path integral description, we will discuss the classical phase space, the constraints and the resulting gauge freedom of our system in section \ref{sect:pathintegrals}. The analysis in section \ref{sect:pathintegrals} confirms that indeed one can use the gauge freedom of the gravitational boundary conditions to fix
\begin{equation}
\beta=\gamma=0 \,.\label{gaugefixxx}
\end{equation}

With this constraint the gravitational matrix elements become
\begin{equation}
    \braket{\phi\rvert\theta} \equiv R_{\theta\,\i \i}(e^{2\phi H}) = \bra{\theta\,\i} e^{2\phi H} \ket{\theta\,\i}\,.
\end{equation}
and the Hamiltonian \eqref{cascas} on this physical function space simplifies to
\begin{equation}
    (q-q^{-1})^2 \mathbf{H} = q T^\phi_{\log q}+(q^{-1}-qe^{-2\phi}) T^\phi_{-\log q}\,,\label{3.43}
\end{equation}
such that the Schr\"odinger equation $\mathbf{H}\cdot f(g)=E\, f(g)$ becomes
\begin{equation}
    q \braket{\phi+\log q\rvert \theta}+(q^{-1}-qe^{-2\phi})\braket{\phi-\log q\rvert \theta}=(q-q^{-1})^2E(\theta) \braket{\phi\rvert \theta}\,.\label{differenceeq}
\end{equation}
As a difference equation, this Schr\"odinger equation only relates the wavefunction at points with discrete separation in $\phi$. Therefore there is an (uncountable) infinite degeneracy for each $E(\theta)$ in the solution space. Indeed, for any solution $\braket{\phi\rvert \theta}$ of \eqref{differenceeq}, the function $\braket{\phi\rvert \theta} f_\text{periodic}(\phi)$, with $f_\text{periodic}(\phi+\log q)=f_\text{periodic}(\phi)$, is also a solution to \eqref{differenceeq}. 
This is a general feature of working with difference equations instead of differential equations, and requires additional input in our model to address. 

In our case, the resolution is that $\braket{\phi\rvert \theta} f_\text{periodic}(\phi)$ and $\braket{\phi\rvert \theta}$ are actually physically indistinguishable. Indeed, as we will see in the next section \ref{sect:haar}, the natural inner product on $\Uqsu$ we will use (the Haar measure) essentially samples the wavefunctions at discrete locations only, which in our $\beta=\gamma=0$ constraint system boils down to the equidistant sampling
\begin{equation}
    \phi=-n\log q\,. \label{3.46}
\end{equation}
In physics terms this means that the wavefunctions  $\braket{\phi\rvert \theta} f_\text{periodic}(\phi)$ and $\braket{\phi\rvert \theta}$ correspond to states which are related by the addition of a null state, i.e. they have identical overlaps (inner products) with all physical wavefunctions. One can think of this as a reduction of the space of coordinate states $\bra{\phi}$ to the physical Hilbert space $\bra{n}$. Graphically we can summarize the redundancy as follows
\begin{equation}
    \begin{tikzpicture}[baseline={([yshift=-.5ex]current bounding box.center)}, scale=0.7]
\pgftext{\includegraphics[scale=1]{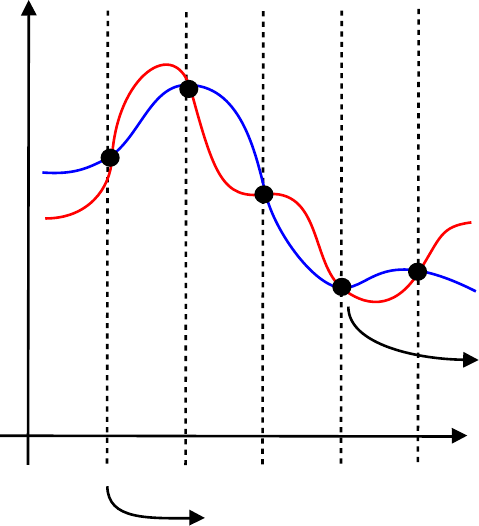}} at (0,0);
    \draw (4.5,0.6) node {\color{red}$\braket{\phi\rvert\theta}$ solution};
    \draw (3.5,-2.6) node {inner product sampling \eqref{3.46}};
     \draw (5.8,-0.2) node {\color{blue}$\braket{\phi\rvert\theta}$ equivalent solution};
     \draw (2.8,-1.8) node {$\phi$};
     \draw (5.1,-1) node {$\braket{n\rvert\theta}$ physical data};
  \end{tikzpicture}
\end{equation}
To match with the physics of DSSYK, this step is crucial, since the discretized coordinate $n$ plays the role of the integer number of chords \cite{Lin:2022rbf,Berkooz:2022mfk} which represents a discretization of bulk spacetime. In our framework, this discretization arises from quantum mechanics. We will also see this in section \ref{sect:pathintegrals}, where we write \emph{classical} actions for the \emph{continuous} fields $\g, \phi$ and $\beta$.\footnote{A useful analogy is that momentum quantization for a non-relativistic particle in a periodic potential is also a quantum effect.}

With this input, the difference equation \eqref{differenceeq} has a unique solution for each $E(\theta)$ up to an ``initial'' condition which we will take to be
\begin{equation}
    \braket{0\rvert\theta}=1\,.
\end{equation} 

Now we can finally solve the Schr\"odinger equation for the physical data, namely the values of the wavefunctions $\braket{n\rvert \theta}$ at the discrete sampling points. From \eqref{differenceeq} one finds
\begin{equation}
    q\braket{n-1\rvert\theta}+(q^{-1}-q^{2n +1}) \braket{n+1\rvert \theta}=(q-q^{-1})^2 E(\theta)\braket{n\rvert \theta}\,,\label{3.47}
\end{equation}
with solution for the gravitational matrix element\footnote{The recursion relation for the $q$-Hermite polynomials $H_n(x\rvert q^2)$ is
\begin{equation}
    (1-q^{2n})H_{n-1}(x\rvert q^2)+H_{n+1}(x\rvert q^2)=2\cos(\theta) H_{n}(x\rvert q^2)\,.\label{redrecrel}
\end{equation}
}
\begin{equation}
    \boxed{\braket{n\rvert \theta}=R_{\theta\, \i\i}(n)=\frac{q^n}{(q^2;q^2)_n}H_n(\cos(\theta)\rvert q^2)}\,,\quad E(\theta)=\frac{1}{(q-1/q)^2}2\cos(\theta)\,.\label{waefunanswer}
\end{equation}
In this reduced recursion relation \eqref{redrecrel}, the Hamiltonian $\mathbf{H}$ reduces to the transfer matrix $T$ of DSSYK, which is discussed very nicely in \cite{Berkooz:2018qkz,Lin:2022rbf}.\footnote{There are minor differences between $\mathbf{H}$ and $T$, having to do with the normalization of the states $\bra{n}$ - which in our case is determined by the Haar measure in section \ref{sect:haar}. These differences are physically irrelevant, as we will recover the DSSYK amplitudes (the physical data) in section \ref{sect:correlators}. One should include an overall factor in our $\mathbf{H}$ if we want to precisely match $\mathbf{H}$ eigenvalues with the DSSYK energies, see footnote \ref{fn:9}.} These wavefunctions $\braket{n\rvert \theta}$ are known to be the eigenfunctions of $T$ in DSSYK \cite{Berkooz:2018qkz,Lin:2022rbf}.

The functions \eqref{waefunanswer} are the bulk wavefunctions, obtained in the chords language in \cite{Berkooz:2018qkz,Lin:2022rbf}. What we are shooting for eventually is to identify them as the WdW wavefunctions of a bulk gravitational theory \eqref{fig:DSSYKduality2} with a first-order formulation that reduces to quantum mechanics on $\Uqsu$:
\begin{equation}
    \begin{tikzpicture}[baseline={([yshift=-.5ex]current bounding box.center)}, scale=0.7]
\pgftext{\includegraphics[scale=1]{dssyk8v2.pdf}} at (0,0);
    \draw (0,1) node{$\phi$};
    \draw (7,1) node{wavefunctions $\braket{\phi\rvert \theta}=R_{\theta\,\i\i}(\phi)$ \eqref{waefunanswer}};
  \end{tikzpicture}\label{fig:339}
\end{equation}
\subsection{Inner product }\label{sect:haar}
In order to compute observables and construct a Hilbert space, we ought to define an inner product on states
\begin{equation}
    \braket{\psi_1\rvert\psi_2}=\int (\d g)_q\,\braket{\psi_1\rvert g}\braket{g\rvert\psi_2}\,.\label{ip}
\end{equation}
For classical Lie groups the natural inner product to be used is the Haar measure $\d g$, which is uniquely defined (up to a prefactor) by the property that it is left-right invariant, i.e. for any group element $h$
\begin{equation}
    \int \d g\, f(h^{-1}\,g)=\int \d g\,f(g)=\int\d g\,f(g\,h)\,.
\end{equation}
For quantum groups we will work with the infinitesimal version of this constraint. In particular, we will fix the Haar measure $(\d g)_q$ on $\Uqsu$ by imposing
\begin{equation}
    \int (\d g)_q\, L_E\cdot f(g)=0\,,\quad \int (\d g)_q\, L_F\cdot f(g)=0\,,\label{measurecon}
\end{equation}
with identical constraints for $R_E$ and $R_F$. Consistency with the algebra \eqref{eq:qalg} then implies furthermore
\begin{equation}
    \int (\d g)_q\,L_K^2\cdot f(g)=1\,.\label{kcon}
\end{equation}
Our treatment here was inspired by \cite{olshanetsky2002unitary}.
For SL$(2,\mathbb{R})$ with Gauss decomposition \eqref{gausssclassical}, imposing these constraints indeed gives the known Haar measure $\d g=\d\g\,e^{2\phi}\d\phi\,\d\b$. 

We claim that the correct Haar integration on $\Uqsu$ for our model is
\begin{equation}
    \int (\d g)_q=\int_0^\infty (\d\g)_{q^{-2}}\int_0^\infty (\d e^{2\phi})_{q^2}\, \int_0^\infty (\d\b)_{q^2}\,.\label{haar}
\end{equation}
Here the integrals are so-called Jackson integrals:
\begin{equation}
    \int_0^\infty (\d x)_a\,f(x) \equiv (1-a)\sum_{n=-\infty}^{+\infty} a^{-n}\,f(a^{-n})\,,\label{integration}
\end{equation}
and more particularly we define the action of our Haar integral on some function $f(g)=f(\g,\phi,\b)$ \eqref{taylor} as\footnote{This definition in general means evaluating the inner products in \eqref{ip} can become tedious, as one has to reorder the product of the bra and ket wavefunctions, before computing the Jackson integrals. We won't need to deal with this because we will be setting $\beta=\gamma=0$. The notation of \eqref{3.55} can be viewed as a `formal' power series focusing on how ordering issues can be treated consistently for the most general function of the non-commutative variables.}
\begin{equation}
     f(\gamma,\phi,\beta)=\sum_{n,m,p}f_{n m p}\, \gamma^n \phi^m \beta^p\quad \to \quad \sum_{n, m, p}f_{n m p} \int_0^\infty (\d\g)_{q^{-2}}\,\g^n\int_0^\infty (\d e^{2\phi})_{q^2}\,\phi^m \int_0^\infty (\d\b)_{q^2}\,\beta^p\,.\label{3.55}
\end{equation}
The Jackson integrals are Riemann sums that have the property mentioned around \eqref{3.46} of sampling the integrand at the discrete locations mentioned there.

We now check that our Haar integral \eqref{integration} is left- and right-invariant in the sense of \eqref{measurecon}. One needs two properties of the Jackson integral which follow elementary from the definition \eqref{integration}. First, there is the rescaling relation
\begin{equation}
    \int_0^\infty (\d x)_a\,f(a^n x)=a^{-n}\int_0^\infty (\d x)_a\,f(x)\,,\label{scale}
\end{equation}
from which the second relation follows, namely that the Jackson integral is the inverse of the $q$-derivative $(d/dx)_a$, meaning the Jackson integral of $(d/dx)_a f(x)$ vanishes. With this property it is obvious that
\begin{equation}
    \int (\d g)_q\, L_F\cdot f(g)=0\,,\quad \int (\d g)_q\, R_E\cdot f(g)=0\,,
\end{equation}
since $L_E=-(d/d\g)_{q^{-2}}$ \eqref{LF}. Similarly, reordering the Taylor series of $R_F\cdot f(\g,\phi,\b)$ as in \eqref{3.55} cancels the $T^\phi_{-\log q}$ prefactor in \eqref{RA}, thus one finds that $R_F$ acts purely as $(d/d\b)_{q^2}$ on the $\beta$ dependence in the Jackson integral, so this vanishes. To verify the $L_E$ constraint \eqref{measurecon} one writes \eqref{LE}
\begin{align}
    -L_E\cdot f(\g,\phi,\b)&=\sum_{n,m,p}f_{n m p}\, \gamma^n\,e^{-2\phi} \phi^m\, \left(\frac{d}{d\beta}\right)_{\hspace{-0.1cm}q^{2}} \beta^p+\g\,\frac{f(\g,\phi+\log q,\beta)-f(q^2 \g,\phi-\log q,\beta)}{q-q^{-1}}\,.
\end{align}
The Haar integral \eqref{3.55} of the first term vanishes because the $\beta$ dependence is a total derivative. The two other contributions can then be shown to cancel by using the rescaling relation \eqref{scale} for both the $\gamma$ and $\phi$ dependence. The other constraints including \eqref{kcon} can be quite simply checked using the same tools. 

It is interesting to try to reverse this logic and \emph{derive} the Haar measure \eqref{3.55} from these constraints. Usually, the Haar measure is determined up to an overall multiplicative constant. 
However even when taking this into account, the solution is not unique (unlike for classical groups). The integration we have presented in equation \eqref{3.55} is the ``roughest'' sampling which is a solution; all other solutions are more ``dense'' samplings $(d \g)_{q^{-2/a}}(d e^{2\phi})_{q^{2/b}}(d \b)_{q^{2/c}}$ for integers $a,b,c$, which even includes the limiting classical measure $\d\g\,e^{2\phi}\d\phi\,\d\b$. In our set-up, we choose the ``roughest'' sampling as the correct inner product, to avoid infinitely degenerate Hilbert spaces as discussed around \eqref{3.46}.\footnote{It is interesting to note that for SL$_q(2,\mathbb{R})$ (or better SL$_q^+(2,\mathbb{R})$) one should use the continuous limit of the Haar measure instead. We comment on the difference and how that case addresses the infinite degeneracy problem in the conclusion section \ref{sect:5.4}.}

After gauge-fixing $\beta=\g=0$ \eqref{gaugefixxx} the Haar integral \eqref{3.55} implements the inner product as
\begin{equation}
    \braket{\theta_1\rvert \theta_2}=\int_0^\infty (d e^{2\phi})_{q^2}\,\braket{\theta_1\rvert\phi}\braket{\phi\rvert \theta_2}=\sum_{n=-\infty}^{+\infty}q^{-2 n}\braket{\theta_1\rvert n }\braket{n\rvert \theta_2}\,,\label{ipfinal}
\end{equation}
where precisely the sampling \eqref{3.46} rolls out and where $\braket{n\rvert \theta_2}$ is the wavefunction \eqref{waefunanswer}.

\subsection{Some amplitudes}\label{sect:correlators}
In order to compute the overlap $\braket{\theta_1\rvert \theta_2}$ we must still determine the left eigenvector $\braket{\theta_1\rvert n }$ of the Casimir \eqref{3.43}. Indeed, if we impose that 
\begin{equation}
\mathbf{H}\ket{\theta_2}=2\cos(\theta_2)\ket{\theta_2}\,,\quad \bra{\theta_1}\mathbf{H}=2\cos(\theta_1)\bra{\theta_1},
\end{equation}
then we are guaranteed orthogonality 
\begin{equation}
    \braket{\theta_1\rvert \theta_2} \, \sim \,  \delta(\theta_1-\theta_2)\,.
\end{equation}
This would be completely standard were it not for the fact that $\mathbf{H}$ in \eqref{3.43} is a non-Hermitian operator. This is perhaps most evident when we write $T^\phi_{\log q}$ as in \eqref{shiftop}, but see also later \eqref{4.20} which is manifestly ``complex''.\footnote{Again it is interesting to compare to SL$_q(2,\mathbb{R})$ in which case these translation operators are in the imaginary direction and thus Hermitian. The Casimir is Hermitian, and the left-and right wavefunctions are just complex conjugates \cite{Fan:2021bwt}.\label{fnliou}} 

The result is that the left-eigenvectors are not just (the complex conjugate of) the right-eigenvectors
\begin{equation}
    \braket{n\rvert \theta}\neq \braket{\theta\rvert n}^*\,,
\end{equation}
a well-known fact in the condensed matter literature \cite{yao2018edge} when one tries to diagonalize non-Hermitian Hamiltonians. Since we have an inner product \eqref{ipfinal} we can derive the equation for $\braket{n\rvert \theta_1}$:
\begin{align}
    \braket{\theta_1\rvert \mathbf{H} \rvert \theta_2}&=\sum_{n=-\infty}^{+\infty}q^{-2 n+1}\braket{\theta_1\rvert n }\braket{n-1\rvert\theta_2}+\sum_{n=-\infty}^{+\infty}(q^{-2 n-1}-q)\braket{\theta_1\rvert n}\braket{n+1\rvert \theta_2}\nonumber\\&=\sum_{n=-\infty}^{+\infty}q^{-2 n-1}\braket{\theta_1\rvert n+1 }\braket{n\rvert\theta_2}+\sum_{n=-\infty}^{+\infty}(q^{-2 n+1}-q)\braket{\theta_1\rvert n-1}\braket{n\rvert \theta_2}\nonumber\\&=2\cos(\theta_1)\sum_{n=-\infty}^{+\infty}q^{-2 n}\braket{\theta_1\rvert n }\braket{n\rvert \theta_2}\,,
\end{align}
from which we deduce the Schr\"odinger equation for the left wavefunction:
\begin{equation}
    q^{-1}\braket{\theta_1\rvert n+1}+q(1-q^{2n})\braket{\theta_1\rvert n-1}=2\cos(\theta_1)\braket{\theta_1\rvert n}\,,
\end{equation}
with unique solution
\begin{equation}
     \braket{\theta\rvert n}=q^n H_n(\cos(\theta)\rvert q^2)\,.
\end{equation}
Note the (subtle) difference in prefactor with the right wavefunction \eqref{waefunanswer}. This difference is important because now one indeed finds explicitly orthogonal states (this final sum is computed in \cite{Berkooz:2018qkz}, and again we dropped overall constants not fixed by the current arguments)
\begin{align}
    \braket{\theta_1\rvert \theta_2}&=\sum_{n=0}^{+\infty}q^{-2 n} \braket{\theta_1\rvert n} \braket{n\rvert \theta_2}=\sum_{n=0}^{+\infty}\frac{1}{(q^2;q^2)_n}H_n(\cos(\theta_1)\rvert q^2)H_n(\cos(\theta_2)\rvert q^{2})=\frac{\delta(\theta_1-\theta_2)}{(e^{\pm 2\i\theta_1};q^2)_\infty}\,,
\end{align}
which reproduces the known DSSYK spectral density \cite{Berkooz:2018qkz,Cotler:2016fpe}
\begin{equation}
    \rho(\theta)=(e^{\pm 2\i\theta};q^2)_\infty\,.
\end{equation}
The disk partition function of the putative bulk dual is then computed as follows
\begin{equation}
    \text{Tr}(e^{-\beta H_\text{DSSYK}})=\bra{0}e^{-\beta H}\ket{0}=\int_0^\pi \d\theta\,\rho(\theta)\,e^{-\beta \,\frac{1}{(q-1/q)^2}2\cos(\theta)} \braket{0\rvert\theta}\braket{\theta\rvert0}= \int_0^\pi \d\theta\,\rho(\theta)\,e^{-\beta \,\frac{1}{(q-1/q)^2}2 \cos(\theta)}\,,\label{3.69}
\end{equation}
again in line with the DSSYK results of \cite{Berkooz:2018qkz,Cotler:2016fpe}. The state $n=0$ reflects the fact that in some eventual bulk calculation we start our bulk evolution from a point, evaluating the representation matrix element on the identity $g=\mathbf{1}$, where indeed \eqref{UqGE} puts $n=0$ \cite{Blommaert:2018oro}. In the chord language it means we start from a no-chords state \cite{Lin:2022rbf} (a different way of saying we contract the bulk Cauchy slice to a point):
\begin{equation}
    \bra{0}e^{-\beta H}\ket{0}=\quad \begin{tikzpicture}[baseline={([yshift=-.5ex]current bounding box.center)}, scale=0.7]
 \pgftext{\includegraphics[scale=1]{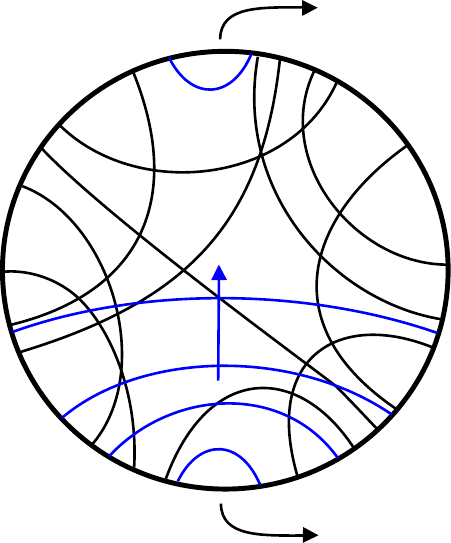}} at (0,0);
    \draw (2.6,-1) node {$\beta$};
    \draw (5.3,-2.65) node {initial configuration $n=0$ or $g=\mathbf{1}$};
     \draw (5.1,2.65) node {final configuration $n=0$ or $g=\mathbf{1}$};
    \draw (-3.2,-1.5) node {\color{blue}time flow};
  \end{tikzpicture}
\end{equation}

One of the elements that is still missing is how operators in DSSYK are embedded in representation theory and quantum mechanics on $\Uqsu$. We recall from \eqref{gravmatintinter} that our Hilbert space consists of states $\ket{\theta\, \i\i}$ obtained by diagonalizing the currents \eqref{constraints}. The only operators which make sense in the constrained system are operators $\mathcal{O}_\Delta$ which do not violate these constraints, meaning that $\mathcal{O}_\Delta\ket{\theta\,\i\i}$ is still an eigenstate of $-q^{-L_H}L_F$ and of $q^{R_H}R_E$ with unchanged eigenvalue. So $\mathcal{O}_\Delta$ must not be charged under the constrained currents
\begin{equation}
    -q^{-L_H}L_F=0\,,\quad q^{R_H}R_E=0\,.\label{371}
\end{equation}
Any such operator can be decomposed into operators with fixed Casimir eigenvalue. We can find operators with these eigenvalues by considering discrete series representation matrix elements of $\Uqsu$ (for which the Casimir can be found in (see (G.8) in \cite{Jafferis:2022wez})\footnote{The states $\ket{\theta}$ correspond with continuous series representations $\Delta=1/2+\i s$ with $\theta=-2\log q\, s$.}
\begin{equation}
\mathcal{O}_\Delta=\bra{\Delta\,0} g\ket{\Delta\, 0}=R_{\Delta\, 0 0}(g)\,.
\end{equation}
With the constraints \eqref{371} the Casimir equation \eqref{cascas} boils down to
\begin{equation}
    q T^\phi_{\log q}+q^{-1}T^\phi_{-\log q}=2\cosh(2\log q (\Delta-1/2))\,,\label{3.75}
\end{equation}
where we parametrized the Casimir eigenvalue as:
\begin{equation}
    L_C=\frac{1}{(q-1/q)^2}\,2\cosh(2\log q (\Delta-1/2))\,,\quad \Delta>0\,.
\end{equation}
The matrix elements become manifestly independent of $\gamma$ and $\beta$: $R_{\Delta\, 0 0}(g)=R_{\Delta\, 0 0}(\phi)$. Since we are ultimately only interested in inner products of such operators, and since such inner products \eqref{ipfinal} sample only at $\phi=-n\log q$ we are only interested in the operator at these points $R_{\Delta\, 0 0}(n)$. It is easy to see that a solution to \eqref{3.75} is:
\begin{equation}
\boxed{\mathcal{O}_\Delta=R_{\Delta\, 00}(n)=q^{2 n \Delta}}\,,\label{3.76}
\end{equation}
The matrix element of this operator insertion matches the DSSYK calculation of a bilocal operator \cite{Berkooz:2018jqr,Berkooz:2018qkz,Lin:2022rbf}
\begin{align}
    \bra{\theta_1}\mathcal{O}_\Delta\ket{\theta_2}&=\sum_{n=0}^{+\infty}q^{-2 n} \bra{\theta_1} q^{2 n \Delta}\ket{n} \braket{n\rvert \theta_2}\nonumber\\&=\sum_{n=0}^{+\infty}\frac{q^{2n\Delta}}{(q^2;q^2)_n}H_n(\cos(\theta_1)\rvert q^2)H_n(\cos(\theta_2)\rvert q^2)=\frac{(q^{4\Delta};q^2)_\infty}{(q^{2\Delta\pm 2\i \theta_1\pm 2\i \theta_2};q^2)_\infty}\,.
\end{align}
Therefore the putative gravitational two-point function becomes
\begin{align}
    \bra{0}e^{-\beta_1 H}\mathcal{O}_\Delta e^{-\beta_2 H}\ket{0}&=\int_0^\pi \d\theta_1\,\rho(\theta_1)\,e^{-\beta_1\,\frac{1}{(q-1/q)^2}2\cos(\theta_1)}\int_0^\pi \d\theta_2\,\rho(\theta_2)\,e^{-\beta_2\,\frac{1}{(q-1/q)^2}2\cos(\theta_2)}\bra{\theta_1}\mathcal{O}_\Delta\ket{\theta_2}\nonumber\\&=\begin{tikzpicture}[baseline={([yshift=-.5ex]current bounding box.center)}, scale=0.7]
 \pgftext{\includegraphics[scale=1]{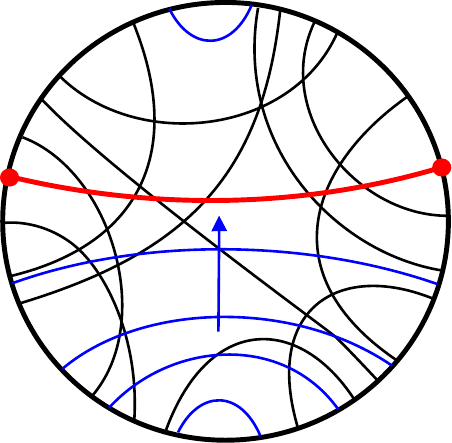}} at (0,0);
    \draw (2.6,-1) node {$\beta_1$};
    \draw (-2.3,1.6) node {$\beta_2$};
    \draw (-2.9,0.5) node {\color{red}$\mathcal{O}_\Delta$};
    \draw (2.9,0.5) node {\color{red}$\mathcal{O}_\Delta$};
  \end{tikzpicture}\label{3.79}
\end{align}
in which one simply inserts the previously obtained building blocks. This equation matches \eqref{eq:2pt}. 

These results pave the way for identifying a bulk description of these operators. For instance, in BF theory you obtain this operator by inserting a Wilson line in the path integral \cite{Blommaert:2018oro}:
\begin{equation}
    R_{\Delta\,00}\bigg(\mathcal{P}\exp\bigg(\int_{x_1}^{x_2} A\bigg) \bigg)\,,
\end{equation}
which in turn for SL$(2,\mathbb{R})$ (see for instance \cite{Iliesiu:2019xuh}) is the path integral of a particle with $m^2=\Delta(\Delta-1)$ propagating through the JT gravity bulk.

In summary, we have shown that DSSYK amplitudes follow from quantum mechanics on SU$_q(1,1)$ with constraints \eqref{constraints}. This embedding allows us to write up Schwarzian-type path integral descriptions in section \ref{sect:pathintegrals}, and helps in identifying a first order (gauge theory-like) bulk gravitational description of DSSYK.

\section{q-Liouville and q-Schwarzian boundary actions for DSSYK}\label{sect:pathintegrals}
In this section, we will construct classical dynamical (phase space) actions involving ordinary commutative fields, which upon quantization give rise to the non-commutative Hamiltonian quantum systems described in section \ref{Sec:rep} (which as we showed describe DSSYK). This classical dynamical system is a consistent description of what one would call a ``particle on $\Uqsu$'' (with constraints). The resulting actions can then be used in a path integral description to compute the same amplitudes which we computed in section \ref{Sec:rep} (there from a Hamiltonian point of view).

\subsection{Non-commutative coordinates from canonical quantization}\label{sect:4.1}
We will ad hoc construct a classical phase space, which upon quantization produces the Hamiltonian systems of section \ref{Sec:rep}. We consider a classical 6-dimensional phase space with coordinates $\g,\p,\b,p_\g,p_\p,p_\b$ equipped with the symplectic two-form
\begin{equation}\label{measure}
    \omega=\d \g\wedge \d p_\g +\d\b\wedge \d p_\b +\d(\phi-\i\log q\,(\g p_\g+\b\,p_\b))\wedge \d p_\phi\,.
\end{equation}
These coordinates are not Darboux (canonical) coordinates of the symplectic manifold. However, one can find canonical coordinates by transferring to new coordinates $\g,\varphi,\b,p_\g,p_\varphi,p_\b$ where we only change the $\phi$-coordinate as:
\begin{equation}
    \varphi=\phi-\i\log q\, (\g p_\g+\b p_\b)\,,\quad  p_\varphi=p_\phi\,.\label{canonicalcoo}
\end{equation}

This symplectic structure implies the usual Poisson brackets
\begin{equation}
    \{\g,p_\g \}=1,\quad \{\b,p_\b\}=1, \quad \{\p,p_\p \}=1\,,
\end{equation}
supplemented by
\begin{equation}
    \{\g,\phi\}=\i\log q\,\g\,,\quad \{p_\g,\phi\}=-\i\log q\,p_\g\,,\quad \{\b,\phi\}=\i\log q\,\b\,,\quad \{p_\b,\phi\}=-\i\log q\,p_\b\,.
\end{equation}
Now we perform canonical quantization by the usual prescription:
\begin{equation}
    p_\g \to -i\hbar\, \frac{d}{d\g }, \quad p_\varphi \to -i\hbar\,\frac{d}{d\varphi}, \quad p_\b \to -i\hbar \,\frac{d}{d\b}\,.\label{4.5}
\end{equation}
The point that we want to make is that ordinary canonical quantization, after the change of coordinates \eqref{canonicalcoo}, gives rise quantum mechanically to non-commutative coordinates $\g,\phi,\b$ with algebra \eqref{cooalg}, and the algebra of derivatives \eqref{dercomm1} and \eqref{dercomm2}.
In particular canonical quantization \eqref{4.5} implies
\begin{equation}
    \comm{\g}{p_\g}=\i\hbar\,,\quad \comm{\varphi}{p_\varphi}=\i\hbar\,,\quad \comm{\b}{p_\b}=\i\hbar\,.\label{4.7}
\end{equation}
When one defines according to \eqref{cooalg} the new coordinate $\p=\varphi+\log q\, (\g \partial_\g+\b \partial_\b)$, one finds immediately from \eqref{4.7} the coordinate algebra
\begin{equation}
    \comm{\g}{\phi}=-\log q\,\g\,,\quad  \comm{\b}{\phi}=-\log q\,\b\,,\quad\comm{\b}{\g}=0\,,
\end{equation}
which are precisely the relations \eqref{cooalg} for the non-commutative coordinates of $\Uqsu$. The algebra with the derivatives \eqref{dercomm1} and \eqref{dercomm2} follows in the same manner. Here we have rescaled $q$ in the following manner $\log q_\text{quant}=\hbar \log q_\text{class}$. Such scaling is required to describe the classical phase space underlying the quantum Hamiltonian system of section \ref{Sec:rep}. 
From here on, we put $\hbar=1$ in all equations to streamline notation.

This procedure should come as no surprise to readers familiar with quantum mechanics on non-commutative spaces \cite{Morariu:2001dv}. The usual approach is to consider the non-commutative algebra of coordinates (and their derivatives) and find the Darboux basis of canonical coordinates. The result is an ``ordinary'' system with commutative coordinates, and eigenstates of the Hamiltonian can be expressed as wavefunctions of those ordinary commutative coordinates.

\subsection{Path integral description of a particle on SU$_q(1,1)$}
To construct the classical action for particle on SU$_q(1,1)$, one first needs to construct the corresponding classical symmetry generators on phase space. Consider the following set of such classical currents
\begin{equation}\label{poissongen}
\begin{aligned}
    &h_L=\frac{1}{2}p_\phi-\g p_\g\,.
    \\&f_L=-\frac{e^{-2\i \log q\,\g p_\g}-1}{ 2\i\log q\,\g},
    \\& e_L=e^{-2\phi}\,e^{i\log q\,(-p_\phi+2\g p_\g)}\,\frac{e^{2\i \log q\,\b p_\b}-1}{2\i\log q\,\b}-\g\,e^{\i\log q\, p_\phi}\,\frac{e^{2\i \log q\,(\g p_\g-p_\phi)}-1}{2\i\log q}.
\end{aligned}
\end{equation}
There are the classical limits of the left-regular realization of the quantum algebra $\Uqsual$ \eqref{LF}, \eqref{LH} and \eqref{LE} obtained by taking $a_L=\i \hbar L_A$ and taking $\hbar\to 0$. These satisfy the Poisson brackets:
\begin{equation}
    \{h_L,e_L\}=e_L\,,\quad  \{h_L,f_L\}=-f_L\,,\quad  \{e_L,f_L\}=\frac{q^{2\i h_L}-q^{-2\i h_L}}{2\i\log q}\,,\label{chargealgebra}
\end{equation}
which is indeed the classical equivalent of the $\Uqsual$ algebra \eqref{eq:qalg}. Similarly the currents
\begin{equation}\label{poissongenright}
\begin{aligned}
    &h_R=-\frac{1}{2}p_\phi+\b p_\b\,.
    \\&f_R=-e^{-2\phi}\,\frac{e^{2\i \log q\,\g p_\g}-1}{2\i\log q\,\g}-\b\,\frac{e^{2\i \log q\,(p_\phi-\b p_\b)}-1}{2\i\log q},
    \\& e_R=-\,e^{-\i\log q\, p_\phi}\,\frac{e^{2\i \log q\,\b p_\b}-1}{ 2\i\log q\,\b}.
\end{aligned}
\end{equation}
satisfy \eqref{chargealgebra} and are the classical limits of the right regular realization or $\Uqsual$ \eqref{RA}.

To obtain a system which has these currents generating symmetries, we should choose a Hamiltonian which has vanishing Poisson brackets with all of them. There is only one independent possibility:
\begin{align}
   \mathbf{H} &=e f+\bigg( \frac{\sin( \log q\,h)}{\log q}\bigg)^2\nonumber\\&=-\frac{1}{4}\frac{e^{\i\log q\,p_\phi}}{(\log q)^2}-\frac{1}{4}\frac{e^{-\i\log q\,p_\phi}}{(\log q)^2}-\frac{1}{4}\,e^{-2\phi} \,e^{-\i\log q\, p_\phi}\frac{1}{\beta \gamma}\frac{1}{(\log q)^2}(e^{2\i \log q\,\b p_\b}-1)(e^{2\i \log q\,\g p_\g}-1)\,,\label{hamiltonian}
\end{align}
which is the classical limit of the $\Uqsu$ Casimir \eqref{3.24}. As a consistency check, one can take the classical limit $q\to 1^-$ of all currents as well as the Hamiltonian to recover the particle on SL$(2,\mathbb{R})$, discussed in section \ref{sect:JTreview}. Using the canonical coordinates \eqref{canonicalcoo}, it is elementary to write up the Feynman phase space path integral description for this model:
\begin{align}
    \int \dpi \g \dpi \varphi \dpi \b & \dpi p_\g \dpi p_\varphi \dpi p_\b\exp\bigg(\i\int \d t\,\bigg(p_\varphi\,\varphi'+p_\b\, \b'+p_\g\, \g'+\frac{1}{4}\frac{e^{\i\log q\,p_\varphi}}{(\log q)^2}+\frac{1}{4}\frac{e^{-\i\log q\,p_\varphi}}{(\log q)^2}\nonumber\\& \qquad\qquad\qquad+\frac{1}{4}\,e^{-2\varphi} \,e^{-\i\log q\, p_\varphi}\frac{1}{\beta \gamma}\frac{1}{(\log q)^2}(e^{-2\i \log q\,\b p_\b}-1)(e^{-2\i \log q\,\g p_\g}-1)\bigg)\bigg)\,,\label{pathintegralbeforeconstraints}
\end{align}
which should be considered the generalization of \eqref{2.4} to $\Uqsu$.

This system is periodic in the momentum variable $p_\varphi$ with period $\frac{2\pi}{\log q}$, consistent with the quantized system effectively being defined on some $\varphi$-lattice with spacing $\log q$.\footnote{In contrast for Liouville gravity the $e^{\i\log q\,p_\varphi}$ factors are replaced by real exponentials resulting in no $\varphi$-periodicity.} Likewise, the system is periodic in $p_{\log \beta} = \beta p_\beta$ and $p_{\log \g} = \g p_\g$, consistent with an effective lattice spacing $2 \log q$ for the coordinates $\log \beta$ and $\log \g$ in the quantum theory, as indeed suggested by the Haar measure \eqref{haar}.

This theory will result in the Hamiltonian system described in section \ref{Sec:rep} upon quantizing. Indeed, (as we already learned in section \ref{sect:4.1}) the coordinates $\g, \phi, \b$ and their derivatives acquire the correct non-commutative structure upon canonical quantization.  Furthermore, the currents $a_L,a_R$ become the quantum generators $L_A,R_A$. 

\subsection{Constraints and the q-Liouville path integral}
Let us now mimic the analysis for JT gravity on an interval with two boundaries of section \ref{sect:JTreview}, resulting in a generalization of the Liouville action \eqref{liouvilleaction}. For this we need to impose the gravitational boundary conditions \eqref{constraints}, which we reproduce here for convenience
\begin{equation}
    q^{-L_H}L_F=\i \frac{q^{1/2}}{q-q^{-1}}\,,\quad q^{R_H} R_E= -\i \frac{q^{1/2}}{q-q^{-1}}\,.\label{4.15}
\end{equation}
Here we want to impose these as classical constraints on phase space. The associated classical constraints can be written as 
\begin{equation}
    \psi_L=\frac{1}{\g}(e^{-3\i \log q\, \g p_\g}-e^{-\i \log q\, \g p_\g})\,e^{\i \log q\, p_\varphi/2}=\i\,,\quad 
    \psi_R=\frac{1}{\beta}(e^{\i\log q\, \beta p_\b}-e^{-\i\log q\, \beta p_\b})\,e^{-\i \log q\,p_\varphi/2}=-\i \, .
\end{equation}
By construction (as the classical limit of \eqref{4.15}), or via explicitly computing Poisson brackets, one sees that these constraints have vanishing Poisson brackets with one another and with the Hamiltonian
\begin{equation}
   \{\psi_L,\mathbf{H}\}=0\,,\quad \{\psi_R,\mathbf{H}\}=0\,,\quad\{\psi_L,\psi_R\}=0\,,
\end{equation}
so they are indeed first-class constraints consistent with time evolution. The constrained Hamiltonian becomes
\begin{equation}
    \mathbf{H} \stackrel{?}{=} -\frac{1}{4}\frac{e^{\i\log q\,p_\varphi}}{(\log q)^2}-\frac{1}{4}\frac{e^{-\i\log q\,p_\varphi}}{(\log q)^2}+\frac{1}{4}\,e^{-2\varphi} \frac{e^{\i\log q (- p_\varphi+\g p_\g-\b p_\b)}}{(\log q)^2}\,.\label{4.18}
\end{equation}
Of course \cite{dirac2001lectures} since we are imposing constraints on a Hamiltonian system we should also check whether there are pure gauge modes, and what our physical coordinates are. Those physical coordinates have vanishing Poisson brackets with $\psi_L$ and $\psi_R$. Besides $p_\varphi$ there is \emph{one} additional physical combination
\begin{equation}
    \{\varphi+\i \log q (\beta p_\beta -\g p_\g)/2,\psi_L\}= \{\varphi+\i \log q (\beta p_\beta -\g p_\g)/2,\psi_R\}=0\,.
\end{equation}
This makes sense, as the constrained Hamiltonian also only depends on $p_\varphi$ and this second coordinate. However, the remaining two coordinates (in the 4d reduced classical phase space) are gauge modes and non-physical. They should be gauge-fixed, and one simple gauge choice is to fix $\beta=\g=0$ \eqref{gaugefixxx}.

Renaming the physical coordinates as $p_\phi$ and $\phi$ (for notational comfort) one recovers a constrained simple Hamiltonian:\footnote{With the gauge-choice $\g=\b=0$ this is actually precisely what the physical coordinates become.}
\begin{equation}
    -4 (\log q)^2 \mathbf{H} = (1- e^{-2\phi})\,e^{-\i\log q\,p_\phi}+e^{\i\log q\,p_\phi}\,.\label{4.20}
\end{equation}
This is the transfer matrix of DSSYK \cite{Berkooz:2018qkz,Berkooz:2018jqr,Lin:2022rbf} and our equation \eqref{3.43}. The path integral becomes
\begin{equation}
    \boxed{\int\dpi \phi \dpi p_\phi\exp\bigg(\i\int\d t\bigg(p_\phi \,\phi' +\,\frac{1}{4}\frac{e^{-\i\log q\,p_\phi}}{(\log q)^2}+\frac{1}{4}\frac{e^{\i\log q\,p_\phi}}{(\log q)^2} - e^{-2\phi} \,\frac{1}{4}\frac{e^{-\i\log q\,p_\phi}}{(\log q)^2}\bigg)\bigg)} \,. \label{q-liouville}
\end{equation}
This path integral describes the two-sided dynamics from the boundary perspective.

By diagonalizing the Hamiltonian of this dynamical system, we obtain the wavefunctions \eqref{waefunanswer} but without the factor $q^n$. This was to be expected, and is similar to the difference between $K_{2 \i E^{1/2}}(2e^{-\phi})$ and $e^{-\phi}K_{2 \i E^{1/2}}(2e^{-\phi})$ in the JT story, as the Haar measure \eqref{haar} has a factor $q^{-2 n}$ whereas the path integral naturally comes with a flat measure.

\subsection{The q-Schwarzian path integral}
Finally we reproduce the analogue of the Schwarzian phase space path integral \eqref{sch4d} for which we only impose one constraint $\psi_L=\i$. Again, we can gauge-fix to $\g=0$ which reduces the Hamiltonian to\footnote{One technical comment is that this depends slightly on the choice of $\alpha_1$ and $\alpha_2$ discussed in footnote \ref{fn:19}. An independent calculation of the wavefunctions \eqref{4.31} of this q-Schwarzian theory should show which choice matches with DSSYK. The only effect would be changing the prefactor of $\i \log q\,p_\phi$ in the exponential in the final term.}
\begin{align}
   \mathbf{H} = -\frac{1}{4}\frac{e^{\i\log q\,p_\phi}}{(\log q)^2}-\frac{1}{4}\frac{e^{-\i\log q\,p_\phi}}{(\log q)^2}+\frac{1}{2\log q}\,e^{-2\phi} \,e^{-\i\log q\, 3 p_\phi/2}\,\frac{e^{2\i\log q\,\b p_\b}-1}{2\i\log q\, \b}\,.
\end{align}
The generalization of the Schwarzian path integral \eqref{sch4d} to the DSSYK context is then
\begin{align}
    \int \dpi \phi \dpi \b & \dpi p_\phi \dpi p_\b\exp\bigg(\i\int \d t\,\bigg(p_\varphi\,\varphi'+p_\b\, \b'+\frac{1}{4}\frac{e^{\i\log q\,p_\varphi}}{(\log q)^2}+\frac{1}{4}\frac{e^{-\i\log q\,p_\varphi}}{(\log q)^2}\nonumber\\& \qquad\qquad\qquad\qquad\qquad\qquad\qquad+\frac{1}{2\log q}\,e^{-2\phi} \,e^{-\i\log q\, 3 p_\phi/2}\,\frac{e^{2\i\log q\,\b p_\b}-1}{2\i\log q\, \b}\bigg)\bigg)\,.\label{4.30}
\end{align}
Notice indeed that for $q\to 1$ this reproduces \eqref{sch4d}. Unlike \eqref{sch4d}, one cannot immediately integrate over any of the variables to produce a simpler description.

Whilst directly computing this path integral looks daunting, as we have shown in section \ref{Sec:rep} one can solve this quantum problem 
simply by canonical quantization using the (constrained) Casimir equation of the underlying $\Uqsu$ quantum group. The resulting wavefunctions carry an extra free label $s$ as only one of the generators was diagonalized:
\begin{equation}
    \braket{\phi,\b\rvert \theta\,s}=R_{\theta\,\i s}(\phi,\b). \label{4.31}
\end{equation}
This picture corresponds to the ``angular'' slicing discussed for JT gravity and BF gauge theory in \cite{Blommaert:2018iqz}.
The one-sided wavefunctions \eqref{4.31} will have a bulk interpretation as having support on the black hole horizon at the ``inner'' endpoint \cite{Blommaert:2018iqz,Mertens:2022ujr}, with a frozen particle on $\Uqsu$ describing edge modes $s$ on the horizon. It would be interesting to make this more explicit.

Finally, we remark that that whereas the two-sided $q$-Liouville action \eqref{q-liouville} can be derived directly from a chords description \cite{Lin:2022rbf}, the same cannot be said about the one-sided $q$-Schwarzian description \eqref{4.30}, for which one really must start from the embedding in the full particle on $\Uqsu$ description \eqref{pathintegralbeforeconstraints}. 

\section{Concluding remarks}
We conclude this work with several comments. 
\label{s:conclu}

First we'll propose a candidate gravitational path integral dual to our story (and hence to DSSYK). Then we will compare our realizations of section \ref{Sec:rep} with the ones of \cite{Berkooz:2022mfk}. Finally we briefly discuss some potential generalizations.

\subsection{Towards a gravitational dual of DSSYK}\label{sect:discbulkdual}
For the bulk gravitational dual we propose the following first-order formulation, most commonly known as a Poisson-sigma (gauge theoretical) model
\begin{equation}
    \int \dpi A^B\dpi \chi_B\,\exp\bigg(\i\int_{\mathcal{M}}  \bigg(A^B\wedge d\chi_B+\frac{1}{2}P_{B C}(\chi) A^B\wedge A^C  \bigg)-\i \int \d t\, H(\chi)\bigg)\,,\label{PSMbos}
\end{equation}
consisting of one-forms $A^B$ and scalars $\chi_B$. The target space is 3-dimensional (the indices $B,C$ can take three values), and we will label the components as $A^0, A^1, A^H$ and  $\chi_0,\chi_1,\chi_H$. The Poisson matrix takes the form
\begin{equation}
    P_{H 0}(\chi)=\chi_1\,,\quad P_{H 1}(\chi)=\chi_0\,,\quad P_{0 1}(\chi)=-\frac{\sin(2\log q \chi_H)}{2\log q}\,.
\end{equation}
We choose the boundary term as
\begin{equation}
    H(\chi)=\chi_0^2-\chi_1^2-\frac{\cos(2\log q \chi_H)}{2\log q}\,.\label{hhhhh}
\end{equation}
which forces upon us the boundary equation of motion
\begin{equation}
    A^B_t=\frac{d}{d\chi_B} H(\chi)\,,\label{5.4}
\end{equation}
which should be considered the non-linear generalization of \eqref{bc}.

One can show that this model is topological with only six total degrees of freedom \cite{Cattaneo:2001bp} in classical phase space, because one can integrate out the bulk values of $A_t$ and mod out by redundancies associated with this constraint. The Hamiltonian of the theory comes only from the boundary term in \eqref{PSMbos}. One furthermore finds that half of the classical phase space variables are (suppose the boundary is at $x=0$) $\chi_B(0)$, with the following Poisson brackets \cite{Cattaneo:2001bp}
\begin{equation}
    \{\chi_H(0),\chi_0(0)\}=\chi_1(0)\,,\quad \{\chi_H(0),\chi_1(0)\}=\chi_0(0)\,,\quad\,\{\chi_0(0),\chi_1(0)\}=-\frac{\sin(2\log q \chi_H(0))}{2\log q}\,.
\end{equation}
These are precisely the Poisson brackets of our $\Uqsu$ currents $e, f$ and $h$ \eqref{chargealgebra} with the identifications
\begin{equation}
    \chi_H(0)=h\,,\quad \chi_0(0)=\frac{e+f}{2}\,,\quad \chi_1(0)=\frac{e-f}{2}\,,
\end{equation}
and moreover the Hamiltonian due to \eqref{hhhhh} is precisely the classical $\Uqsu$ Hamiltonian \eqref{hamiltonian}
\begin{equation}
    H=e f-\frac{\cos(2\log q h)}{2\log q}\,.
\end{equation}
This guarantees that the quantization of this system matches with our boundary path integral \eqref{pathintegralbeforeconstraints}. One can rewrite \eqref{PSMbos} as a dilaton gravity model by introducing $A^B=(e^0,e^1,\omega)$ and $\chi_B=(X_0,X_1,\Phi)$
\begin{align}
\label{dilgrav1}
\exp\bigg( \i \int_{\mathcal{M}}  \bigg(\Phi\, d\omega + X_a d e^a -\frac{1}{2} V(\Phi)\,e^0\wedge e^1+X_0\,\omega\wedge e^1+X_1\,\omega\wedge e^0\bigg)
+ \i S_{\text{bdy}}\bigg)\,,
\end{align}
which after integrating out the torsion constraints attains the second-order form\footnote{One then also deduces from \eqref{5.4} the boundary conditions
\begin{equation}
    \omega_t+\frac{\sin(2\log q\,\Phi)}{\log q}=0\,,
\end{equation}
generalizing the JT boundary conditions $\omega_t+2\Phi=0$ relating the length (or curvature) and boundary value of the dilaton. If the boundary value of the dilaton were fixed for instance to $\Phi=\pi/2 \log q$, this conditions says we are at the location of some static observer (pode/antipode) in the $R=2$ region where $\omega_t=0$.}
\begin{align}
\label{dilgrav}
\exp\bigg(\frac{\i}{2}\int_{\mathcal{M}}  d^2 x\sqrt{-g}\, (\Phi R + V(\Phi))+\i \int_{\partial \mathcal{M}} \d t\sqrt{-h} \Phi K \bigg)\,,\quad  V(\Phi)=\frac{\sin(2\log q\, \Phi)}{\log q}\,,
\end{align}
as claimed in the introduction around \eqref{1.2}.\footnote{We are being slightly schematic here, since we did not take into account the boundary term \eqref{hhhhh}, nor did we investigate the translation of the constraints \eqref{constraints} into gravitational boundary conditions. We leave a more detailed analysis to the future \cite{wip}.}

Remarkably this very model was also suggested as a candidate bulk description in section 4.3 of \cite{Goel:2022pcu} and in \cite{HVerlindetalk} based on completely different arguments using the $(G,\Sigma)$ formulation of the SYK model \cite{Cotler:2016fpe}. 

We want to make two more comments on this dilaton gravity model. Firstly, we note the structural similarity with our boundary actions \eqref{q-liouville}, \eqref{4.30}. In particular, note the periodicity in $\Phi$ of the dilaton potential. This stems from the periodicity in our boundary Hamiltonian, which ended up discretizing spacetime into chords on the quantum mechanical level. See the discussion around \eqref{3.46}.

Secondly, the classical solution of the sine dilaton gravity model has on-shell Ricci scalar\footnote{In DSSYK, one typically considers the coupling constant $-1/\log q=N/p^2$ of order $1$ \cite{Cotler:2016fpe}. Upon rescaling the dilaton by $\log q$ one sees that this is indeed also the coupling constant in our gravitational theory \eqref{dilgrav}. This suggests the theory should be under semiclassical control when considering $-\log q\ll 1$. Indeed \cite{Goel:2023svz}, the DSSYK amplitudes have non-trivial and interesting semiclassical approximations when taking $-\log q\ll 1$ but keeping the energy finite (if the energy is taken to zero simultaneously, one recovers JT gravity). One would hope to reproduce these results from dilaton gravity semiclassics using \eqref{dilgrav}. For $q$ or order $1$, quantum fluctuations will be large.}
\begin{equation}
R = - 2 \cos( 2 \log q\, \Phi)\,.
\end{equation}
Choosing a radial coordinate in the bulk as $r = \Phi$ \cite{Witten:2020ert}, we see that the radial region $r\sim 0$ is essentially AdS$_2$ with a curvature $R=-2$. Perhaps more tantalizing is the fact that for $r\sim -\frac{\pi}{2\log q}$ one finds a radial region of approximately constant positive curvature $R=+2$, dS$_2$.\footnote{This is not the first example of en embedding of dS$_2$ regions in AdS$_2$ (see e.g. \cite{Anninos:2017hhn,Anninos:2020cwo,Anninos:2022qgy}), but it does seem to be the first one derived from a UV complete boundary description.} So this model provides us with an opportunity to prove in more detail the dictionary between DSSYK and dS$_2$ physics proposed in \cite{Susskind:2021esx,Susskind:2022bia,Lin:2022nss,Rahman:2022jsf,Susskind:2022dfz}. This is exciting as no (top-down) models of dS quantum gravity are well understood to date, therefore this deserves significantly more study.

The corresponding classical solution for the metric is of the form
\begin{equation}
\d s^2 = -\frac{1}{f(r)}\, \d t^2 + f(r)\,\d r^2\,,\quad f(r)=\frac{2 (\log q)^2}{\cos( 2 \log q\, r_h)-\cos( 2 \log q\, r)}\,,
\end{equation}
which has a black hole horizon at $r_h$ and a cosmological horizon at $\pi/\log q-r_h$ .The curvature at both horizons is $R=-2\cos (2 \log q\, r_h)$. Depending on $r_h$ (set by the total energy in the spacetime), we have two cases:
\begin{enumerate}
    \item For $r_h>\frac{\pi}{4\log q}$ both horizons and the intermediate regions have positive cosmological constant $R>0$. In the regime $r_h\sim -\pi/2\log q$ the entire space between the horizons has constant curvature $R=+2$, and one would recover dS$_2$ JT gravity. In light of \cite{Susskind:2021esx,Susskind:2022bia,Lin:2022nss,Rahman:2022jsf,Susskind:2022dfz} it would make sense if this would correspond with boundary conditions where $\beta=0$. We leave a further investigation to future work \cite{wip}.
    \item For $r_h<\frac{\pi}{4\log q}$ the horizons are in a region with $R<0$, however there is still a positively curved region $R>0$ in between the horizons. The AdS$_2$ JT black hole is recovered by $r\ll 1/\log q$ whilst zooming in on the near horizon region $r_h\ll 1/\log q$. This corresponds with zooming in on low energies (deep IR in the bulk).
\end{enumerate}
These regimes of interest are illustrated graphically:
\begin{equation}
    \begin{tikzpicture}[baseline={([yshift=-.5ex]current bounding box.center)}, scale=0.7]
 \pgftext{\includegraphics[scale=1]{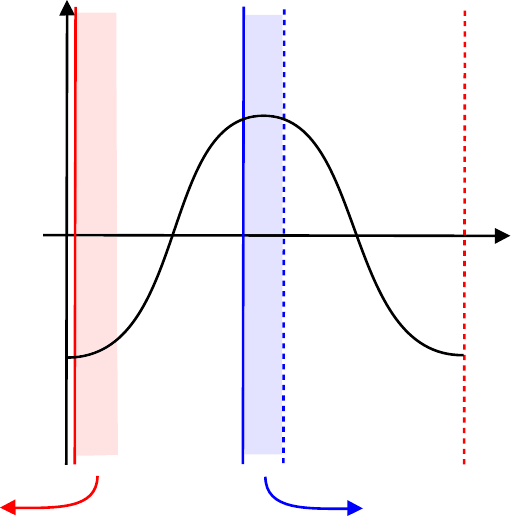}} at (0,0);
    \draw (2.5,-0.3) node {$\Phi$};
    \draw (1.15,1.15) node {$R$};
    \draw (-1.6,1.9) node {\color{red}$2$};
    \draw (0.1,1.9) node {\color{blue}$1$};
    \draw (3.5,-2.5) node {\color{blue}approximately dS$_2$};
    \draw (-5.2,-2.5) node {\color{red}approximately AdS$_2$};
  \end{tikzpicture}
\end{equation}
Here we indicated the black hole horizon $\Phi=r_h$ (full lines) and the cosmological horizon (dotted lines).

It is worthwhile to contrast this discussion with the semi-classical geometry distilled in \cite{Goel:2023svz} from the single Liouville field $g$ describing the double-scaled regime of the $(G,\Sigma)$ formulation of SYK \cite{Cotler:2016fpe}. That Liouville geometry leads to AdS$_2$ which gets corrected by loop effects to even more negative values of the Ricci scalar (using the Ricci scalar computed with the quantum expectation value of the metric).

This seems at odds with our current geometrical interpretation. However, this Liouville geometry is not the same as the metric in the dilaton-gravity description. Indeed, this puzzle has been encountered before in the context of Liouville gravity and the minimal string \cite{StanfordSeiberg,Mertens:2020hbs}, where the actual metric $g_{\mu\nu}$ and dilaton field $\Phi$ are orthogonal linear combinations of the Liouville field $g$ and the matter sector field.

\subsection{Factorization across bulk entangling surfaces}
A further question one can ask is whether we can factorize the bulk Hilbert space using this (quantum) group theoretic approach \cite{Harlow:2018tqv}. In \cite{Blommaert:2018iqz}, such an argument was presented for JT gravity, and in \cite{Mertens:2022ujr,Mertens:2022aou} this was extended to pure AdS$_3$ gravity. As alluded to at the end of section \ref{sect:pathintegrals}, technically, one needs to understand more general representation matrix elements that represent one-sided wavefunctions which connect the holographic boundary to the black hole horizon. In this work, we have computed the two-sided wavefunction $R_{\theta \, \i \i}(g)$, whereas the one-sided wavefunction $R_{\theta \, \i s}(g)$ and the interior wavefunction $R_{\theta \, ss}(g)$ were not studied. Finding expressions for these seems to be within reach by solving the Casimir eigenvalue problem. In particular, understanding the physical meaning of such a factorization map in this non-local context of DSSYK would be interesting.

\subsection{Comparison with Liouville gravity and beyond}\label{sect:5.4}
It was argued in \cite{StanfordSeiberg,Mertens:2020hbs,Fan:2021bwt} that Liouville gravity (and the minimal string) can be formulated in terms of a dilaton gravity model like \eqref{dilgrav}, but with a sinh dilaton potential, whose amplitudes are governed by the quantum group SL$^+_q(2,\mathbb{R})$. There is hence a clear parallel to this DSSYK story. However, there are also noteworthy technical differences.

For the quantum group relevant to Liouville gravity the Haar measure used when defining the inner product is continuous (see section 2.3 of \cite{Fan:2021bwt}), unlike in our case where is is discrete \eqref{haar}. Nevertheless the Casimir equation is still a difference equation, which leads to the question how the ``infinite degeneracy problem'' discussed around \eqref{differenceeq} is resolved in that case. It turns out that the two-boundary wavefunctions of Liouville gravity are the simultaneous solution of \emph{two} incommensurate Casimir difference equations, \eqref{3.24} and its dual related by $b\to 1/b$ \cite{Mertens:2022aou} with $q=e^{\pi i b^2}$.\footnote{Note that the analysis of section \ref{sect:pathintegrals} does not include this (quantum) modular duality at the level of the classical action, which is as expected. This mirrors an old subtlety in 2d Liouville CFT, where the classical action has no obvious $b\to1/b$ symmetry, but it is still present in quantum amplitudes (and exploited in the conformal bootstrap when using Teschner's trick \cite{Teschner:1995yf}). See for instance \cite{ORaifeartaigh:1998hjm,ORaifeartaigh:2000dkf,Teschner:2001rv}.} These two equations lead to a unique solution and hence no infinite degeneracy (but for a different reason than in the DSSYK case). Also the Casimir operator is Hermitian for Liouville gravity, so left-and right-eigenfunctions are related simply by complex conjugation.

It is relatively straightforward to modify the boundary action analysis of section \ref{sect:pathintegrals} to the situation of Liouville gravity. Given these three datapoints (JT, DSSYK, Liouville gravity), it is tempting to wonder whether one can generalize our boundary action description further to arbitrary dilaton potentials $V(\Phi)$.

A technical issue worth investigating is the range of validity of the Gauss-Euler decomposition \eqref{UqGE} for the application to DSSYK. Indeed, it is known that in the classical case the Gauss-Euler coordinates do not cover the full SL$(2,\mathbb{R})$ group manifold. However, we observed in earlier work \cite{Blommaert:2018iqz,Blommaert:2018oro} that we produce the correct results in JT gravity using this decomposition. The conundrum in that case was resolved by realizing that JT gravity is not just SL$(2,\mathbb{R})$ BF gauge theory. In particular, we argued it to be related to the positive semi-group SL$^+(2,\mathbb{R})$. The latter has a Gauss-Euler decomposition that is contained within the Gauss-Euler patch of the full SL$(2,\mathbb{R})$ group manifold, resolving this puzzle.
For Liouville gravity, since its description is based on the $q$-deformation of this positive semi-group, it is apparent that the same resolution is true. However, for DSSYK it is not so obvious why it works a priori. The fact that our results match with the known amplitudes of DSSYK shows a posteriori that the same statement is true. It would be interesting to understand this from first principles.

\subsection{Generalizations to supersymmetric models}\label{sect:5.5}
The solution of DSSYK amplitudes was extended to the supersymmetric SYK \cite{Fu:2016vas} in \cite{Berkooz:2020xne}. Whereas the $\mathcal{N}=1$ version did not receive as much attention as the bosonic model, enough is known to allow for a similar computation of the amplitudes, which then begs for a similar interpretation in terms of quantum (super)group theory. The group theoretical structure governing these amplitudes is OSp$_q(1 \vert 2 , \mathbb{R})$ where $0<q<1$.\footnote{There does not seem to be notation to distinguish this from the other real forms of the complex group OSp$_q(1 \vert 2)$. In terms of the quantum algebra, the real forms were classified \cite{JLukierski_1992} and parallel the classification of the real forms of SL$_q(2)$. This real form was denoted there as $U_q(\mathfrak{osp}(1\vert 2,\mathbb{R}))$ (option (10a) in their Table 1).}

A natural conjecture for the Gauss decomposition of the quantum group element of OSp$_q(1 \vert2 , \mathbb{R})$ is:
\begin{equation}
\label{eq:guesss}
g \, \stackrel{?}{=} \, e^{2\theta_{\mm} F^-}e_{q^{-2}}^{\gamma E^-} \, e^{2\phi H} \, e_{q^2}^{\beta E^+} e^{2\theta_{\+} F^+}\,,\quad 0<q<1\,,
\end{equation}
in terms of the $3 \vert 2$ generators $H,E^\pm \vert F^\pm$ of the U$_q(\mathfrak{osp}(1|2, \mathbb{R}))$ quantum superalgebra:
\begin{align}
\label{salgeb}
q^H F^\pm = q^{\pm\frac{1}{2}}F^{\pm} q^H , \qquad \{F^+, F^-\} = \frac{q^{2H} -q^{-2H}}{q-q^{-1}}, \qquad \{F^\pm, F^\pm\} = \pm \frac{1}{2} E^{\pm}\,,
\end{align}
and the non-commutative supernumbers $\phi,\beta,\gamma \vert \theta_{\+},\theta_{\mm}$ with algebra
\begin{alignat}{4}
\label{eq:comm}
&e^\phi\gamma = q \gamma e^\phi\,, \quad &&e^\phi \beta= q \beta e^\phi\,, \quad &&[\beta,\gamma]=0\,, \nonumber\\
&e^\phi \theta_{\+} = q \theta_{\+} e^\phi\,, \quad &&e^\phi \theta_{\mm} = q \theta_{\mm} e^\phi\,, \quad &&[\theta_{\pmt}, \beta] = 0 = [\theta_{\pmt}, \gamma]\,, \nonumber\\
&\theta_{\+}\theta_{\mm} = - \theta_{\mm}\theta_{\+}\,, \quad &&\left(\theta_{\pmt}\right)^2 = 0
\end{alignat}
where the second and last lines are new. We prove these (anti)-commutation relations in Appendix \ref{app:detailssusy}.

Notice that the exponentials in \eqref{eq:guesss} for the fermionic generators did not get $q$-deformed because their series expansion truncates after the linear term in any case. It would be interesting to prove the expression \eqref{eq:guesss} directly, and then follow the main line of this work to find the analogous decomposition of the $\mathcal{N}=1$ DSSYK amplitudes.

Generalizing to $\mathcal{N}=2$ requires OSp$_q(2 \vert2 , \mathbb{R})$. It would likewise be interesting to match the $q$-representation theory of this quantum group with $\mathcal{N}=2$ DSSYK amplitudes.

\section*{Acknowledgments}
We thank Alejandro Cabo-Bizet, Damian Galante, Akash Goel, Henry Lin, Henry Maxfield, Vladimir Narovlansky, Xiao-Liang Qi and Douglas Stanford for useful discussions. AB was supported by the ERC-COG Grant NP-QFT No. 864583 and by INFN Iniziativa Specifica GAST. TM acknowledges financial support from the European Research Council (grant BHHQG-101040024). Funded by the European Union. Views and opinions expressed are however those of the author(s) only and do not necessarily reflect those of the European Union or the European Research Council. Neither the European Union nor the granting authority can be held responsible for them.

\appendix
\section{Quantum group SU$_q(1,1)$ and its Gauss decomposition }\label{app:gauss}
In the main text, we studied the representation theory of quantum group SU$_q(1,1)$. Here we give a bit more background, and comments on its relation with the quantum algebra $U_q(\mathfrak{su}(1,1))$.

\subsection{Hopf duality and the Gauss decomposition}
In this subsection, we provide some background and details on \eqref{UqGE} which we hope is sufficient to guide the reader through the actual proof \cite{Fronsdal:1991gf}. To do this, it is important to clarify the interrelation between the two different Hopf algebras $\Uqsu$ and $\Uqsual$.

The quantum group $\Uqsu$ (which in some literature is also called the coordinate Hopf algebra $\mathcal{O}_q(SU(1,1))$) is a Hopf algebra generated by four generators $a,b,c,d$, modulo the following relations\cite{Klimyk:1997eb}
\begin{equation}\label{a4}
    ab=qba\,, \quad cd=qdc\,, \quad ac=qca\,, \quad bd=qdb\,, \quad bc=cb\,, \quad ad-da=(q-q^{-1})bc\,,
\end{equation}
\begin{equation}\label{qdet}
    {\det}_q=ad-qbc=1.
\end{equation}
The second line is called the quantum determinant, and one can check that it commutes with all the generators of the algebra. The co-product on this Hopf algebra is defined as
\begin{equation}
\begin{aligned}
    & \Delta(a)=a\otimes a' + b \otimes c',\quad \Delta(b)= a\otimes b'+b\otimes d' ,
    \\& \Delta(c)=c\otimes a' + d \otimes c' ,\quad \Delta(d)=c\otimes b' + d \otimes d'
\end{aligned}\label{coprod}
\end{equation}
and by construction looks like how one would perform the ordinary matrix product of $(2\times 2)$-matrices.
For our purpose, it is convenient to use the related coordinates for the same algebra:
\begin{equation}\label{transag}
    a=e^{\phi}\,, \quad b=e^{\phi}\beta\,, \quad c=\gamma e^{\phi}\,, \quad d=e^{-\phi} +\gamma e^{\phi}\beta\,.
\end{equation}
These relations trivially satisfy the $q$-determinant condition. The algebra \eqref{a4} reduces to \eqref{cooalg}:
\begin{equation}
    \comm{\g}{\phi}=-\log q\,\g\,,\quad  \comm{\g}{\phi}=-\log q\,\g\,,\quad\comm{\b}{\g}=0\,.
\end{equation}

An a priori unrelated Hopf algebra is $\Uqsual$ \cite{Klimyk:1997eb,Berkooz:2018jqr,Lin:2023trc}, which is the algebra generated by three generators $E,F,H$, modulo the following relations
\begin{equation}
\label{eq:usal}
[H,E]=E\,,\quad [H,F]=-F\,,\quad [E,F] = \frac{q^{2H}-q^{-2H}}{q-q^{-1}}\,.
\end{equation}
It can be thought of as a deformation of the universal enveloping algebra of $\mathfrak{su}(1,1)$. The co-product on this Hopf algebra is defined as:
\begin{equation}
    \begin{aligned}
        \Delta(E)=E\otimes K +1\otimes E \,,\quad \Delta(F)=F\otimes 1 +K^{-1}\otimes F \,,\quad \Delta(K)=K\otimes K\,,
    \end{aligned}
\end{equation}
with $K=q^{2H}$.

Now, these two Hopf algebras are actually closely related in a way that mimics the relation between the Lie algebra and Lie group. The relation is through the mathematical construction of Hopf duality. We first discuss this in a general setting and come back to the concrete case of $\Uqsu$ and $\Uqsual$ below.

Let us denote a basis for the two Hopf algebras by $P_\alpha$ and $X^\alpha$ respectively, where $\alpha$ runs through all basis elements.\footnote{E.g. for a universal enveloping algebra, we can think of $\alpha$ as running through the Poincar\'e-Birkhoff-Witt basis.}
The dual pairing of Hopf algebras is defined \cite{Klimyk:1997eb} by a bilinear mapping $\left\langle\, .\,,\,.\, \right\rangle$ for which the dual basis elements of both algebras $P_\alpha$ and $X^\alpha$ are related as $\langle P_\alpha,X^\beta \rangle = \delta_\alpha^\beta$. Moreover, it has to satisfy the additional duality properties
\begin{equation}
\label{eq:dual}
\langle \Delta(P_\alpha),X^a \otimes X^b\rangle = \langle P_\alpha, X^aX^b\rangle, \qquad \langle P_\alpha P_\beta, X^a\rangle = \langle P_a \otimes P_b, \Delta(X^a) \rangle,
\end{equation}
where one defines the operation $\left\langle\, .\,,\,.\, \right\rangle$ on the tensor product as: $\left\langle \, . \, \otimes \, .\,,\,.\,  \otimes \, .\,  \right\rangle \equiv \left\langle \, .\,,\,.\,  \right\rangle \left\langle \, .\,,\,.\,  \right\rangle$. The duality properties \eqref{eq:dual} encode that multiplication and co-multiplication of both Hopf algebras get swapped under duality. Indeed, writing generically
\begin{alignat}{3}
\label{eq:exp1}
\Delta(P_\gamma) &= \sum_{\alpha,\beta} E^{\alpha\beta}_\gamma P_\alpha \otimes P_\beta, \qquad &&P_\alpha P_\beta && = \sum_{\gamma} F_{\alpha\beta}^\gamma P_\gamma,\\
\label{eq:exp2}
X^\alpha X^\beta &= \sum_c H^{\alpha\beta}_\gamma X^\gamma, \qquad
&&\Delta(X^\alpha) &&= \sum_{\beta,\gamma} G^{\alpha}_{\beta\gamma} X^\beta \otimes X^\gamma,
\end{alignat}
for structure coefficients $E^{\alpha\beta}_\gamma, F_{\alpha\beta}^\gamma, H^{\alpha\beta}_\gamma$ and $G^{\alpha}_{\beta\gamma}$, 
one finds that \eqref{eq:dual} leads to the equalities:
\begin{equation}
\label{eq:dualsolve}
H^{\alpha\beta}_\gamma = E^{\alpha\beta}_{\gamma}, \quad G^\alpha_{\beta\gamma} = F^{\alpha}_{\beta\gamma},
\end{equation}
matching the co-product of one algebra to the product of the other and vice versa.

Now, for matrix quantum groups such as $\Uqsu$ one defines the co-product \eqref{coprod} to match with how ordinary matrix multiplication would occur between 2 matrices with entries $a,b,c,d$ and $a',b',c',d'$, which mutually commute. Denoting the basis elements of this coordinate Hopf algebra as $P_\alpha$, the corresponding elements of the product matrix have the form
\begin{equation}
\label{eq:matprod}
\sum_{\alpha,\beta}E^{\alpha\beta}_\gamma P_\alpha P_\beta',
\end{equation}
where the structure coefficients $E^{\alpha\beta}_\gamma$ \eqref{eq:exp1} of the co-product were used. The prime denotes an entry taken from the second matrix in the matrix multiplication. 
For example for $\Uqsu$ with co-product \eqref{coprod}, e.g. the basis element $a^2b$ of the product matrix becomes $(aa'+bc')^2(ab'+bd')$.

We now define two objects $g_1,g_2$ as:
\begin{equation}
g_1 \equiv \sum_\alpha P_\alpha X^\alpha, \qquad g_2 \equiv \sum_\alpha P_\alpha' X^\alpha,
\end{equation}
and read this as an expansion in ``coordinates'' $P_\alpha,P_\alpha'$ and ``generators'' $X^\alpha$. Their product is of the form:
\begin{equation}
\label{eq:repgrou}
g_1 \, g_2 = \sum_{\alpha,\beta} P_\alpha P_\beta' X^\alpha X^\beta = \sum_{\gamma} \left( \sum_{\alpha,\beta }E^{\alpha\beta}_{\gamma} P_\alpha P_\beta' \right) X^\gamma, 
\end{equation}
where we used the product of the $X^\alpha$ ``generator'' algebra \eqref{eq:exp2}. The coordinates of the product matrix $g_1\,g_2$, $\sum_{\alpha,\beta }E^{\alpha\beta}_{\gamma} P_\alpha P_\beta'$, have precisely the form of \eqref{eq:matprod} and hence the $g_i$'s form a (co-)representation of the matrix quantum group.

We now go through the above construction with the following logic \cite{Fronsdal:1991gf}. Given a representation and a Hopf algebra generated by the $X^\alpha$, we can directly bootstrap the dual Hopf algebra and basis $P_\alpha$ (using the duality properties \eqref{eq:dual}). From this we can write down explicitly the $g_i$ which hence form a (co-)representation of the dual matrix quantum group.

Let us explicitly apply this strategy to the case at hand of $\Uqsu$ and $\Uqsual$. The Hopf algebra $\Uqsual$ is spanned by the basis elements  $X_\alpha=F^nH^kE^m$ where $\alpha$ labels the multi-index $(n,k,m)$. The dual Hopf algebra $\Uqsu$ is spanned by $\gamma^a (2\phi)^b \beta^c$. We need to change basis and find the (a priori unknown) dual basis $P_\alpha$, satisfying $\langle P_\alpha,X^\beta \rangle = \delta_\alpha^\beta$. Their precise form can be found by using the precise quantum algebra relations \eqref{eq:usal} and the duality conditions \eqref{eq:dual} leading to \cite{Fronsdal:1991gf}:
\begin{equation}
P_\alpha = C_{nkm}\gamma^n (2\phi)^k \beta^m, \qquad  C_{nkm}=\frac{1}{[n]_{q^{-2}}!}\frac{1}{k!}\frac{1}{[m]_{q^2}!}.
\end{equation}

This results in the decomposition \eqref{UqGE}:
\begin{equation}
\label{eq:gapp}
g \, \equiv \, \sum_\alpha P_\alpha X^\alpha \, = \, e_{q^{-2}}^{\gamma F} \, e^{2\phi H} \, e_{q^2}^{\beta E}\,,\qquad 0<q<1\,,
\end{equation}
where the ($q$-)exponentials are automatically obtained upon summing over all indices $(n,k,m)$. The first equality (definition) shows that $g$ is a specific element of both dual Hopf algebras $\Uqsu$ and $\Uqsual$ (depending on one's perspective on what to call ``coordinates'' and ``generators''), and not directly a generalization of a Lie group.

However, because of the above general property \eqref{eq:repgrou}, $g$ can also be thought of as a (co-)representation of the matrix quantum group $\Uqsu$, which is made more suggestive by the Gauss-like decomposition in the second equality. 
The object $g$ \eqref{eq:gapp} has many good properties, for example when we take $q\to 1$, it goes back to the undeformed Gauss decomposition. 
Most importantly for our application, this expression for $g$ allows us to construct representation theory of $\Uqsual$ on $\Uqsu$, because as shown in the main text, the generators of $\Uqsual$ acting on $g$ can be represented by $q$-calculus\cite{Klimyk:1997eb} on the non-commutative coordinates $(\gamma,\phi,\beta)$. 

\subsection{Gauss decomposition for pragmatists}
To give people a practical feeling how this Gauss decomposition and Hopf algebra can be viewed as a generalization of Lie groups into ``noncommutative'' geometries, we talk about the two-dimensional fundamental (co-)representation, where the quantum group SU$_q(1,1)$ is written as two-by-two matrices \cite{Klimyk:1997eb,jaganathan2000introduction}
\begin{equation}
    g=\begin{pmatrix}a & b 
    \\ c& d\end{pmatrix}
\end{equation}
where the entries $a,b,c,d$ are non-commutative entries satisfying \eqref{a4} plus an additional condition on the quantum determinant \eqref{qdet}.
 
The non-trivial statement in this definition is that this composes well under standard matrix multiplication: if $g_1$ and $g_2$ are defined with non-commutative entries $a_1, a_2$ etc. as above and
\begin{equation}
    g_1\cdot g_2=\begin{pmatrix}a_{12} & b_{12} 
    \\ c_{12}& d_{12}\end{pmatrix}\,,
\end{equation}
then $a_{12}, b_{12}, c_{12}, d_{12}$ satisfy the same algebra \eqref{a4} as  $a, b, c$ and $d$.

As in the classical case, it is convenient to parameterize the quantum group manifold in different ways. In particular, we choose
\begin{equation}
    g=\begin{pmatrix}e^\phi & e^\phi \beta \\\gamma e^\phi & e^{-\phi}+\g e^\phi \b\end{pmatrix}
\end{equation}
where the non-commutative coordinates $(\g, \phi, \beta)$ satisfy \eqref{cooalg}, which one can check is equivalent to \eqref{a4} and \eqref{qdet}. And this is actually the basis transformation we take in \eqref{transag}.

Now we connect SU$_{q}(1,1)$ to the $q$-deformed algebra $U_q(\mathfrak{su}(1,1))$. In the defining two-dimensional representation we have
\begin{equation}
    E=\begin{pmatrix}
        0&1\\0&0
    \end{pmatrix}\,,\quad F=\begin{pmatrix}
        0&0\\1&0
    \end{pmatrix}\,,\quad H=\frac{1}{2}\begin{pmatrix}
        1&0\\0&-1
    \end{pmatrix}\,,\label{a1}
\end{equation}
which indeed satisfy the algebra \eqref{eq:qalg}, and the quantum algebra and quantum group are related by the Gauss decomposition
\begin{equation}
    g\,=\,\begin{pmatrix}e^\phi & e^\phi \beta \\\gamma e^\phi & e^{-\phi}+\g e^\phi \b\end{pmatrix} \, = \,e_{q^{-2}}^{\gamma F} e^{2\phi H} e_{q^2}^{\beta E} \, = \, \begin{pmatrix}1 & 0 \\ \gamma  & 1\end{pmatrix} \begin{pmatrix}e^{\phi} & 0 \\0  & e^{-\phi}\end{pmatrix} \begin{pmatrix}1 & \beta \\0  & 1\end{pmatrix}\,.\label{a2}
\end{equation}
In the defining representation, this identity is somewhat trivial since both the quantum algebra and the $q$-exponentials are the same as in the undeformed $q\to 1$ case. 

Checking that the middle entry of \eqref{a2} (the Gauss decomposition) is the correct generalization to other representations takes more work 
\cite{Fronsdal:1991gf,Bonechi:1993sn,Morozov:1994ab,Jagannathan:1994cm,VanDerJeugt:1995yn}, but the idea is simple. For any finite dimension $n$, one can construct $(n \times n)$-dimensional matrices $E, F$ and $H$ which form a representation of the quantum algebra \eqref{eq:qalg} (see e.g. \cite{Klimyk:1997eb}). Then one checks that the resulting $(n \times n)$-dimensional matrix $g$ with entries that are polynomials in $\g, e^{\phi}$ and $\beta$ still composes properly under matrix multiplication, using the algebra of coordinates \eqref{cooalg}.\footnote{Which means checking it implies the co-product \eqref{coprod} with \eqref{transag}.} This fixes the Taylor series in $F$, $H$ and $E$ of the middle entry in \eqref{a2} and eventually leads to the $q$-exponentials.\footnote{In fact, since $F$ and $E$ are lower- resp. upper triangular, the Taylor series of the $q$-exponentials in the dimension $n$ irrep truncates at the $n^{th}$ term, allowing a clean ab initio determination of the series of the $q$-exponential order by order as one increases $n$.} The case $n=3$ is rather pedagogically detailed in \cite{jaganathan2000introduction}.

\subsection{Real forms and complex conjugation}
Finally a comment about $\Uqsu$ versus SL$_{q}(2,\mathbb{R})$. By definition, $\Uqsu$ corresponds to the real form of the aforementioned quantum group with $0<q<1$ real and SL$_{q}(2,\mathbb{R})$ is the real form of the quantum group with $\abs{q}=1$. Classically, there is a distinction between SU$(1,1)$ and SL$(2,\mathbb{R})$ by the way their elements behave under complex conjugation. This carries over to the $q$-deformed setup, and in fact determines the range of $q$ in these definitions.

In particular for $\Uqsu$ one defines ``complex conjugation'', by introducing a $*$ operation as:
\begin{equation}
    a^*=d\,, \quad b^*=qc\,,\quad c^*=q^{-1} b\,, \quad d^*=a\,.\label{a6}
\end{equation}
The non-trivial point is that this definition is consistent with the algebra \eqref{a4} only when $q^*=q$.
The defining 2d representation has the following property in common with the classical SU$(1,1)$:
\begin{equation}
    g^{\dagger} J\, g= J\,, \quad J=\begin{pmatrix}1 & 0 
    \\ 0& -1\end{pmatrix}\,,\quad g=\begin{pmatrix}a & b 
    \\ q^{-1}b^*& a^*\end{pmatrix}\,,\quad g^{\dagger}= \begin{pmatrix}a^* & q^{-1}b 
    \\ b^*& a\end{pmatrix}\,.
\end{equation}
Indeed, one finds
\begin{equation}
    g^{\dagger} J\, g=\begin{pmatrix}
        aa^*-q^{-2}bb^*&a^*b-q^{-1} b a^*\\b^*a-q^{-1}a b^*&b^*b-aa^*
    \end{pmatrix}=\begin{pmatrix}1 & 0 
    \\ 0& -1\end{pmatrix}=J\,,
\end{equation}
where in the second step we eliminated the conjugates using \eqref{a6} and then used \eqref{a4} to simplify the result. This definition of ``complex conjugation'' means in terms of our coordinates
\begin{equation}
    g^\dagger=\begin{pmatrix}e^{-\phi}+\g e^\phi \b & q^{-1}e^\phi \beta \\q\gamma e^\phi & e^{\phi}\end{pmatrix}=\begin{pmatrix}e^{-\phi}+\b e^\phi \g & \beta e^\phi  \\ e^\phi\gamma & e^{\phi}\end{pmatrix}=\begin{pmatrix}1 & \beta \\ 0  & 1\end{pmatrix} \begin{pmatrix}e^{-\phi} & 0 \\0  & e^{\phi}\end{pmatrix} \begin{pmatrix}1 & 0 \\\g  & 1\end{pmatrix}\,,
\end{equation}
which means one should not think of $\beta, e^{\phi}$ and $\g$ as simply ``real coordinates'', unlike for the quantum group SL$_{q}(2,\mathbb{R})$. This discussion about ``complex conjugation'' plays no role in the main text, we present it just to resolve potential confusions the reader might have about the distinction between $\Uqsu$ versus SL$_{q}(2,\mathbb{R})$.\footnote{Representation theory of $\Uqsl$ is actually far more common in gravitational contexts, see for instance \cite{Fan:2021bwt,Mertens:2022aou} for relations with the minimal string and the ubiquitous appearance of $\Uqsl$ in the context of 3d gravity \cite{Mertens:2022ujr,Mertens:2017mtv,Mertens:2018fds,Jackson:2014nla,Ponsot:1999uf,Ponsot:2000mt}.}

\section{Details on $\mathcal{N}=1$ quantum supergroup OSp$_q(1 \vert 2 ,\mathbb{R})$}
\label{app:detailssusy}
The U$_q(\mathfrak{osp}(1|2, \mathbb{R}))$ quantum superalgebra is generated by three elements $H\rvert F^+,F^-$ with
\begin{align}
\label{salgeba}
q^H F^\pm = q^{\pm\frac{1}{2}}F^{\pm} q^H \,, \qquad \{F^+, F^-\} = \frac{q^{2H} -q^{-2H}}{q-q^{-1}}\,.
\end{align}
The bosonic generators $E^\pm$ are defined through $\{F^\pm, F^\pm\} = \pm \frac{1}{2} E^{\pm}$, and are obsolete in the universal enveloping algebra, which is why one usually does not write them explicitly. They are important when parametrizing the (Hopf) dual quantum supergroup, as we show below.

One readily checks that the matrix generators in the fundamental representation satisfy \eqref{salgeba}
\begin{gather}
\label{defrep}
H = \left[\begin{array}{cc|c} 
1/2 & 0 & 0 \\
0 & -1/2 & 0 \\
\hline
0 & 0 & 0
\end{array} \right], \quad
F^- = \left[\begin{array}{cc|c} 
0 & 0 & 0 \\
0 & 0 & -q^{-1/2}/2 \\
\hline
1/2 & 0 & 0
\end{array} \right], \quad F^+ = \left[\begin{array}{cc|c} 
0 & 0 & 1/2 \\
0 & 0 & 0 \\
\hline
0 & q^{1/2}/2 & 0
\end{array} \right],
\end{gather}
The associated non-deformed supergroup OSp$(1\vert 2 ,\mathbb{R})$ has the fundamental representation:
\begin{equation}
\label{ospdef}
g = \left[\begin{array}{cc|c}
a & b & \alpha \\
c & d & \gamma \\ \hline
\beta & \delta & e
\end{array}\right],
\end{equation}
with relations
\begin{equation}
    \label{rela}
ad - bc - \delta\beta = 1\,, \quad  e^2 + 2\gamma\alpha = 1\,,\quad c\alpha - a\gamma - \beta e = 0\,, \quad  d\alpha - b\gamma - \delta e = 0\,.
\end{equation}

Classically, a group element can be written in terms of the Gauss-Euler decomposition as (see e.g. \cite{Fan:2021wsb} for a recent reference):\footnote{We have taken care not to swap the ordering of any of the variables. This is of course allowed for the classical supergroup (keeping in mind the Grassmann nature of some of the variables), but it is convenient to write it already in a way that is directly generalizable to non-commutative variables in the quantum group scenario.}
\begin{align}
\label{eq:GEsusy}
g= e^{2\theta_{\mm} F^-}e^{\gamma E^-}e^{2\phi H}e^{\beta E^+}e^{2\theta_{\+}F^+} 
= \left[\begin{array}{cc|c}
e^\phi & e^\phi \beta & e^\phi\theta_{\+} \\
\gamma e^\phi & e^{-\phi} + \gamma e^\phi \beta - \theta_{\mm}\theta_{\+} & \gamma e^\phi \theta_{\+} - q^{-1/2}\theta_{\mm} \\ \hline
\theta_{\mm} e^\phi & \theta_{\mm}e^\phi \beta + q^{1/2}\theta_{\+} & 1 + \theta_{\mm} e^\phi\theta_{\+}
\end{array}\right]\,.
\end{align}

The quantum supergroup OSp$_q(1 \vert 2 ,\mathbb{R})$ was defined in \cite{Kulish:1989sv}, where in particular the antipode $S(g)$ was written down:
\begin{equation}
S(g)
= \left[\begin{array}{cc|c}
d & -q^{-1}b & -q^{-1/2}\delta \\
-qc & a & q^{1/2}\beta \\ \hline
q^{1/2}\gamma & -q^{-1/2}\alpha & e
\end{array}\right], \qquad g \cdot S(g) = S(g) \cdot g = \mathbb{1}.
\end{equation}
Writing out the relations $g \cdot S(g) = S(g) \cdot g = \mathbb{1}$, one gets the $q$-deformed version of the relations \eqref{rela}:
\begin{alignat}{4}
\label{eqs:relqsusy}
&ad-qbc + q^{1/2}\alpha \gamma =1, \quad &&-q^{-1}ab +ba -q^{-1/2}\alpha^2=0, \quad &&-q^{-1/2}a \delta + q^{1/2} b \beta + \alpha e = 0, \nonumber \\
&cd -qdc +q^{1/2} \gamma^2 =0, \quad
&&-q^{-1}cb + da -q^{-1/2} \gamma \alpha = 1, \quad &&-q^{-1/2} c\delta +q^{1/2} d \beta + \gamma e =0, \nonumber \\
&\beta d -q \delta c + q^{1/2}e\gamma =0, \quad &&-q^{-1}\beta b + \delta a - q^{-1/2}e \alpha =0, \quad &&-q^{-1/2}\beta\delta + q^{1/2}\delta \beta +e^2 = 1, \nonumber \\
&da -q^{-1}bc-q^{-1/2}\delta \beta =1, \quad &&db-q^{-1}bd -q^{-1/2} \delta^2 = 0, \quad &&d\alpha -q^{-1}b\gamma -q^{-1/2}\delta e =0, \nonumber \\
&-qca +ac + q^{1/2} \beta^2 =0, \quad && -qcb +ad + q^{1/2}\beta \delta =1, \quad && -q c\alpha + a\gamma +q^{1/2}\beta e =0, \nonumber \\
&q^{1/2}\gamma a -q^{-1/2}\alpha c + e\beta =0, \quad &&q^{1/2} \gamma b -q^{-1/2} \alpha d +e\delta = 0, \quad &&q^{1/2}\gamma \alpha -q^{-1/2}\alpha \gamma +e^2 = 1.
\end{alignat}
These relations are not all independent, but can be conveniently solved in terms of non-commutative variables $\phi,\gamma,\beta \vert \theta_{\mm},\theta_{\+}$.
Using the parametrization \eqref{eq:GEsusy}, it is straightforward (but tedious) to show that the above relations \eqref{eqs:relqsusy} are satisfied if we impose the non-commutativity relations:
\begin{alignat}{4}
\label{eq:comma}
&e^\phi\gamma = q \gamma e^\phi, \qquad &&e^\phi \beta= q \beta e^\phi, \qquad &&[\beta,\gamma]=0, \nonumber\\
&e^\phi \theta_{\+} = q \theta_{\+} e^\phi, \quad &&e^\phi \theta_{\mm} = q \theta_{\mm} e^\phi, \qquad &&[\theta_{\pmt}, \beta] = 0 = [\theta_{\pmt}, \gamma], \nonumber\\
&\theta_{\+}\theta_{\mm} = - \theta_{\mm}\theta_{\+}, \qquad &&\left(\theta_{\pmt}\right)^2 = 0.
\end{alignat}
This is the resulted stated in section \ref{sect:5.5}.

\bibliographystyle{ourbst}
\bibliography{Refs}

\providecommand{\href}[2]{#2}\begingroup\raggedright\begin{thebibliography}{100}

\bibitem{Sachdev:1992fk}
S.~Sachdev and J.~Ye, ``{Gapless spin fluid ground state in a random, quantum Heisenberg magnet},'' \href{http://dx.doi.org/10.1103/PhysRevLett.70.3339}{{\em Phys. Rev. Lett.} {\bfseries 70} (1993) 3339},
\href{http://arxiv.org/abs/cond-mat/9212030}{{\ttfamily arXiv:cond-mat/9212030 [cond-mat]}}.

\bibitem{kitaev2015simple}
A.~Kitaev, ``A simple model of quantum holography,'' {\em Entanglement in Strongly-Correlated Quantum Matter} (2015) 38.

\bibitem{Maldacena:2016hyu}
J.~Maldacena and D.~Stanford, ``{Remarks on the Sachdev-Ye-Kitaev model},'' \href{http://dx.doi.org/10.1103/PhysRevD.94.106002}{{\em Phys. Rev. D} {\bfseries 94} no.~10, (2016) 106002}, \href{http://arxiv.org/abs/1604.07818}{{\ttfamily arXiv:1604.07818 [hep-th]}}.

\bibitem{Lin:2022rbf}
H.~W. Lin, ``{The bulk Hilbert space of double scaled SYK},'' \href{http://dx.doi.org/10.1007/JHEP11(2022)060}{{\em JHEP} {\bfseries 11} (2022) 060}, \href{http://arxiv.org/abs/2208.07032}{{\ttfamily arXiv:2208.07032 [hep-th]}}.

\bibitem{Klimyk:1997eb}
A.~Klimyk and K.~Schmudgen, {\em {Quantum groups and their representations}}.
\newblock Springer Berlin, Heidelberg, 1997.

\bibitem{Jackiw:1984je}
R.~Jackiw, ``{Lower Dimensional Gravity},''
\href{http://dx.doi.org/10.1016/0550-3213(85)90448-1}{{\em Nucl. Phys.} {\bfseries B252} (1985) 343--356}.

\bibitem{Teitelboim:1983ux}
C.~Teitelboim, ``{Gravitation and Hamiltonian Structure in Two Space-Time Dimensions},''
\href{http://dx.doi.org/10.1016/0370-2693(83)90012-6}{{\em Phys. Lett.} {\bfseries 126B} (1983) 41--45}.

\bibitem{Maldacena:2016upp}
J.~Maldacena, D.~Stanford, and Z.~Yang, ``{Conformal symmetry and its breaking in two dimensional Nearly Anti-de-Sitter space},'' \href{http://dx.doi.org/10.1093/ptep/ptw124}{{\em PTEP} {\bfseries 2016} no.~12, (2016) 12C104}, \href{http://arxiv.org/abs/1606.01857}{{\ttfamily arXiv:1606.01857 [hep-th]}}.

\bibitem{Engelsoy:2016xyb}
J.~Engels\"oy, T.~G. Mertens, and H.~Verlinde, ``{An investigation of AdS$_{2}$ backreaction and holography},'' \href{http://dx.doi.org/10.1007/JHEP07(2016)139}{{\em JHEP} {\bfseries 07} (2016) 139}, \href{http://arxiv.org/abs/1606.03438}{{\ttfamily arXiv:1606.03438 [hep-th]}}.

\bibitem{Jensen:2016pah}
K.~Jensen, ``{Chaos in AdS$_2$ Holography},'' \href{http://dx.doi.org/10.1103/PhysRevLett.117.111601}{{\em Phys. Rev. Lett.} {\bfseries 117} no.~11, (2016) 111601}, \href{http://arxiv.org/abs/1605.06098}{{\ttfamily arXiv:1605.06098 [hep-th]}}.

\bibitem{Mertens:2022irh}
T.~G. Mertens and G.~J. Turiaci, ``{Solvable Models of Quantum Black Holes: A Review on Jackiw-Teitelboim Gravity},'' \href{http://arxiv.org/abs/2210.10846}{{\ttfamily arXiv:2210.10846 [hep-th]}}.

\bibitem{Gross:2017hcz}
D.~J. Gross and V.~Rosenhaus, ``{The Bulk Dual of SYK: Cubic Couplings},'' \href{http://dx.doi.org/10.1007/JHEP05(2017)092}{{\em JHEP} {\bfseries 05} (2017) 092}, \href{http://arxiv.org/abs/1702.08016}{{\ttfamily arXiv:1702.08016 [hep-th]}}.

\bibitem{Das:2017pif}
S.~R. Das, A.~Jevicki, and K.~Suzuki, ``{Three Dimensional View of the SYK/AdS Duality},'' \href{http://dx.doi.org/10.1007/JHEP09(2017)017}{{\em JHEP} {\bfseries 09} (2017) 017}, \href{http://arxiv.org/abs/1704.07208}{{\ttfamily arXiv:1704.07208 [hep-th]}}.

\bibitem{Das:2017hrt}
S.~R. Das, A.~Ghosh, A.~Jevicki, and K.~Suzuki, ``{Three Dimensional View of Arbitrary $q$ SYK models},'' \href{http://dx.doi.org/10.1007/JHEP02(2018)162}{{\em JHEP} {\bfseries 02} (2018) 162}, \href{http://arxiv.org/abs/1711.09839}{{\ttfamily arXiv:1711.09839 [hep-th]}}.

\bibitem{Goel:2021wim}
A.~Goel and H.~Verlinde, ``{Towards a String Dual of SYK},'' \href{http://arxiv.org/abs/2103.03187}{{\ttfamily arXiv:2103.03187 [hep-th]}}.

\bibitem{Cotler:2016fpe}
J.~S. Cotler, G.~Gur-Ari, M.~Hanada, J.~Polchinski, P.~Saad, S.~H. Shenker, D.~Stanford, A.~Streicher, and M.~Tezuka, ``{Black Holes and Random Matrices},'' \href{http://dx.doi.org/10.1007/JHEP05(2017)118}{{\em JHEP} {\bfseries 05} (2017) 118}, \href{http://arxiv.org/abs/1611.04650}{{\ttfamily arXiv:1611.04650 [hep-th]}}. [Erratum: JHEP 09, 002 (2018)].

\bibitem{Berkooz:2018jqr}
M.~Berkooz, M.~Isachenkov, V.~Narovlansky, and G.~Torrents, ``{Towards a full solution of the large N double-scaled SYK model},'' \href{http://dx.doi.org/10.1007/JHEP03(2019)079}{{\em JHEP} {\bfseries 03} (2019) 079}, \href{http://arxiv.org/abs/1811.02584}{{\ttfamily arXiv:1811.02584 [hep-th]}}.

\bibitem{Berkooz:2018qkz}
M.~Berkooz, P.~Narayan, and J.~Simon, ``{Chord diagrams, exact correlators in spin glasses and black hole bulk reconstruction},'' \href{http://dx.doi.org/10.1007/JHEP08(2018)192}{{\em JHEP} {\bfseries 08} (2018) 192}, \href{http://arxiv.org/abs/1806.04380}{{\ttfamily arXiv:1806.04380 [hep-th]}}.

\bibitem{Mertens:2017mtv}
T.~G. Mertens, G.~J. Turiaci, and H.~L. Verlinde, ``{Solving the Schwarzian via the Conformal Bootstrap},'' \href{http://dx.doi.org/10.1007/JHEP08(2017)136}{{\em JHEP} {\bfseries 08} (2017) 136}, \href{http://arxiv.org/abs/1705.08408}{{\ttfamily arXiv:1705.08408 [hep-th]}}.

\bibitem{Yang:2018gdb}
Z.~Yang, ``{The Quantum Gravity Dynamics of Near Extremal Black Holes},'' \href{http://dx.doi.org/10.1007/JHEP05(2019)205}{{\em JHEP} {\bfseries 05} (2019) 205}, \href{http://arxiv.org/abs/1809.08647}{{\ttfamily arXiv:1809.08647 [hep-th]}}.

\bibitem{Mertens:2018fds}
T.~G. Mertens, ``{The Schwarzian theory \textemdash{} origins},'' \href{http://dx.doi.org/10.1007/JHEP05(2018)036}{{\em JHEP} {\bfseries 05} (2018) 036}, \href{http://arxiv.org/abs/1801.09605}{{\ttfamily arXiv:1801.09605 [hep-th]}}.

\bibitem{Kitaev:2018wpr}
A.~Kitaev and S.~J. Suh, ``{Statistical mechanics of a two-dimensional black hole},'' \href{http://dx.doi.org/10.1007/JHEP05(2019)198}{{\em JHEP} {\bfseries 05} (2019) 198}, \href{http://arxiv.org/abs/1808.07032}{{\ttfamily arXiv:1808.07032 [hep-th]}}.

\bibitem{Blommaert:2018oro}
A.~Blommaert, T.~G. Mertens, and H.~Verschelde, ``{The Schwarzian Theory - A Wilson Line Perspective},'' \href{http://dx.doi.org/10.1007/JHEP12(2018)022}{{\em JHEP} {\bfseries 12} (2018) 022}, \href{http://arxiv.org/abs/1806.07765}{{\ttfamily arXiv:1806.07765 [hep-th]}}.

\bibitem{Iliesiu:2019xuh}
L.~V. Iliesiu, S.~S. Pufu, H.~Verlinde, and Y.~Wang, ``{An exact quantization of Jackiw-Teitelboim gravity},'' \href{http://dx.doi.org/10.1007/JHEP11(2019)091}{{\em JHEP} {\bfseries 11} (2019) 091}, \href{http://arxiv.org/abs/1905.02726}{{\ttfamily arXiv:1905.02726 [hep-th]}}.

\bibitem{Saad:2019pqd}
P.~Saad, ``{Late Time Correlation Functions, Baby Universes, and ETH in JT Gravity},'' \href{http://arxiv.org/abs/1910.10311}{{\ttfamily arXiv:1910.10311 [hep-th]}}.

\bibitem{Jafferis:2022wez}
D.~L. Jafferis, D.~K. Kolchmeyer, B.~Mukhametzhanov, and J.~Sonner, ``{JT gravity with matter, generalized ETH, and Random Matrices},'' \href{http://arxiv.org/abs/2209.02131}{{\ttfamily arXiv:2209.02131 [hep-th]}}.

\bibitem{Susskind:2022bia}
L.~Susskind, ``{De Sitter Space, Double-Scaled SYK, and the Separation of Scales in the Semiclassical Limit},'' \href{http://arxiv.org/abs/2209.09999}{{\ttfamily arXiv:2209.09999 [hep-th]}}.

\bibitem{Bhattacharjee:2022ave}
B.~Bhattacharjee, P.~Nandy, and T.~Pathak, ``{Krylov complexity in large q and double-scaled SYK model},'' \href{http://dx.doi.org/10.1007/JHEP08(2023)099}{{\em JHEP} {\bfseries 08} (2023) 099}, \href{http://arxiv.org/abs/2210.02474}{{\ttfamily arXiv:2210.02474 [hep-th]}}.

\bibitem{Okuyama:2022szh}
K.~Okuyama, ``{Hartle-Hawking wavefunction in double scaled SYK},'' \href{http://dx.doi.org/10.1007/JHEP03(2023)152}{{\em JHEP} {\bfseries 03} (2023) 152}, \href{http://arxiv.org/abs/2212.09213}{{\ttfamily arXiv:2212.09213 [hep-th]}}.

\bibitem{Susskind:2023hnj}
L.~Susskind, ``{De Sitter Space has no Chords. Almost Everything is Confined.},'' \href{http://dx.doi.org/10.22128/jhap.2023.661.1043}{{\em JHAP} {\bfseries 3} no.~1, (2023) 1--30}, \href{http://arxiv.org/abs/2303.00792}{{\ttfamily arXiv:2303.00792 [hep-th]}}.

\bibitem{Mukhametzhanov:2023tcg}
B.~Mukhametzhanov, ``{Large p SYK from chord diagrams},'' \href{http://dx.doi.org/10.1007/JHEP09(2023)154}{{\em JHEP} {\bfseries 09} (2023) 154}, \href{http://arxiv.org/abs/2303.03474}{{\ttfamily arXiv:2303.03474 [hep-th]}}.

\bibitem{Berkooz:2023cqc}
M.~Berkooz, Y.~Jia, and N.~Silberstein, ``{Parisi's hypercube, Fock-space frustration and near-AdS$_2$/near-CFT$_1$ holography},'' \href{http://arxiv.org/abs/2303.18182}{{\ttfamily arXiv:2303.18182 [hep-th]}}.

\bibitem{Okuyama:2023bch}
K.~Okuyama and K.~Suzuki, ``{Correlators of double scaled SYK at one-loop},'' \href{http://dx.doi.org/10.1007/JHEP05(2023)117}{{\em JHEP} {\bfseries 05} (2023) 117}, \href{http://arxiv.org/abs/2303.07552}{{\ttfamily arXiv:2303.07552 [hep-th]}}.

\bibitem{Lin:2022nss}
H.~Lin and L.~Susskind, ``{Infinite Temperature's Not So Hot},'' \href{http://arxiv.org/abs/2206.01083}{{\ttfamily arXiv:2206.01083 [hep-th]}}.

\bibitem{Susskind:2021esx}
L.~Susskind, ``{Entanglement and Chaos in De Sitter Space Holography: An SYK Example},'' \href{http://dx.doi.org/10.22128/jhap.2021.455.1005}{{\em JHAP} {\bfseries 1} no.~1, (2021) 1--22}, \href{http://arxiv.org/abs/2109.14104}{{\ttfamily arXiv:2109.14104 [hep-th]}}.

\bibitem{Berkooz:2020xne}
M.~Berkooz, N.~Brukner, V.~Narovlansky, and A.~Raz, ``{The double scaled limit of Super--Symmetric SYK models},'' \href{http://dx.doi.org/10.1007/JHEP12(2020)110}{{\em JHEP} {\bfseries 12} (2020) 110}, \href{http://arxiv.org/abs/2003.04405}{{\ttfamily arXiv:2003.04405 [hep-th]}}.

\bibitem{Berkooz:2022mfk}
M.~Berkooz, M.~Isachenkov, M.~Isachenkov, P.~Narayan, and V.~Narovlansky, ``{Quantum groups, non-commutative AdS$_{2}$, and chords in the double-scaled SYK model},'' \href{http://dx.doi.org/10.1007/JHEP08(2023)076}{{\em JHEP} {\bfseries 08} (2023) 076}, \href{http://arxiv.org/abs/2212.13668}{{\ttfamily arXiv:2212.13668 [hep-th]}}.

\bibitem{Goel:2023svz}
A.~Goel, V.~Narovlansky, and H.~Verlinde, ``{Semiclassical geometry in double-scaled SYK},'' \href{http://dx.doi.org/10.1007/JHEP11(2023)093}{{\em JHEP} {\bfseries 11} (2023) 093}, \href{http://arxiv.org/abs/2301.05732}{{\ttfamily arXiv:2301.05732 [hep-th]}}.

\bibitem{Fukuyama:1985gg}
T.~Fukuyama and K.~Kamimura, ``{Gauge Theory of Two-dimensional Gravity},''
\href{http://dx.doi.org/10.1016/0370-2693(85)91322-X}{{\em Phys. Lett.} {\bfseries 160B} (1985) 259--262}.

\bibitem{Isler:1989hq}
K.~Isler and C.~A. Trugenberger, ``{A Gauge Theory of Two-dimensional Quantum Gravity},''
\href{http://dx.doi.org/10.1103/PhysRevLett.63.834}{{\em Phys. Rev. Lett.} {\bfseries 63} (1989) 834}.

\bibitem{Chamseddine:1989yz}
A.~H. Chamseddine and D.~Wyler, ``{Gauge Theory of Topological Gravity in (1+1)-Dimensions},''
\href{http://dx.doi.org/10.1016/0370-2693(89)90528-5}{{\em Phys. Lett.} {\bfseries B228} (1989) 75--78}.

\bibitem{Jackiw:1992bw}
R.~Jackiw, ``{Gauge theories for gravity on a line},'' \href{http://dx.doi.org/10.1007/BF01017075}{{\em Theor. Math. Phys.} {\bfseries 92} (1992) 979--987}, \href{http://arxiv.org/abs/hep-th/9206093}{{\ttfamily arXiv:hep-th/9206093}}.

\bibitem{Grumiller:2017qao}
D.~Grumiller, R.~McNees, J.~Salzer, C.~Valc\'arcel, and D.~Vassilevich, ``{Menagerie of AdS$_{2}$ boundary conditions},'' \href{http://dx.doi.org/10.1007/JHEP10(2017)203}{{\em JHEP} {\bfseries 10} (2017) 203}, \href{http://arxiv.org/abs/1708.08471}{{\ttfamily arXiv:1708.08471 [hep-th]}}.

\bibitem{Gonzalez:2018enk}
H.~A. Gonz\'alez, D.~Grumiller, and J.~Salzer, ``{Towards a bulk description of higher spin SYK},'' \href{http://dx.doi.org/10.1007/JHEP05(2018)083}{{\em JHEP} {\bfseries 05} (2018) 083}, \href{http://arxiv.org/abs/1802.01562}{{\ttfamily arXiv:1802.01562 [hep-th]}}.

\bibitem{Blommaert:2018iqz}
A.~Blommaert, T.~G. Mertens, and H.~Verschelde, ``{Fine Structure of Jackiw-Teitelboim Quantum Gravity},'' \href{http://dx.doi.org/10.1007/JHEP09(2019)066}{{\em JHEP} {\bfseries 09} (2019) 066}, \href{http://arxiv.org/abs/1812.00918}{{\ttfamily arXiv:1812.00918 [hep-th]}}.

\bibitem{Saad:2019lba}
P.~Saad, S.~H. Shenker, and D.~Stanford, ``{JT gravity as a matrix integral},'' \href{http://arxiv.org/abs/1903.11115}{{\ttfamily arXiv:1903.11115 [hep-th]}}.

\bibitem{Goel:2022pcu}
A.~Goel, {\em {Investigations of Holographic Duality in Two Dimensions}}.
\newblock PhD thesis, Princeton U., 2022.

\bibitem{HVerlindetalk}
H.~Verlinde, ``{Duality between SYK and 2+1 dimensional de Sitter}.'' Talks given at the QGQC5 conference, UC Davis, August 2019, the Franqui Symposium, Brussels, November 2019, at ‘Quantum Gravity on Southern Cone’, Argentina, December 2019, and ‘SYK models and Gauge Theory’ workshop at Weizmann Institute, December 2019.

\bibitem{Cattaneo:2001bp}
A.~S. Cattaneo and G.~Felder, ``{Poisson sigma models and deformation quantization},'' \href{http://dx.doi.org/10.1142/S0217732301003255}{{\em Mod. Phys. Lett. A} {\bfseries 16} (2001) 179--190}, \href{http://arxiv.org/abs/hep-th/0102208}{{\ttfamily arXiv:hep-th/0102208}}.

\bibitem{Ikeda:1993aj}
N.~Ikeda and K.~I. Izawa, ``{General form of dilaton gravity and nonlinear gauge theory},'' \href{http://dx.doi.org/10.1143/PTP.90.237}{{\em Prog. Theor. Phys.} {\bfseries 90} (1993) 237--246}, \href{http://arxiv.org/abs/hep-th/9304012}{{\ttfamily arXiv:hep-th/9304012}}.

\bibitem{Ikeda:1993fh}
N.~Ikeda, ``{Two-dimensional gravity and nonlinear gauge theory},'' \href{http://dx.doi.org/10.1006/aphy.1994.1104}{{\em Annals Phys.} {\bfseries 235} (1994) 435--464}, \href{http://arxiv.org/abs/hep-th/9312059}{{\ttfamily arXiv:hep-th/9312059}}.

\bibitem{Rahman:2022jsf}
A.~A. Rahman, ``{dS JT Gravity and Double-Scaled SYK},'' \href{http://arxiv.org/abs/2209.09997}{{\ttfamily arXiv:2209.09997 [hep-th]}}.

\bibitem{Susskind:2022dfz}
L.~Susskind, ``{Scrambling in Double-Scaled SYK and De Sitter Space},'' \href{http://arxiv.org/abs/2205.00315}{{\ttfamily arXiv:2205.00315 [hep-th]}}.

\bibitem{wip}
A.~Blommaert, T.~G. Mertens, and S.~Yao, ``{Work in progress},''.

\bibitem{Blommaert:2023wad}
A.~Blommaert, T.~G. Mertens, and S.~Yao, ``{The q-Schwarzian and Liouville gravity},'' \href{http://arxiv.org/abs/2312.00871}{{\ttfamily arXiv:2312.00871 [hep-th]}}.

\bibitem{Fan:2021bwt}
Y.~Fan and T.~G. Mertens, ``{From quantum groups to Liouville and dilaton quantum gravity},'' \href{http://dx.doi.org/10.1007/JHEP05(2022)092}{{\em JHEP} {\bfseries 05} (2022) 092}, \href{http://arxiv.org/abs/2109.07770}{{\ttfamily arXiv:2109.07770 [hep-th]}}.

\bibitem{yao2018edge}
S.~Yao and Z.~Wang, ``Edge states and topological invariants of non-hermitian systems,'' {\em Physical review letters} {\bfseries 121} no.~8, (2018) 086803.

\bibitem{Coussaert:1995zp}
O.~Coussaert, M.~Henneaux, and P.~van Driel, ``{The Asymptotic dynamics of three-dimensional Einstein gravity with a negative cosmological constant},'' \href{http://dx.doi.org/10.1088/0264-9381/12/12/012}{{\em Class. Quant. Grav.} {\bfseries 12} (1995) 2961--2966}, \href{http://arxiv.org/abs/gr-qc/9506019}{{\ttfamily arXiv:gr-qc/9506019}}.

\bibitem{dirac2001lectures}
P.~A.~M. Dirac, {\em Lectures on quantum mechanics}, vol.~2.
\newblock Courier Corporation, 2001.

\bibitem{Bagrets:2016cdf}
D.~Bagrets, A.~Altland, and A.~Kamenev, ``{Sachdev\textendash{}Ye\textendash{}Kitaev model as Liouville quantum mechanics},'' \href{http://dx.doi.org/10.1016/j.nuclphysb.2016.08.002}{{\em Nucl. Phys. B} {\bfseries 911} (2016) 191--205}, \href{http://arxiv.org/abs/1607.00694}{{\ttfamily arXiv:1607.00694 [cond-mat.str-el]}}.

\bibitem{Harlow:2018tqv}
D.~Harlow and D.~Jafferis, ``{The Factorization Problem in Jackiw-Teitelboim Gravity},'' \href{http://dx.doi.org/10.1007/JHEP02(2020)177}{{\em JHEP} {\bfseries 02} (2020) 177}, \href{http://arxiv.org/abs/1804.01081}{{\ttfamily arXiv:1804.01081 [hep-th]}}.

\bibitem{Lin:2023trc}
H.~W. Lin and D.~Stanford, ``{A symmetry algebra in double-scaled SYK},'' \href{http://dx.doi.org/10.21468/SciPostPhys.15.6.234}{{\em SciPost Phys.} {\bfseries 15} no.~6, (2023) 234}, \href{http://arxiv.org/abs/2307.15725}{{\ttfamily arXiv:2307.15725 [hep-th]}}.

\bibitem{Fronsdal:1991gf}
C.~Fronsdal and A.~Galindo, ``{The Dual of a quantum group},'' \href{http://dx.doi.org/10.1007/BF00739590}{{\em Lett. Math. Phys.} {\bfseries 27} (1993) 59--72}.

\bibitem{jaganathan2000introduction}
R.~Jaganathan, ``{An Introduction to quantum algebras and their applications},'' \href{http://arxiv.org/abs/math-ph/0003018}{{\ttfamily arXiv:math-ph/0003018}}.

\bibitem{Mertens:2022aou}
T.~G. Mertens, ``{Quantum exponentials for the modular double and applications in gravity models},'' \href{http://dx.doi.org/10.1007/JHEP09(2023)106}{{\em JHEP} {\bfseries 09} (2023) 106}, \href{http://arxiv.org/abs/2212.07696}{{\ttfamily arXiv:2212.07696 [hep-th]}}.

\bibitem{brown1986central}
J.~D. Brown and M.~Henneaux, ``Central charges in the canonical realization of asymptotic symmetries: an example from three dimensional gravity,'' {\em Communications in Mathematical Physics} {\bfseries 104} (1986) 207--226.

\bibitem{mostafazadeh2002pseudo}
A.~Mostafazadeh, ``{Pseudo-Hermiticity versus PT symmetry: the necessary condition for the reality of the spectrum of a non-Hermitian Hamiltonian},'' {\em Journal of Mathematical Physics} {\bfseries 43} no.~1, (2002) 205--214.

\bibitem{Ip}
I.~C.-H. Ip, ``{Representation of the quantum plane, its quantum double and harmonic analysis on $GL_q^+(2,R)$},'' \href{http://dx.doi.org/10.1007/s00029-012-0112-4}{{\em Selecta Mathematica New Series} {\bfseries Vol 19 (4)} (2013) 987--1082}, \href{http://arxiv.org/abs/1108.5365}{{\ttfamily arXiv:1108.5365 [math.QA]}}.

\bibitem{cmp/1103908695}
F.~A. Berezin and V.~N. Tolstoy, ``{The group with Grassmann structure ${\rm UOSP}(1.2)$},'' \href{http://dx.doi.org/cmp/1103908695}{{\em Communications in Mathematical Physics} {\bfseries 78} no.~3, (1980) 409--428}.

\bibitem{sevostyanov1999quantum}
A.~Sevostyanov, ``{Quantum deformation of Whittaker modules and Toda lattice},'' \href{http://arxiv.org/abs/math/9905128}{{\ttfamily arXiv:math/9905128 [math.QA]}}.

\bibitem{Kharchev:2001rs}
S.~Kharchev, D.~Lebedev, and M.~Semenov-Tian-Shansky, ``{Unitary representations of U(q) (sl(2, R)), the modular double, and the multiparticle q deformed Toda chains},'' \href{http://dx.doi.org/10.1007/s002200100592}{{\em Commun. Math. Phys.} {\bfseries 225} (2002) 573--609}, \href{http://arxiv.org/abs/hep-th/0102180}{{\ttfamily arXiv:hep-th/0102180}}.

\bibitem{backgammon1967functions}
H.~Jacquet, ``Whittaker functions associated with chevalley groups,'' {\em Bulletin of the math{\'e}matic society of France} {\bfseries 95} (1967) 243--309.

\bibitem{hashizume1979whittaker}
M.~Hashizume, ``Whittaker models for real reductive groups,'' {\em Japanese journal of mathematics. New series} {\bfseries 5} no.~2, (1979) 349--401.

\bibitem{hashizume1982whittaker}
M.~Hashizume, ``Whittaker functions on semisimple lie groups,'' {\em Hiroshima Mathematical Journal} {\bfseries 12} no.~2, (1982) 259--293.

\bibitem{witten1989quantum}
E.~Witten, ``Quantum field theory and the jones polynomial,'' {\em Communications in Mathematical Physics} {\bfseries 121} no.~3, (1989) 351--399.

\bibitem{elitzur1989remarks}
S.~Elitzur, G.~Moore, A.~Schwimmer, and N.~Seiberg, ``Remarks on the canonical quantization of the chern-simons-witten theory,'' {\em Nuclear Physics B} {\bfseries 326} no.~1, (1989) 108--134.

\bibitem{witten1991quantum}
E.~Witten, ``On quantum gauge theories in two dimensions,'' {\em Communications in Mathematical Physics} {\bfseries 141} no.~1, (1991) 153--209.

\bibitem{Witten:1992xu}
E.~Witten, ``{Two-dimensional gauge theories revisited},'' \href{http://dx.doi.org/10.1016/0393-0440(92)90034-X}{{\em J. Geom. Phys.} {\bfseries 9} (1992) 303--368}, \href{http://arxiv.org/abs/hep-th/9204083}{{\ttfamily arXiv:hep-th/9204083}}.

\bibitem{Cordes:1994fc}
S.~Cordes, G.~W. Moore, and S.~Ramgoolam, ``{Lectures on 2-d Yang-Mills theory, equivariant cohomology and topological field theories},'' \href{http://dx.doi.org/10.1016/0920-5632(95)00434-B}{{\em Nucl. Phys. B Proc. Suppl.} {\bfseries 41} (1995) 184--244}, \href{http://arxiv.org/abs/hep-th/9411210}{{\ttfamily arXiv:hep-th/9411210}}.

\bibitem{Blommaert:2018oue}
A.~Blommaert, T.~G. Mertens, and H.~Verschelde, ``{Edge dynamics from the path integral \textemdash{} Maxwell and Yang-Mills},'' \href{http://dx.doi.org/10.1007/JHEP11(2018)080}{{\em JHEP} {\bfseries 11} (2018) 080}, \href{http://arxiv.org/abs/1804.07585}{{\ttfamily arXiv:1804.07585 [hep-th]}}.

\bibitem{Cotler:2020ugk}
J.~Cotler and K.~Jensen, ``{AdS$_{3}$ gravity and random CFT},'' \href{http://dx.doi.org/10.1007/JHEP04(2021)033}{{\em JHEP} {\bfseries 04} (2021) 033}, \href{http://arxiv.org/abs/2006.08648}{{\ttfamily arXiv:2006.08648 [hep-th]}}.

\bibitem{Eberhardt:2022wlc}
L.~Eberhardt, ``{Off-shell Partition Functions in 3d Gravity},'' \href{http://arxiv.org/abs/2204.09789}{{\ttfamily arXiv:2204.09789 [hep-th]}}.

\bibitem{Blommaert:2018rsf}
A.~Blommaert, T.~G. Mertens, H.~Verschelde, and V.~I. Zakharov, ``{Edge State Quantization: Vector Fields in Rindler},'' \href{http://dx.doi.org/10.1007/JHEP08(2018)196}{{\em JHEP} {\bfseries 08} (2018) 196}, \href{http://arxiv.org/abs/1801.09910}{{\ttfamily arXiv:1801.09910 [hep-th]}}.

\bibitem{olshanetsky2002unitary}
M.~A. Olshanetsky and V.-B. Rogov, ``Unitary representations of the quantum lorentz group and quantum relativistic toda chain,'' {\em Theoretical and mathematical physics} {\bfseries 130} no.~3, (2002) 299--322.

\bibitem{Morariu:2001dv}
B.~Morariu and A.~P. Polychronakos, ``{Quantum mechanics on the noncommutative torus},'' \href{http://dx.doi.org/10.1016/S0550-3213(01)00294-2}{{\em Nucl. Phys. B} {\bfseries 610} (2001) 531--544}, \href{http://arxiv.org/abs/hep-th/0102157}{{\ttfamily arXiv:hep-th/0102157}}.

\bibitem{Mertens:2022ujr}
T.~G. Mertens, J.~Sim\'on, and G.~Wong, ``{A proposal for 3d quantum gravity and its bulk factorization},'' \href{http://dx.doi.org/10.1007/JHEP06(2023)134}{{\em JHEP} {\bfseries 06} (2023) 134}, \href{http://arxiv.org/abs/2210.14196}{{\ttfamily arXiv:2210.14196 [hep-th]}}.

\bibitem{Witten:2020ert}
E.~Witten, ``{Deformations of JT Gravity and Phase Transitions},'' \href{http://arxiv.org/abs/2006.03494}{{\ttfamily arXiv:2006.03494 [hep-th]}}.

\bibitem{Anninos:2017hhn}
D.~Anninos and D.~M. Hofman, ``{Infrared Realization of dS$_2$ in AdS$_2$},'' \href{http://dx.doi.org/10.1088/1361-6382/aab143}{{\em Class. Quant. Grav.} {\bfseries 35} no.~8, (2018) 085003}, \href{http://arxiv.org/abs/1703.04622}{{\ttfamily arXiv:1703.04622 [hep-th]}}.

\bibitem{Anninos:2020cwo}
D.~Anninos and D.~A. Galante, ``{Constructing AdS$_{2}$ flow geometries},'' \href{http://dx.doi.org/10.1007/JHEP02(2021)045}{{\em JHEP} {\bfseries 02} (2021) 045}, \href{http://arxiv.org/abs/2011.01944}{{\ttfamily arXiv:2011.01944 [hep-th]}}.

\bibitem{Anninos:2022qgy}
D.~Anninos, D.~A. Galante, and S.~U. Sheorey, ``{Renormalisation group flows of deformed SYK models},'' \href{http://dx.doi.org/10.1007/JHEP11(2023)197}{{\em JHEP} {\bfseries 11} (2023) 197}, \href{http://arxiv.org/abs/2212.04944}{{\ttfamily arXiv:2212.04944 [hep-th]}}.

\bibitem{StanfordSeiberg}
D.~Stanford and N.~Seiberg, ``{unpublished},'' 2019.

\bibitem{Mertens:2020hbs}
T.~G. Mertens and G.~J. Turiaci, ``{Liouville quantum gravity -- holography, JT and matrices},'' \href{http://dx.doi.org/10.1007/JHEP01(2021)073}{{\em JHEP} {\bfseries 01} (2021) 073}, \href{http://arxiv.org/abs/2006.07072}{{\ttfamily arXiv:2006.07072 [hep-th]}}.

\bibitem{Teschner:1995yf}
J.~Teschner, ``{On the Liouville three point function},'' \href{http://dx.doi.org/10.1016/0370-2693(95)01200-A}{{\em Phys. Lett. B} {\bfseries 363} (1995) 65--70}, \href{http://arxiv.org/abs/hep-th/9507109}{{\ttfamily arXiv:hep-th/9507109}}.

\bibitem{ORaifeartaigh:1998hjm}
L.~O'Raifeartaigh, J.~M. Pawlowski, and V.~V. Sreedhar, ``{Duality in quantum Liouville theory},'' \href{http://dx.doi.org/10.1006/aphy.1999.5951}{{\em Annals Phys.} {\bfseries 277} (1999) 117--143}, \href{http://arxiv.org/abs/hep-th/9811090}{{\ttfamily arXiv:hep-th/9811090}}.

\bibitem{ORaifeartaigh:2000dkf}
L.~O'Raifeartaigh, J.~M. Pawlowski, and V.~V. Sreedhar, ``{The Two exponential Liouville theory and the uniqueness of the three point function},'' \href{http://dx.doi.org/10.1016/S0370-2693(00)00448-2}{{\em Phys. Lett. B} {\bfseries 481} (2000) 436--444}, \href{http://arxiv.org/abs/hep-th/0003247}{{\ttfamily arXiv:hep-th/0003247}}.

\bibitem{Teschner:2001rv}
J.~Teschner, ``{Liouville theory revisited},'' \href{http://dx.doi.org/10.1088/0264-9381/18/23/201}{{\em Class. Quant. Grav.} {\bfseries 18} (2001) R153--R222}, \href{http://arxiv.org/abs/hep-th/0104158}{{\ttfamily arXiv:hep-th/0104158}}.

\bibitem{Fu:2016vas}
W.~Fu, D.~Gaiotto, J.~Maldacena, and S.~Sachdev, ``{Supersymmetric Sachdev-Ye-Kitaev models},'' \href{http://dx.doi.org/10.1103/PhysRevD.95.026009}{{\em Phys. Rev. D} {\bfseries 95} no.~2, (2017) 026009}, \href{http://arxiv.org/abs/1610.08917}{{\ttfamily arXiv:1610.08917 [hep-th]}}. [Addendum: Phys.Rev.D 95, 069904 (2017)].

\bibitem{JLukierski_1992}
J.~Lukierski and A.~Nowicki, ``{Real forms of U$_q(OSp(1 \vert 2)$) and quantum D=2 supersymmetry algebras},'' \href{http://dx.doi.org/10.1088/0305-4470/25/4/003}{{\em Journal of Physics A: Mathematical and General} {\bfseries 25} (1992) L161}.

\bibitem{Bonechi:1993sn}
F.~Bonechi, E.~Celeghini, R.~Giachetti, C.~M. Perena, E.~Sorace, and M.~Tarlini, ``{Exponential mapping for nonsemisimple quantum groups},'' \href{http://dx.doi.org/10.1088/0305-4470/27/4/023}{{\em J. Phys. A} {\bfseries 27} (1994) 1307--1316}, \href{http://arxiv.org/abs/hep-th/9311114}{{\ttfamily arXiv:hep-th/9311114}}.

\bibitem{Morozov:1994ab}
A.~Morozov and L.~Vinet, ``{Free field representation of group element for simple quantum groups},'' \href{http://dx.doi.org/10.1142/S0217751X9800072X}{{\em Int. J. Mod. Phys. A} {\bfseries 13} (1998) 1651--1708}, \href{http://arxiv.org/abs/hep-th/9409093}{{\ttfamily arXiv:hep-th/9409093}}.

\bibitem{Jagannathan:1994cm}
R.~Jagannathan and J.~Van~der Jeugt, ``{Finite dimensional representations of the quantum group $GL_{p,q}(2)$ using the exponential map from $U_{p,q}(gl(2))$},'' \href{http://dx.doi.org/10.1088/0305-4470/28/10/013}{{\em J. Phys. A} {\bfseries 28} (1995) 2819--2832}, \href{http://arxiv.org/abs/hep-th/9411200}{{\ttfamily arXiv:hep-th/9411200}}.

\bibitem{VanDerJeugt:1995yn}
J.~Van Der~Jeugt and R.~Jagannathan, ``{The Exponential map for representations of $U_{p,q}(gl(2))$},'' \href{http://dx.doi.org/10.1007/BF01688821}{{\em Czech. J. Phys.} {\bfseries 46} (1996) 269}, \href{http://arxiv.org/abs/q-alg/9507009}{{\ttfamily arXiv:q-alg/9507009}}.

\bibitem{Jackson:2014nla}
S.~Jackson, L.~McGough, and H.~Verlinde, ``{Conformal Bootstrap, Universality and Gravitational Scattering},'' \href{http://dx.doi.org/10.1016/j.nuclphysb.2015.10.013}{{\em Nucl. Phys. B} {\bfseries 901} (2015) 382--429}, \href{http://arxiv.org/abs/1412.5205}{{\ttfamily arXiv:1412.5205 [hep-th]}}.

\bibitem{Ponsot:1999uf}
B.~Ponsot and J.~Teschner, ``{Liouville bootstrap via harmonic analysis on a noncompact quantum group},'' \href{http://arxiv.org/abs/hep-th/9911110}{{\ttfamily arXiv:hep-th/9911110}}.

\bibitem{Ponsot:2000mt}
B.~Ponsot and J.~Teschner, ``{Clebsch-Gordan and Racah-Wigner coefficients for a continuous series of representations of U(q)(sl(2,R))},'' \href{http://dx.doi.org/10.1007/PL00005590}{{\em Commun. Math. Phys.} {\bfseries 224} (2001) 613--655}, \href{http://arxiv.org/abs/math/0007097}{{\ttfamily arXiv:math/0007097}}.

\bibitem{Fan:2021wsb}
Y.~Fan and T.~G. Mertens, ``{Supergroup structure of Jackiw-Teitelboim supergravity},'' \href{http://dx.doi.org/10.1007/JHEP08(2022)002}{{\em JHEP} {\bfseries 08} (2022) 002}, \href{http://arxiv.org/abs/2106.09353}{{\ttfamily arXiv:2106.09353 [hep-th]}}.

\bibitem{Kulish:1989sv}
P.~P. Kulish and N.~Y. Reshetikhin, ``{Universal R matrix of the quantum superalgebra osp(2 $\vert$ 1)},'' \href{http://dx.doi.org/10.1007/BF00401868}{{\em Lett. Math. Phys.} {\bfseries 18} (1989) 143--149}.

\end{thebibliography}\endgroup

\end{document}